\documentclass{jfm}

\usepackage{graphicx}
\usepackage{newtxtext}
\usepackage{newtxmath}
\usepackage{amsmath}
\usepackage{enumerate}
\usepackage{tikz}
\usepackage{natbib}
\usepackage[justification=justified]{caption}
\usepackage{hyperref}
\usetikzlibrary{calc,patterns,angles,quotes}

\definecolor{celestialblue}{rgb}{0.29, 0.59, 0.82}

\hypersetup{
    colorlinks = true,
    urlcolor   = blue,
    citecolor  = black,
}

\newcommand{\RomanNumeralCaps}[1]

\title{Linear Stability Analysis of Oblique Couette-Poiseuille flows}

\author{Muhammad Abdullah\aff{1},
  George I. Park\aff{1} \corresp{\email{gipark@seas.upenn.edu}}}

\affiliation{\aff{1} Department of Mechanical Engineering, University of Pennsylvania, PA, 19104, US}

\begin{document}
\maketitle

\begin{abstract}

We perform a detailed numerical study of modal and non-modal stability in oblique Couette-Poiseuille profiles, which are among the simplest examples of three-dimensional boundary layers. Through a comparison with the Orr-Sommerfeld operator for the aligned case, we show how an effective wall speed succinctly characterizes modal stability. Large-scale parameter sweeps reveal that the misalignment between the pressure gradient and wall motion is, in general, destabilizing. For flows that are sufficiently oblique, the instability is found to depend exclusively on the direction of wall motion and not on its speed, a conclusion supported, in part, by the perturbation energy budget and the evolution of the critical layers. Closed forms for the critical parameters in this regime are derived using a simple analysis. Finally, a modified long-wavelength approximation is developed, and the resulting asymptotic eigenvalue problem is used to show that there is no cutoff wall speed for unconditional stability whenever the angle of wall motion is non-zero, in stark contrast to the aligned case. From a non-modal perspective, pseudo-resonance is examined through the resolvent and the $\epsilon$-pseudospectra. An analysis of the unforced initial value problem shows that the maximum energy gain is highly dependent on both the magnitude and direction of the wall velocity. However, the strongest amplification is always achieved for configurations that are only weakly skewed. Finally, the optimal perturbations appear to develop via a lift-up effect induced by an Orr-like mechanism.

\end{abstract}

\begin{keywords}
Authors should not enter keywords on the manuscript, as these must be chosen by the author during the online suboldsymbolission process and will then be added during the typesetting process (see \href{https://www.cambridge.org/core/journals/journal-of-fluid-mechanics/information/list-of-keywords}{Keyword PDF} for the full list).  Other classifications will be added at the same time.
\end{keywords}

{\bf MSC Codes }  {\it(Optional)} Please enter your MSC Codes here

\section{Introduction}

Shear flows demonstrate rich dynamical behavior and underpin a variety of technological applications, ranging from microfluidics and turbo-machinery to large-scale aerodynamics and meteorology. Quantifying the transition to turbulence in these flows is a multi-faceted problem, and despite much concentrated effort in the past few decades, a unified description is yet to be agreed upon. The state-of-the-art on the topic can be found in references such as \cite{Kerswell_2005, MANNEVILLE2015345, Barkley_2016, eckhardt_review1, avila_2023}. Unfortunately, the insight afforded by prevailing theories is rather limited since a vast majority, primarily for the sake of simplification, focus on flows that are statistically two-dimensional, with a mean flow direction independent of the wall-normal coordinate. In contrast, most practical flow scenarios suffer from non-equilibrium effects that introduce anisotropy and symmetry-breaking, inducing a \textit{three-dimensional boundary layer} described by a skewing of the mean velocity vectors and a flow direction that varies as a non-constant function along the wall-normal axis \citep{flack}. In this work, we systematically examine the stability of a relatively underexplored class of three-dimensional internal flows that are both simple in their description and physically representative in their context.

Usually, the investigation of instability in a fluid system derives from the Navier-Stokes equations (NSEs), appropriately linearized around some time-invariant state of interest. The classical (modal) approach focuses on the eigenvalues of the associated linear operator, and the flow is considered unstable with respect to some dimensionless quantity -- usually a Reynolds number $Re$ -- if there exists a mode with a positive growth rate. These disturbances are capable of achieving exponential-in-time amplification, a type of so-called primary instability, before saturating due to non-linear phenomena. The result is either transition or the development of a steady, modified, base flow susceptible to secondary instabilities. To varying degrees of success, this framework has been applied to numerous laminar profiles, such as rectilinear pressure-driven flow \citep{thomas_ppf, Gage_Reid_1968, orszag_1971, Zhang_Lashgari_Zaki_Brandt_2013}, plane-Couette flow \citep{drazin_reid_2004, couette_rarefied}, classic Rayleigh-Bénard convection \citep{rb_conv, chandrasekhar}, Hagen-Poiseuille flow \citep{Salwen_Cotton_Grosch_1980, schmidstability}, and the asymptotic suction boundary layer \citep{Hughes_Reid_1965_asbl, FRANSSON_ALFREDSSON_2003_asbl}. 
Contextualizing these calculations against experimental observations, however, is not always straightforward, particularly because the most unstable disturbance, as predicted by modal theory, can only be realized in asymptotic time. On the other hand, significant non-modal energy growth can occur on much shorter timescales and is, therefore, not captured \citep{tref_pseudospec, tref_pseudo_siam, schmidstability, schmid_annrev}. A potential model for this behavior lies in the non-normality of the linearized NSE operator and its, in general, non-orthogonal eigenfunctions. In particular, within a basis expansion, the contribution of these modes can grow or decay at different rates, allowing for a transient amplification of energy. In many canonical flows and their variants, this non-modal growth has been shown to be substantial, even in linearly stable parameter regimes \citep{farrell_1988, reddyschmidhenn, Schmid_Henningson_1994, tc_flow_transient, tc_meseguer, MESEGUER2003178, Liu_Liu_2012}.


A flow that has arguably received limited attention in the general fluids literature is the plane Couette-Poiseuille (PCP) flow, generated by the interaction between a pressure gradient and the prescribed motion of one or both boundaries. PCP configurations are standard in geo-physical fluid mechanics (for example, in modeling asthenospheric counterflows; see \cite{Turcotte_Schubert_2002}), flow in ducts \citep{ducts}, and tribology. Comprehensive stability analyses for PCP flows are somewhat sparse in number, and the first few treatments can be found in \cite{potter_1966, Reynolds_Potter_1967, hains_pcp}. Described by a wall speed $\xi$ often made non-dimensional with the Poiseuille maximum, the superposition of a (parallel) Couette component with an otherwise pressure-driven flow is, in general, stabilizing, at least in terms of a critical Reynolds number $Re_c$ below which modal instability is absent. Furthermore, beyond a threshold value for $\xi$, the base flow is capable of achieving complete modal stability against infinitesimal perturbations, $Re_c\to\infty$. \cite{cowley_smith_1985}, using a weakly non-linear analysis, determined this ``cutoff" velocity to be $\xi\approx0.7$. From a non-modal perspective, \cite{bergstrom} showed that the peak in transient energy amplification can depend heavily on the relative influence of the Poiseuille and Couette components. Modifications to the geometry and rheology of PCP flows have also been considered, and their linear response analyzed, for example, in \cite{nouar, GUHA_2010, CHOKSHI201517, SAMANTA2020103160}. More recently, using a zero-mean advection velocity setup, \cite{klotz} experimentally probed the sub-critical transition in PCP flows. Uniform turbulence originating from a natural transition was observed beyond $Re\approx 780$ (with the Reynolds number based on the wall velocity), which stands in reasonable agreement with the results of \cite{Tsanis_Leutheusser_1988}.

Despite their individual merits, we note that the previous literature on the transitional regime in PCP flows is somewhat restrictive in its applicability. For convenience in modeling or due to experimental limitations, the pressure gradient and wall velocity vectors are almost always assumed to be perfectly co-incident. Although this uni-directional assumption enables a fairly tractable analysis, it breaks down in more practical scenarios, e.g., wind-ocean interactions, where the direction of the pressure gradient in the bulk flow need not be aligned with that of the wind shear. In these cases, a \textit{crossflow} must be taken into account, and the flow angle cannot be assumed to be constant, potentially affecting, among other aspects, the onset of instability. Therefore, the primary goal of this work is to contribute to this gap.

We note that linear instability in three-dimensional boundary layers has been the subject of extensive prior investigation, with excellent reviews published in \cite{reed_saric} and \cite{reed_saric_white}. In most flow situations, the introduction of mean three-dimensionality allows for multiple competing mechanisms for transition. For example, in flows over swept wings, local pressure gradients induce a near-wall crossflow component that is inflectional and, therefore, unstable to the inviscid amplification of the so-called crossflow modes \citep{gray1952nature, greg_stuart_walker, BIPPES1999363}. These modes are co-rotating and, following non-linear saturation, incite breakdown to turbulence via high-frequency secondary instabilities \citep{WHITE_SARIC_2005}. Such crossflow instabilities are also common in, for example, the boundary layers that form on rotating disks \citep{federov, kobayashi_1980, mujeeb_crossflow}. In particular, using von Kármán's similarity solution for a swirling flow over an infinitely long rotating disk, \cite{mwo} determined the critical Reynolds number associated with these crossflow modes to be $Re_c\approx 170$. Taking into account the streamline curvature and Coriolis effects, the authors refined this estimate in the same work to $Re_c\approx 290$, which was in better agreement with their experimental results. Later, \cite{Malik_1986} and \cite{bala_and_malik} found a second minimum in the neutral stability curve at $Re\approx450$, which they associated with a viscous-type instability. Other classic studies on the transition of three-dimensional boundary layers include those of swept cylinders \citep{Poll_1985, KOHAMA1988151, Itoh_1996}, rotating cones \citep{Kobayashi1981LinearST, Kobayashi_Izumi_1983, Kobayashi_Kohama_Kurosawa_1983}, and the Ekman layer \citep{ekman_lilly, melander1983, ekman_comp}. On the topic of non-modal disturbances, \cite{corbett_bottaro_2001} found that swept boundary layers subjected to adverse or favorable pressure gradients were prone to larger transient growth than unswept cases. The authors also determined that, similar to two-dimensional flows, streamwise-elongated streaks comprised the most dangerous initial perturbation. More recently, \cite{Hack_Zaki_2014} probed the effects of time-harmonic spanwise wall motion on transitional boundary layers using direct numerical simulation. They observed that the wall motion could either suppress or accelerate transition. Using a frozen-in-phase as well as a Floquet approach, \cite{Hack_Zaki_2015} attributed this to the competition between the non-modal amplification of boundary layer streaks and the inviscid growth of inflectional modes introduced by the spanwise Stokes profile.

Interestingly, studies on skewed \textit{turbulent} Couette-Poiseuille flows seem not to be uncommon, though almost every work so far has focused exclusively on wall motion that is purely orthogonal to the pressure gradient. In this community, such cases fall into the category of ``shear-driven" or ``viscous-induced" three-dimensional boundary layers. \cite{COLEMAN1996333, sandham, LE2000480}, for example, explored the variation in turbulent statistics of a two-dimensional channel flow after the sudden imposition of a spanwise wall motion. \cite{kannepalli_piomelli_2000} displaced only a finite section of the wall, focusing on the contrast between the initial response to the perturbation and the subsequent relaxation to a two-dimensional equilibrium turbulence. More recently, \cite{holstad_pcp} investigated near-wall coherent structures in a turbulent Couette flow skewed by a spanwise pressure gradient. A common topic of emphasis within these studies seems to be the counter-intuitive structural changes that occur between two-dimensional and three-dimensional (both equilibrium and non-equilibrium) boundary layers. For example, although the addition of mean shear in the two-dimensional case is known to enhance turbulence, it usually leads to a reduction in turbulent stresses in the three-dimensional setting. Furthermore, Townsend's structure parameter is also often quoted to decrease, indicating a decline in the efficiency of turbulent kinetic energy production by the mean flow. For relevant reviews on the topic, we redirect the reader to \cite{olcmen, eaton, flack}. Given the intricate nature of three-dimensional boundary layers and our limited understanding of their physics, it is hoped that this work will supplement and invigorate ongoing investigations in this area.

We structure the remainder of the paper as follows. Section \ref{sec:problem_formulation} introduces our base flow and develops our analysis frameworks. Sections \ref{sec:modal_analysis_section} and \ref{sec:nm_analysis} explore, respectively, modal and non-modal perturbations. Section \ref{sec:conclusion} offers conclusions and a discussion of relevant future work.

\section{Problem Formulation}\label{sec:problem_formulation}

\subsection{Governing Equations and Base Profiles}\label{ssec:gov_eqs}

\begin{figure}
    \centering
 \begin{tikzpicture}[scale = 1.5]

\coordinate (A) at (1.5, 2, 0.5);
\coordinate (B) at (2.35, 2, 1.45);
\coordinate (C) at (2.5, 2, 0.5);
    
\draw[thick,dashed, fill = gray!35] (0,0,-0.15) -- (4,0,-0.15) -- (4,0,2) -- (0,0,2) -- cycle;
\draw[thick,dashed, fill = gray!35] (0,2,-0.15) -- (4,2,-0.15) -- (4,2,2) -- (0,2,2) -- cycle;
\draw[thick, -latex] (0,2,-0.4) -- node[midway, above] {$\widetilde{x}$} (0.75,2,-0.4);
\draw[thick, -latex] (-0.15,2,-0.3) -- node[midway, above left] {$\widetilde{z}$} (-0.15,2,0.85);
\draw[thick, -latex] (0.05,2.2,0) -- node[midway, left] {$\widetilde{y}$} (0.05, 2.8,0);

{\draw[thick, -, celestialblue] (2,2,2) to[out = 330, in = 30] (1, 0, 2) node[xshift = 1.85cm, yshift = 0.25cm, midway] {};
\draw[thick, -, celestialblue, dashed] (1,2,2) -- (1, 0,2);
\draw[dashed] (A) -- node[midway, right, xshift = -0.85cm, yshift = -0.45cm] {$U_w$} (C);
\draw[thick, -latex] (A) -- (B);
\pic [draw, -, "$\theta$", angle eccentricity=1.5] {angle = B--A--C};

\draw[thick, -latex] (2, 0.75, 1) -- node[above, midway] {$-\mathrm{d}p/\mathrm{d}x$} (3.25, 0.75, 1);
\draw[thick, -latex, celestialblue] (1,1.75,2) -- (2.25,1.75,2);
\draw[thick, -latex, celestialblue] (1,1.5,2) -- (2.275,1.5,2);
\draw[thick, -latex, celestialblue] (1,1.25,2) -- (2.195,1.25,2);
\draw[thick, -latex, celestialblue] (1,1,2) -- (2.075,1,2);
\draw[thick, -latex, celestialblue] (1,0.765,2) -- (1.875,0.765,2);
\draw[thick, -latex, celestialblue] (1,0.5,2) -- (1.635,0.5,2);
\draw[thick, -latex, celestialblue] (1,0.25,2) -- (1.35,0.25,2);}
\draw[thick, -, dashed] (4, 0, -0.15)node[right] {$\widetilde{y}=-h$} -- (4, 2, -0.15) node[right] {$\widetilde{y}=h$};
\draw[thick, -, dashed] (0, 0, 2) -- (0, 2, 2);
\draw[thick, -, dashed] (0, 0, -0.15) -- (0, 1.15, -0.15);
\draw[thick, -, dashed] (4, 0, 2) -- (4, 2, 2);

\draw[thick, -, celestialblue] (4, 0, -0.15) -- (4, 2, 2);
\draw[thick, -, celestialblue, dashed] (4, 0, -0.15) -- (4, 2, -0.15);
\draw[thick, -latex, celestialblue] (4,1.75,-0.15) -- (4,1.75,1.73125);
\draw[thick, -latex, celestialblue] (4,1.5,-0.15) -- (4,1.5,1.4625);
\draw[thick, -latex, celestialblue] (4,1.25,-0.15) -- (4,1.25,1.19375);
\draw[thick, -latex, celestialblue] (4,1,-0.15) -- (4,1,0.925);
\draw[thick, -latex, celestialblue] (4,0.75,-0.15) -- (4,0.75,0.65625);
\draw[thick, -latex, celestialblue] (4,0.5,-0.15) -- (4,0.5,0.3875);
\draw[thick, -latex, celestialblue] (4,0.25,-0.15) -- (4,0.25,0.11875);
    \end{tikzpicture}
    \caption{A sketch of the three-dimensional flow geometry for oblique Couette-Poiseuille flows. The wall at $\widetilde{y} = h$ translates with velocity $U_w$ at an angle $\theta\neq 0$ to the streamwise direction, inducing a three-dimensional shear flow.}
    \label{fig:flow_geometry}
\end{figure}
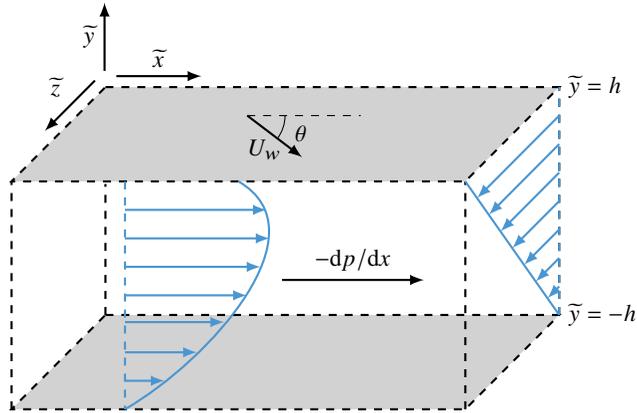

We use the standard equations of motion for an incompressible Newtonian fluid. In the absence of body forces, these can be expressed in dimensional format as follows
\begin{equation}
    \rho\left[\dfrac{\partial\widetilde{\boldsymbol{u}}}{\partial\widetilde{t}} + (\widetilde{\boldsymbol{u}}\cdot\widetilde{\nabla})\widetilde{\boldsymbol{u}}\right] = -\widetilde{\nabla}\widetilde{p} + \mu\widetilde{\nabla}^2\widetilde{\boldsymbol{u}},
\end{equation}
where $\widetilde{\boldsymbol{u}} = \begin{pmatrix} \widetilde{u} & \widetilde{v} & \widetilde{w} \end{pmatrix}^\intercal$ is the Eulerian velocity field, $\widetilde{p}$ the hydrodynamic pressure, $\rho$ the fluid density, and $\mu$ the dynamic viscosity. The flow of interest in this study is illustrated in the schematic presented in Figure \ref{fig:flow_geometry}. Two rigid surfaces, infinite in the wall-parallel directions and located at $\widetilde{y} = \pm h$, confine an incompressible fluid subject to a fixed streamwise pressure gradient $-\mathrm{d}p/\mathrm{d}x > 0$. A crossflow is established by additionally translating the top wall with a constant velocity $U_w$ at an angle $\theta$ with respect to the positive $\widetilde{x}$-axis. The steady laminar velocities in the streamwise and spanwise directions, respectively, are given by
\begin{equation}
    \widetilde{U}\left(\widetilde{y}\right) = \dfrac{h^2}{2\mu}\dfrac{\mathrm{d}p}{\mathrm{d}x}\left(1 - \dfrac{\widetilde{y}^2}{h^2}\right) + \dfrac{U_w}{2}\left(1+\dfrac{\widetilde{y}}{h}\right)\cos\theta,\qquad \widetilde{W}\left(\widetilde{y}\right) = \dfrac{U_w}{2}\left(1+\dfrac{\widetilde{y}}{h}\right)\sin\theta.
\end{equation}
The resulting system is, therefore, a viscous-induced three-dimensional boundary layer, for which the flow angle, defined as $\phi(y) = \tan^{-1}(\widetilde{W}/\widetilde{U})$, changes with the wall-normal direction. These configurations are herein referred to as \textit{oblique Couette-Poiseuille flows} (OCPfs) and, to our knowledge, have not received prior treatment in the stability literature, despite being among the simplest three-dimensional flows capable of retaining homogeneity in the streamwise and spanwise directions. Respectively, $\widetilde{U}$ and $\widetilde{W}$ are Couette-Poiseuille and Couette profiles, their relative strengths modulated by the direction of wall movement. In the limit $U_w\to 0$, standard Poiseuille flow is recovered. On the other hand, for $\theta\to 0$ and $U_w\neq 0$, the crossflow vanishes and the system reduces to the well-known \textit{aligned} Couette-Poiseuille flow (ACPf), in which the pressure gradient and wall motion coincide exactly.

\begin{figure}
    \centering
    \includegraphics[width=\textwidth]{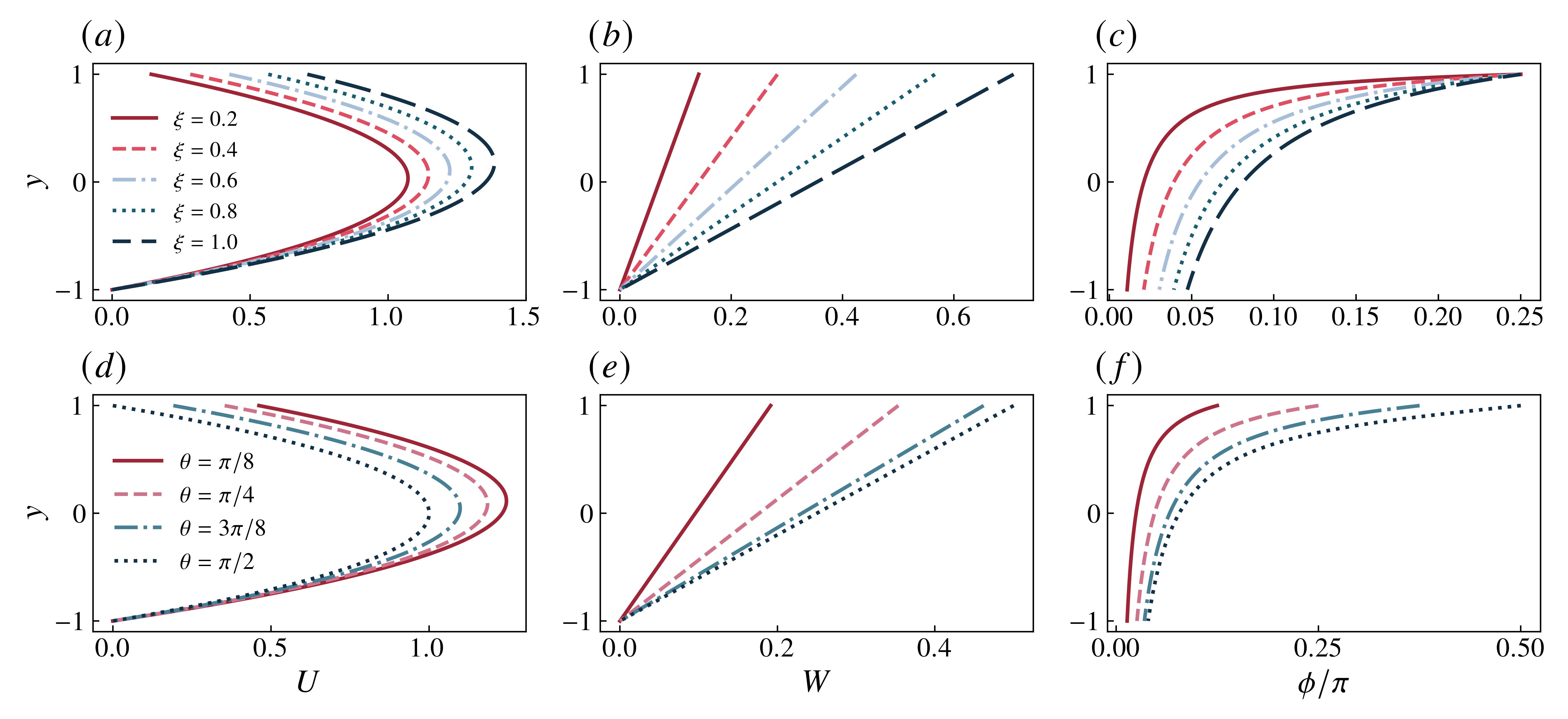}
    \caption{From left to right, plots of the streamwise and spanwise velocities as well as the flow direction $\phi$ (normalized by $\upi$) against the wall-normal coordinate $y$; $(a$--$c)$, $\theta = \upi/4$ and $\xi\in\left\{0.2, 0.4, 0.6, 0.8, 1\right\}$; $(d$--$f)$, $\xi=0.5$ and $\theta\in \left\{\upi/8, \upi/4, 3\upi/8, \upi/2\right\}$. Formally, $\phi$ is singular near the lower wall, where $U$ and $W$ both vanish due to the no-slip condition. However, from l'Hopital's rule, the limit can be computed as $\phi\left(y\to -1\right) = \tan^{-1}\left(\xi\sin\theta/\left(4+\xi\cos\theta\right)\right)$, evidently the angle between the wall shear stresses.}
    \label{fig:base_profiles}
\end{figure}

The parameter space characterizing OCPfs is rather complex, and, as is the case for ACPf, there exist multiple routes to rendering the governing equations non-dimensional. An obvious candidate is $U_p$, the so-called \textit{Poiseuille} velocity scale, which is the streamwise maximum computed in the absence of wall motion. The other option is $U_{\max}$, the ``actual" streamwise maximum, and is preferred if non-equilibrium effects are expected to significantly distort the streamwise profile away from $U_p$. However, in all possible realizations of OCPf, the boundedness of $\cos\theta$ and $\sin\theta$ ensures that $\widetilde{U}$ is $O\left(U_p\right)$. Therefore, to facilitate comparison with the previous literature, we choose to scale with $U_p$. More specifically, the following non-dimensionalization scheme is adopted
\begin{equation}
    \boldsymbol{x} = \dfrac{\widetilde{\boldsymbol{x}}}{h},\qquad \boldsymbol{u} = \dfrac{\widetilde{\boldsymbol{u}}}{U_p},\qquad t = \dfrac{\widetilde{t}}{h/U_p},\qquad p = \dfrac{\widetilde{p}}{\rho U_p^2},
\end{equation}
which yields the dimensionless form of the momentum equations
\begin{align}
\label{eqn:eom}
    \dfrac{\partial \boldsymbol{u}}{\partial t} + \left(\boldsymbol{u}\cdot\nabla\right)\boldsymbol{u} & = -\nabla p + \dfrac{1}{Re}\nabla^2\boldsymbol{u}, \\
\label{eqn:incompressibility_constraint}
    \nabla\cdot\boldsymbol{u} & = 0.
\end{align}
Here, Equation (\ref{eqn:incompressibility_constraint}) represents the incompressibility constraint, and $Re = \rho U_p h/\mu = U_ph/\nu$ is a \textit{Reynolds number}, with $\nu$ being the kinematic viscosity. The base velocity profiles become
\begin{equation}
\label{eqn:nondim_base_profiles}
    U\left(y\right) = 1-y^2 + \dfrac{\xi}{2}\left(1+y\right)\cos\theta, \qquad W\left(y\right) = \dfrac{\xi}{2}\left(1+y\right)\sin\theta,
\end{equation}
where by defining $Re_w = U_wh/\nu$, we can interpret $\xi=Re_w/Re = U_w / U_p$ as the non-dimensional wall-speed. In this setting, the influence of the shear angle on the base profiles becomes more apparent. Suppose that $\xi$ is fixed and $\theta$ is varied; while $W$ maintains its Couette nature, $U$ evolves continuously as a one-parameter homotopy between ACPf and the plane-Poiseuille flow ($\theta = n\upi/2$ for odd $n$). Therefore, it is reasonable to limit attention to pairs $\left(\xi, \theta\right)\in\left[0, 1\right]\times\left[0, 2\upi\right]$, the former due to its physical relevance and the latter due to the periodicity of the base profiles that can be expected to permeate the forthcoming calculations. For select values of the flow parameters, the associated non-dimensional profiles are offered in Figure \ref{fig:base_profiles}.

\subsection{The Linearized System}\label{ssec:linearizing}

This section follows standard monologues on hydrodynamic stability, and we refer the reader to the works of \cite{schmidstability} or \cite{drazin_reid_2004}, for example. In operator format, the Navier-Stokes equations can be rewritten as
\begin{equation}
    \dfrac{\partial \boldsymbol{u}^*}{\partial t} = \mathcal{N}\left(\boldsymbol{u}^*\right),
\end{equation}
where $\mathcal{N}$ is a non-linear function of the state vector $\boldsymbol{u}^* = \begin{pmatrix} \boldsymbol{u} & p \end{pmatrix}^\intercal$. We decompose $\boldsymbol{u}^*$ as $\boldsymbol{u}^* = \boldsymbol{U}^* + {\boldsymbol{u}^\prime}^*$, where $\boldsymbol{U}^*$ is a time-independent base state superposed by a set of infinitesimal fluctuations ${\boldsymbol{u}^\prime}^* = \begin{pmatrix} \boldsymbol{u}^\prime & p^\prime \end{pmatrix}^\intercal$. In particular, we have
\begin{equation}
    \boldsymbol{U}^* = \begin{pmatrix}
        \boldsymbol{U} & P
    \end{pmatrix}^\intercal, \qquad \boldsymbol{U} = \begin{pmatrix}
        U\left(y\right) & 0 & W\left(y\right)
    \end{pmatrix}^\intercal, \qquad \nabla P = \begin{pmatrix}
        -2/Re & \phantom{-}0 & \phantom{-}0
    \end{pmatrix}^\intercal.
\end{equation}
By Taylor expanding $\mathcal{N}$ around $\boldsymbol{U}^*$ and neglecting terms that are $O(\left\lVert {\boldsymbol{u}^\prime}^* \right\rVert^2)$, we obtain a linearized system of evolution equations for the perturbation variables. To reduce computational complexity and the size of the matrices dealt with, the usual procedure here is to eliminate the pressure. This yields a rephrased system based only on fluctuations in the wall-normal velocity/vorticity $\boldsymbol{q} = \begin{pmatrix} v^\prime & \eta^\prime \end{pmatrix}$
\begin{align}
\label{eqn:os}
    \left[\left(\dfrac{\partial}{\partial t} + U\dfrac{\partial}{\partial x} + W\dfrac{\partial}{\partial z}\right)\nabla^2 - \dfrac{\mathrm{d}^2 U}{\mathrm{d}y^2}\dfrac{\partial}{\partial x}- \dfrac{\mathrm{d}^2 W}{\mathrm{d}y^2}\dfrac{\partial}{\partial z}- \dfrac{1}{Re}\nabla^4\right]v' & = 0,\\
    \label{eqn:sq}
       \left[\dfrac{\partial}{\partial t} + U\dfrac{\partial}{\partial x} + W\dfrac{\partial}{\partial z} - \dfrac{1} {Re}\nabla^2\right]\eta'-\dfrac{\mathrm{d}W}{\mathrm{d}y}\dfrac{\partial v'}{\partial x}+\dfrac{\mathrm{d}U}{\mathrm{d}y}\dfrac{\partial v'}{\partial z}&= 0,
\end{align}
where $\nabla^2$ is the usual Laplacian in a Cartesian coordinate system and $\nabla^4\left\langle\cdot\right\rangle\equiv \nabla^2\left(\nabla^2\left\langle\cdot\right\rangle\right)$ is the bi-harmonic operator. Hereon, for notational brevity, we drop the prime notation. Note that, contrary to the case of a purely streamwise base flow for which $W = 0$, the so-called \textit{Squire equation}, Equation (\ref{eqn:sq}), is now forced by mean shear from both the streamwise and spanwise profiles, which are, in general, non-zero. The spatial homogeneity can be exploited via a Fourier Transform
\begin{equation}
  \overline{\boldsymbol{q}}\left(y,t; \alpha, \beta\right) = \iint_{-\infty}^\infty\boldsymbol{q}\left(x,y,z,t\right)e^{-i\left(\alpha x+\beta z\right)}\,\mathrm{d}x\,\mathrm{d}z
\end{equation}
to obtain the canonical form of the \textit{Orr-Sommerfeld-Squire} (OSS) system. Here, $\alpha,\beta\in\mathbb{R}$ are the real-valued wavenumbers in the $x$ and $z$ directions and $\overline{\boldsymbol{q}} = \begin{pmatrix} \overline{v} & \overline{\eta} \end{pmatrix}$ is a block vector of Fourier coefficients. The transformed equations can be compactly written as
\begin{equation}
\label{eqn:ivp}
    \mathsfbi{L}\,\overline{\boldsymbol{q}} = -\dfrac{\partial}{\partial t}\mathsfbi{M}\,\overline{\boldsymbol{q}},
\end{equation}
where, by denoting $\mathcal{D}\equiv\mathrm{d}/\mathrm{d}y$ and $k^2 = \alpha^2+\beta^2$, we have defined
\begin{equation}\label{eqn:orr_somm}
    \mathsfbi{L} = \begin{pmatrix}
        \mathcal{L}_{OS} & 0 \\ i\beta \mathcal{D}U - i\alpha\mathcal{D}W  & \mathcal{L}_{SQ}
    \end{pmatrix},\qquad \mathsfbi{M} = \begin{pmatrix}
        \mathcal{D}^2 - k^2 & 0 \\ 0 & 1
    \end{pmatrix}.
\end{equation}
The \textit{Orr-Sommerfeld} and \textit{Squire} operators, $\mathcal{L}_{OS}$ and $\mathcal{L}_{SQ}$ respectively, are given by
\begin{align}
\label{eqn:os_operator}
    \mathcal{L}_{OS} & = \left(i\alpha U+i\beta W\right)\left(\mathcal{D}^2-k^2\right) - i\alpha \mathcal{D}^2 U-i\beta \mathcal{D}^2 W-\dfrac{1}{Re}\left(\mathcal{D}^2 - k^2\right)^2, \\
    \label{eqn:sq_operator}
    \mathcal{L}_{SQ} & = i\alpha U +i\beta W -\dfrac{1}{ Re}\left(\mathcal{D}^2-k^2\right).
\end{align}
Equation (\ref{eqn:ivp}) forms an initial-value problem for the Fourier-Transformed state vector $\overline{\boldsymbol{q}}$ in wavenumber space, where the associated boundary conditions can be obtained by applying no-slip/impermeability at both walls. Whenever necessary, the velocity-vorticity formulation of the OSS problem can be recast into one for the primitive fluctuations using the transformation
\begin{equation}
\label{c_operator}
    \begin{pmatrix}
        \overline{u} \\ \overline{v} \\ \overline{w}
    \end{pmatrix} = \dfrac{1}{k^2}\begin{pmatrix}
        i \alpha\mathcal{D} & -i\beta\\
        k^2 & \phantom{-}0\\
        i\beta\mathcal{D} & \phantom{-}i\alpha
    \end{pmatrix}\begin{pmatrix}
        \overline{v} \\ \overline{\eta}
    \end{pmatrix}.
\end{equation}
\subsection{Modal Analysis}\label{ssec:modal_analysis}
For a modal or eigenvalue analysis, an additional Fourier Transform is conducted in time
\begin{equation}
\label{eqn:normal_mode}
    \hat{\boldsymbol{q}}\left(y; \alpha, \beta, \omega\right) = \int_{-\infty}^\infty \overline{\boldsymbol{q}}\left(y, t;\alpha, \beta\right)e^{i\omega t}\,\mathrm{d}t
\end{equation}
where $\omega = \omega_r + i \omega_i \in\mathbb{C}$ is the \textit{complex wave frequency}. Equation (\ref{eqn:ivp}) then reduces to a generalized eigenvalue problem described by the linear operator pencil $\left(\mathsfbi{L},\mathsfbi{M}\right)$
\begin{equation}
    \mathsfbi{L}\hat{\boldsymbol{q}} = i\omega\mathsfbi{M}\hat{\boldsymbol{q}},
\end{equation}
with eigenvalues corresponding to $i\omega = i\omega_r - \omega_i$. Note that this is equivalent to solving for the eigensystem of $\mathsfbi{S}^\prime = \mathsfbi{M}^{-1}\mathsfbi{L}$. In general, the spectrum is a function of $\left\{\alpha, \beta, Re, \xi, \theta\right\}$, and exponential amplification occurs over time if $\omega_i > 0$. Consequently, we seek the manifold of marginal stability, designated by
\begin{equation}
    \omega_i\left(\alpha, \beta, Re, \xi, \theta\right) = 0. 
\end{equation}
We note that the presence of a non-zero spanwise velocity in OCPfs prevents an application of Squire's Theorem in its usual form. Although a two-dimensional problem may well be constructed (see, for example, \cite{mack_BL_stability} and \cite{schmidstability}), the ``effective" base velocity depends on both spatial wavenumbers and there is no \textit{a priori} indication of the appropriate search space. Therefore, for a given configuration $\left(\xi, \theta\right)$, since a full stability portrait requires a sweep through the $\left(\alpha, \beta, Re\right)$-space, a numerical approach will inevitably be marred by a lack of resolution. While this is a valid criticism, we point out that most canonical shear flows only become linearly unstable at modest wavenumbers, if at all. Furthermore, in Section \ref{sec:modal_analysis_section}, we demonstrate that from the perspective of modal stability, oblique Couette-Poiseuille flows are essentially continuations of the aligned variant. Therefore, the results of a sufficiently broad numerical search, as conducted here, are likely global.

Before proceeding, we make some key observations. First, as is true for strictly streamwise base flows, the \textit{Squire modes} remain damped. The proof proceeds in the usual way by converting to a formulation involving the \textit{$x$-phase speed}, $c=\omega/\alpha$, multiplying the homogeneous Squire equation by the complex conjugate of the fluctuating normal vorticity, and integrating over $y$. Therefore, for a modal analysis, it suffices to consider only the Orr-Sommerfeld operator, Equation (\ref{eqn:os_operator}). Furthermore, since neither component of the base velocity is inflectional, OCPfs do not admit an inviscid crossflow-like instability as observed, for example, over swept wings or rotating disks. In particular, in the inviscid limit, Rayleigh's criterion can be modified to require the following expression to hold at some wall-normal location
\begin{equation}
\label{eqn:rayleigh}
    \mathcal{D}^2U + \gamma\mathcal{D}^2W = 0,
\end{equation}
where $\gamma = \beta/\alpha$. Although Equation (\ref{eqn:rayleigh}) will, for general flows, vary in wavenumber space, the linearity of $W$ implies that $\mathcal{D}^2W = 0$ for OCPfs. Thus, since $\mathcal{D}^2U = -2$, the instability must be viscous in nature.

\subsection{Non-Modal Analysis}\label{ssec:nonmodal_analysis}
For most shear flows, a spectral analysis of the linearized Jacobian as in Section \ref{ssec:modal_analysis} rarely agrees with experiment. Almost invariably, the transition to turbulence is observed at sub-critical $Re$, that is, below the threshold predicted by modal theory \citep{tref_pseudospec}. This behavior is now well understood to be a consequence of the highly non-normal nature of the Orr-Sommerfeld-Squire (OSS) operator $\mathsfbi{S}^\prime$, which, in turn, arises from the off-diagonal term $\left(i\beta\mathcal{D} U - i\alpha\mathcal{D}W\right)$ driving the Squire equation; see (\ref{eqn:sq}) and (\ref{eqn:orr_somm}). In general oblique Couette-Poiseuille flows, this forcing can evidently comprise both the streamwise and spanwise mean shear.

A non-normal operator such as $\mathsfbi{S}^\prime$ admits eigenfunctions that are non-orthogonal in the underlying Hilbert space. When arbitrary initial states are transformed into the basis of these eigenfunctions, they can suffer from large cross-terms in the induced norm \citep{schmid_annrev}. An immediate consequence is that while a modal analysis might suggest asymptotic decay, energy amplification can still occur over finite time horizons. In shear flows, the transition to turbulent regimes has often been attributed to these transient phenomena, providing a potential explanation for the so-called \textit{bypass transition} \citep{butler_3d_opt_pert}. Furthermore, there is no guarantee that the long-time eigenmode is even realized, in spite of the most careful calibration, since sufficiently strong transient amplification will likely excite non-linear mechanisms in the flow and violate the linear assumption \citep{waleffe_nonnormal, tref_pseudo_siam}.

To explore the implications of non-normality in OCPfs, we first solve the initial-value problem in Equation (\ref{eqn:ivp}) exactly to yield
\begin{equation}
\label{eqn:propagator}
    \overline{\boldsymbol{q}}\left(t\right) = \Phi\left(t, 0\right)\overline{\boldsymbol{q}}_0,
\end{equation}
where $\Phi\left(t,0\right)\equiv e^{-i \mathsfbi{S} t}$ is the \textit{state-transition operator}, $\mathsfbi{S} = i \mathsfbi{S}^\prime$, and $\overline{\boldsymbol{q}}_0$ is the state of the system at the initial time $t=0$. Under appropriate norms in the input and output spaces, the \textit{gain} can be defined as
\begin{equation}
    G\left(\alpha, \beta, Re, \xi, \theta, t\right) = \sup_{\overline{\boldsymbol{q}}_0\neq 0}\dfrac{\left\lVert \overline{\boldsymbol{q}}\right\rVert_{\mathrm{out}}^2}{\left\lVert \overline{\boldsymbol{q}}_0\right\rVert_\mathrm{in}^2},
\end{equation}
where, due to its physical significance, we let $\left\lVert \cdot\right\rVert_\mathrm{out}=\left\lVert \cdot\right\rVert_\mathrm{in}=\left\lVert \cdot\right\rVert_E$ be an energy norm
\begin{equation}
\label{eqn:energy_norm}
    \left\lVert \overline{\boldsymbol{q}}\right\rVert_E^2 = \int_{-1}^1\overline{v}^{\dagger}\overline{v} + \dfrac{1}{k^2}\left(\overline{\eta}^{\dagger}\overline{\eta} + \dfrac{\partial\overline{v}^{\dagger}}{\partial y}\dfrac{\partial\overline{v}}{\partial y}\right)\,\mathrm{d}y\simeq \overline{\boldsymbol{q}}^\dagger\mathsfbi{E}\overline{\boldsymbol{q}}
\end{equation}
over the volume $V$ defined by the Cartesian product $\left(x,y,z\right)\in\left[0,2\upi/\alpha\right]\times\left[-1,1\right]\times \left[0,2\upi/\beta\right]$. In this way, the energy of one full wavelength of a disturbance can be captured; see \cite{butler_3d_opt_pert}. Here, $\left\langle\cdot\right\rangle^\dagger$ denotes a conjugate transpose operation, and the operator $\mathsfbi{E}$ is positive-definite and incorporates the Clenshaw-Curtis quadrature weights \citep{tref_spec_methods}. With a Cholesky decomposition, we may write $\mathsfbi{E} = \mathsfbi{F}^\dagger\mathsfbi{F}$ so that
\begin{equation}
    \left\lVert \overline{\boldsymbol{q}}\right\rVert_E^2 \simeq \overline{\boldsymbol{q}}^\dagger\mathsfbi{F}^\dagger\mathsfbi{F}\overline{\boldsymbol{q}} = \left\lVert \mathsfbi{F}\,\overline{\boldsymbol{q}}\right\rVert_2^2.
\end{equation}
It immediately follows that
\begin{equation}
    G = \sup_{\overline{\boldsymbol{q}}_0\neq 0}\dfrac{\left\lVert\mathsfbi{F}\Phi\left(t,0\right)\overline{\boldsymbol{q}}_0\right\rVert_2^2}{\left\lVert\mathsfbi{F}\,\overline{\boldsymbol{q}}_0\right\rVert_2^2} = \sup_{\overline{\boldsymbol{q}}_0\neq 0}\dfrac{\left\lVert\mathsfbi{F}\Phi\left(t,0\right)\mathsfbi{F}^{-1}\mathsfbi{F}\,\overline{\boldsymbol{q}}_0\right\rVert_2^2}{\left\lVert\mathsfbi{F}\,\overline{\boldsymbol{q}}_0\right\rVert_2^2} = \left\lVert\mathsfbi{F}\Phi\left(t,0\right)\mathsfbi{F}^{-1}\right\rVert_2^2 ,
\end{equation}
which can be computed trivially via the singular value decomposition (note, in fact, that $G = \left\lVert \Phi\left(t,0\right)\right\rVert_E^2$). The associated right and left singular functions represent, respectively, the initial condition and response pair for which the gain at time $t$ is realized.

Intuitively, no energy growth is expected if $G\leq 1$. An equivalent condition can be expressed in terms of the \textit{resolvent} of $\mathsfbi{S}$. Consider an exogenous harmonic forcing profile $\boldsymbol{H}\left(x,y,z,t\right) = \boldsymbol{h}\left(x,y,z\right)e^{-i\zeta t}$ with frequency $\zeta\in\mathbb{C}$ to the linearized system, appropriately transformed into wavenumber space
\begin{equation}
   \overline{\boldsymbol{h}}\left(y; \alpha, \beta\right) = \iint_{-\infty}^\infty\boldsymbol{h}\left(x,y,z\right)e^{-i\left(\alpha x+\beta z\right)}\,\mathrm{d}x\,\mathrm{d}z
\end{equation}
The response can easily be verified to be
\begin{equation}
    \overline{\boldsymbol{q}} = ie^{-i\zeta t}\left(\zeta \mathsfbi{I} - \mathsfbi{S}\right)^{-1}\overline{\boldsymbol{h}},
\end{equation}
where the operator $\mathsfbi{R} \equiv \left(\zeta\mathsfbi{I}-\mathsfbi{S}\right)^{-1}$ is known as the resolvent. From an input-output perspective, $\mathsfbi{R}$ serves as a transfer function between the excitation and its response. The quantity $\mathcal{R} = \left\lVert \mathsfbi{R}\right\rVert_E$ is, therefore, of particular interest here, since for a non-normal system, it can be large even if the forcing is pseudo-resonant, that is, $\zeta\notin\Lambda\left(\mathsfbi{S}\right)$, the spectrum of $\mathsfbi{S}$ \citep{TrefethenEmbree}. Such a paradigm is especially informative for the receptivity of the flow to external disturbances \citep{brandt_lift_up}, and if $\zeta$ is restricted to real values, a physical interpretation of the resolvent is the perturbed operator that can result, for example, from external vibrations or planar imperfections \citep{tref_pseudospec}. By further generalizing to the complex plane, one recovers the $\epsilon$-\textit{pseudospectra}, the set of values defined as
\begin{equation}
\label{eqn:pseudospectra}
    \Lambda_\epsilon\left(\mathsfbi{S}\right) = \left\{\zeta\in\mathbb{C}\colon\mathcal{R}\geq \epsilon^{-1}\right\}.
\end{equation}
For non-normal operators, $\Lambda_\epsilon$ can protrude deep into the upper-half of the complex plane, and the more pronounced this effect, the greater the potential for transient growth irrespective of the presence of linear instability. More rigorously, the Hille-Yosida Theorem states that $G\leq 1$ if and only if the $\epsilon$-pseudospectra lie sufficiently close to the lower half-plane \citep{reddyschmidhenn}. For further details, we refer the reader to this paper, the citations within, and the text of \cite{TrefethenEmbree}.

\subsection{Energy Budget Analysis}\label{ssec:energy_budget}

An investigation of the perturbation energy budget can reveal the mechanism of instability in OCPfs. Throughout this section, the Einstein convention is implied via repeated indices. We define the perturbation energy density $\mathcal{E}$ as
\begin{equation}
    \mathcal{E} = \dfrac{1}{2}\boldsymbol{u}^\dagger\boldsymbol{u} = \dfrac{1}{2}u_i^\dagger u_i = \dfrac{1}{2}\left(\left|u\right|^2 + \left|v\right|^2 + \left|w\right|^2\right).
\end{equation}
By multiplying Equation (\ref{eqn:eom}) throughout by $\boldsymbol{u}^\dagger$ and integrating over $V$, evolution equations for the total energy are recovered
\begin{equation}
\label{eqn:reynolds_orr}
    \dfrac{\mathrm{d}}{\mathrm{d}t}\int_V\mathcal{E}\,\mathrm{d}V =\int_V \dfrac{\mathrm{d}\mathcal{E}}{\mathrm{d}t}\,\mathrm{d}V= -\int_V \dfrac{1}{2}\left(u_i^\dagger u_j+u_i u_j^\dagger\right)\dfrac{\partial U_i}{\partial x_j}\,\mathrm{d}V -\dfrac{1}{Re}\int_V\dfrac{\partial u_i^\dagger}{\partial x_j}\dfrac{\partial u_i}{\partial x_j}\,\mathrm{d}V.
\end{equation}
where we have assumed spatial periodicity of the disturbance field in $x$ and $z$. Under the normal mode ansatz, Equation (\ref{eqn:normal_mode}), the above expression reduces to
\begin{equation}
   2\omega_i\int_{-1}^1\dfrac{1}{2}\hat{u}_i^\dagger\hat{u}_i\,\mathrm{d}y = \mathcal{P}-\varepsilon,
\end{equation}
where we have defined
\begin{align}
        \mathcal{P} & = \underbrace{-\int_{-1}^{1}\dfrac{1}{2}\left(\hat{u}^\dagger\hat{v} + \hat{u}\hat{v}^\dagger\right)\dfrac{\partial U}{\partial y}\,\mathrm{d}y}_{\mathcal{P}_{u}}\underbrace{-\int_{-1}^{1}\dfrac{1}{2}\left(\hat{w}^\dagger\hat{v} + \hat{w}\hat{v}^\dagger\right)\dfrac{\partial W}{\partial y}\,\mathrm{d}y}_{\mathcal{P}_{w}}, \\
        \varepsilon & = \dfrac{1}{Re}\int_{-1}^1\left(\mathcal{D}\hat{u}_i\right)^\dagger\mathcal{D}\hat{u}_i + k^2 \hat{u}_i^\dagger\hat{u}_i\,\mathrm{d}y.
\end{align}
Two contributions to the disturbance kinetic energy can be identified, $\mathcal{P}$, the production against the background shear(s), and $\varepsilon$, the viscous dissipation. The former can be further separated into terms representing the transfer of energy from the base streamwise and spanwise flows, respectively, to the perturbation field through the action of the associated Reynolds stresses, $\tau_{u}$ and $\tau_{w}$. These have been denoted by $\mathcal{P}_{u}$ and $\mathcal{P}_{w}$. In general, (positive) production destabilizes, whereas dissipation stabilizes the disturbance field.


\section{Modal Analysis}\label{sec:modal_analysis_section}
\subsection{Characteristics of the Eigenspectra}\label{ssec:spectra}

\begin{figure}
    \centering
    \includegraphics[width=\textwidth]{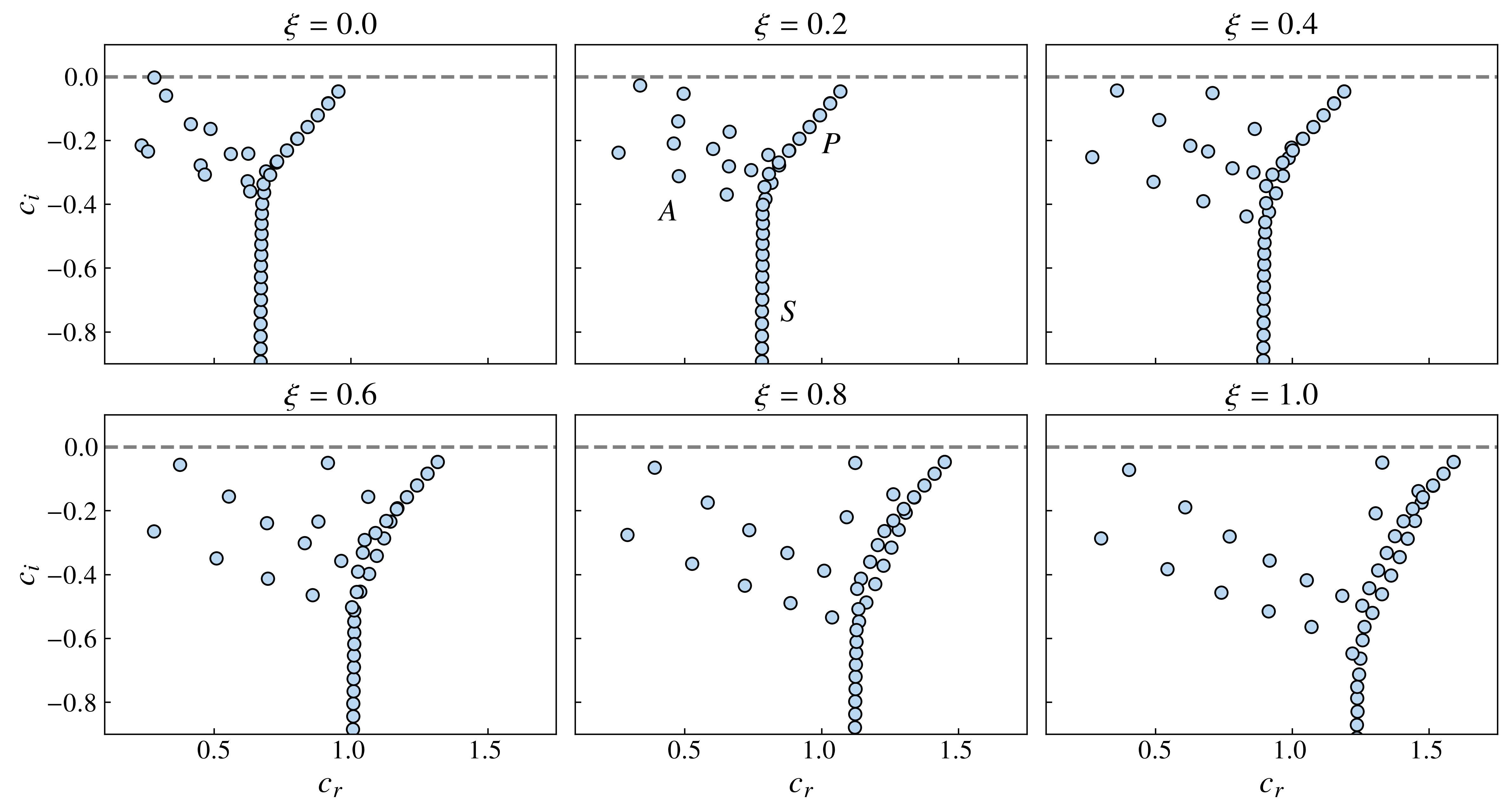}
    \caption{The locus of the eigenspectrum for $\left(\alpha,\beta\right) = \left(1, 0.5\right)$ at $Re = 5700$ and $\theta = \upi/6$ for $\xi\in\left\{0, 0.2, 0.4, 0.6, 0.8, 1\right\}$. The $A$, $P$, and $S$ branches have been appropriately labeled. On each plot, a gray dashed line denotes the stability boundary, $c_i = 0$.}
    \label{fig:ocpf_theta_fixed}
\end{figure}

We begin by investigating the dynamics of the eigenspectra in OCPfs. For a sample wavenumber combination, Figure \ref{fig:ocpf_theta_fixed} illustrates the loci of the first $\approx50$ least stable modes as the non-dimensional wall speed $\xi$ is varied at $\theta = \upi/6$. The results have been presented in terms of $c = \omega/\alpha$. A familiar $Y$-shaped distribution can be observed, with three distinct branches reminiscent of the spectrum for plane-Poiseuille flow (pPf). As the wall speed increases, this structure collectively translates further into the right half-plane, and the most unstable mode monotonically stabilizes. In doing so, the shape of the $S$-branch, comprising the so-called \textit{mean modes} related to the mean velocity, remains relatively undistorted. On the other hand, a sharper change occurs in the $A$-branch -- the \textit{wall modes} -- which separate into two distinct subsets associated, respectively, with each wall. In a somewhat similar manner, starting from its bottom half, the $P$-branch of \textit{center modes} also begins to split into two noticeable sub-branches. Together, these observations are indicative of the increased Couette contribution to the base flow, since the spectra for various flavors of Couette flow are usually scattered symmetrically within two $A$-branches, e.g. \cite{duck_erlebacher_hussaini_1994, schmidstability, Liu_Liu_2012, couette_rarefied}.

In OCPfs, the distribution of this Couette component between the base velocities is directly controlled by the shear angle $\theta$. However, its impact at the level of the Orr-Sommerfeld (OS) equation is rather subtle. Since $W$ is linear, $\mathcal{D}^2W = 0$, and the OS operator, simplified from Equation (\ref{eqn:os_operator}), becomes
\begin{equation}
\label{eqn:reduced_os}
\mathcal{L}_{OS} = \overbrace{\left(i\alpha U+i\beta W\right)\left(\mathcal{D}^2-k^2\right)}^{\mathcal{O}_1} - \overbrace{i\alpha \mathcal{D}^2 U}^{\mathcal{O}_2}-\overbrace{\dfrac{1}{Re}\left(\mathcal{D}^2 - k^2\right)^2}^{\mathcal{O}_3},
\end{equation}
where $U$ and $W$ retain their definitions from Equation (\ref{eqn:nondim_base_profiles}), instantiated with some wall speed $\xi$. We immediately observe, despite the three-dimensionality of the flow, that the spanwise velocity appears only in a single term, $\mathcal{O}_1$, in Equation (\ref{eqn:reduced_os}). In particular, for a spanwise-independent mode, $\beta=0$, the effects of obliqueness in the base flow are, in a sense, ``shut off", since the corresponding OS operator
\begin{equation}
\label{eqn:vanishing_spanwise}
    \mathcal{L}_{OS} = i\alpha U\left(\mathcal{D}^2-\alpha^2\right) - i\alpha \mathcal{D}^2 U-\dfrac{1}{Re}\left(\mathcal{D}^2 - \alpha^2\right)^2
\end{equation}
reduces precisely to that for ACPf under the umbrella of Squire's Theorem (excluding, of course, the factor of $\cos\theta$ in $U$, which can essentially be lumped into the wall speed). To extend this analogy to more general disturbances, a modification must first be made. Consider the generic three-dimensional (that is, \textit{prior} to an application of Squire's result) OS operator for ACPf
\begin{equation}
\label{eqn:os_operator_acpf}
    \mathcal{L}_{OS}^\mathrm{ACPf} = \overbrace{i\alpha U_{\mathrm{ACPf}}\left(\mathcal{D}^2-k^2\right)}^{\mathcal{A}_1} - \overbrace{i\alpha \mathcal{D}^2 U_{\mathrm{ACPf}}}^{\mathcal{A}_2}-\overbrace{\dfrac{1}{Re}\left(\mathcal{D}^2 - k^2\right)^2}^{\mathcal{A}_3},
\end{equation}
where 
\begin{equation}
    U_\mathrm{ACPf} = 1-y^2 + \dfrac{\xi_\mathrm{ACPf}}{2}\left(1+y\right).
\end{equation}
Comparing the two operators in (\ref{eqn:reduced_os}) and (\ref{eqn:os_operator_acpf}) allows us to identify crucial similarities in structure. Specifically, for a constant wave triplet $\left(\alpha,\beta, Re\right)$, while $\mathcal{O}_3\equiv \mathcal{A}_3$ is immediate, $\mathcal{O}_2\equiv\mathcal{A}_2$ follows from the fact that $\mathcal{D}^2U=-2=\mathcal{D}^2 U_{\mathrm{ACPf}}$. Therefore, $\mathcal{L}_{OS}$ and $\mathcal{L}_{OS}^\mathrm{ACPf}$ differ exclusively in their terms $\mathcal{O}_1$ and $\mathcal{A}_1$, respectively. However, since $k^2$ has also been fixed by our choice of wavenumbers, $\mathcal{O}_1 \equiv \mathcal{A}_1$ can be made possible by requiring
\begin{equation}
    i\alpha U_\mathrm{ACPf} = i\alpha U + i\beta W\implies \xi_\mathrm{ACPf} = \xi\left(\cos\theta + \gamma\sin\theta\right),
\end{equation}
\begin{figure}
    \centering
    \includegraphics[width=0.85\textwidth]{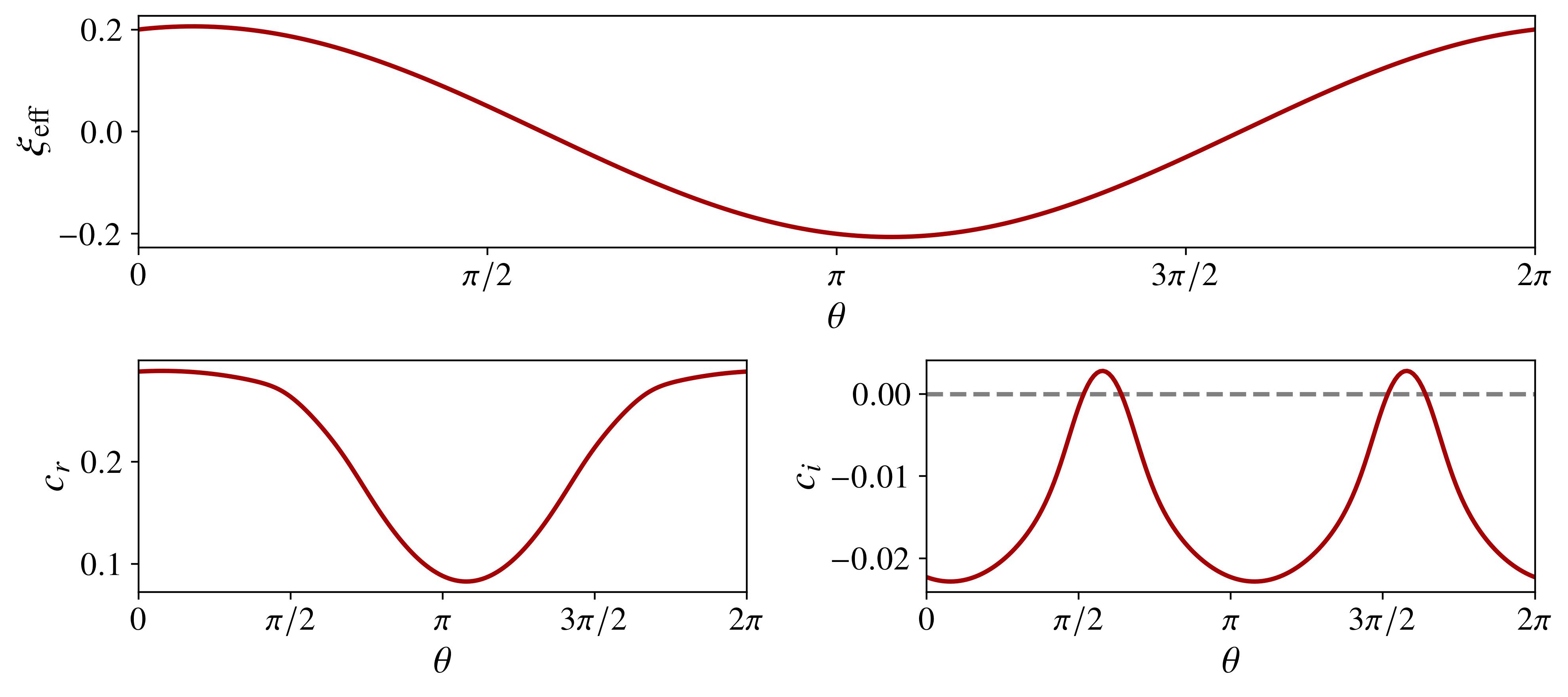}
    \caption{At $Re=10000$ and $\xi=0.2$, the variation with $\xi_\mathrm{eff}$ of the least stable eigenmode for $\left(\alpha, \beta\right) = \left(1,0.25\right)$. The only dashed gray line marks the boundary $c_i=0$. Both components change in tandem with $\xi_\mathrm{eff}$, and when juxtaposed with the information in Figure \ref{fig:ocpf_theta_fixed}, lend weight to $\xi_\mathrm{eff}$ serving as an effective wall speed. Note that $c_i$ is, in fact, $\upi$-periodic, the underlying mechanism being precisely that which allows for symmetric growth rates around $\xi = 0$ for ACPf; see Section \ref{ssec:criticality}.}
    \label{fig:ocpf_xi_fixed}
\end{figure}where $\gamma = \beta/\alpha$. Therefore, for an arbitrary OCPf, the Orr-Sommerfeld problem at any wavenumber pair can be exactly mapped to one for ACPf via the ``effective" wall speed $\xi_{\mathrm{eff}}$
\begin{equation}
\label{eqn:effective_xi}
    \xi_{\mathrm{eff}} \equiv  \xi\left(\cos\theta + \gamma\sin\theta\right).
\end{equation}
With the corollary
\begin{equation}
\label{eqn:relate_ocp_to_acp}
    \omega\left(\alpha,\beta,Re,\xi,\theta\neq0\right) = \omega\left(\alpha,\beta,Re,\xi_{\mathrm{eff}},\theta=0\right),
\end{equation}
we conclude that the stability of any oblique Couette-Poiseuille flow can be prescribed entirely by comparison with the appropriate ACPf configuration(s). A stronger result, and one perhaps in the same spirit as Squire's Theorem, is as follows: if $\mathfrak{O}$ denotes the set of all possible OS operators for OCPf and $\mathfrak{A}$ the equivalent set for ACPf, then $\mathfrak{O}\subseteq\mathfrak{A}$.

The influence of the shear angle on modal behavior can now be made precise. We start by noting that $\xi_\mathrm{eff}$ is $2\upi$-periodic and
\begin{equation}
    -\xi k/\alpha \leq \xi_\mathrm{eff} \leq \xi k/\alpha
\end{equation}
so that it varies strongly even throughout wavenumber space. Mathematically, at a fixed triplet $\left(\alpha, \beta, Re\right)$, its action in $\theta$ seems to be to accentuate or mask the strength of the wall speed. As an example, Figure \ref{fig:ocpf_xi_fixed} shows how changes in $\xi_\mathrm{eff}$ with $\theta$ affect the real and imaginary components of the most unstable eigenvalue for an arbitrarily chosen wavenumber pair. It is evident that the periodicity of $\xi_\mathrm{eff}$ directly translates to that of the spectrum, which itself becomes, at a minimum, $2\upi$-periodic. Furthermore, we found (not shown here; refer to Figure \ref{fig:ocpf_theta_fixed}) that variations in $\xi_\mathrm{eff}$ modified the distribution of the eigenmodes in the complex plane much in the same fashion as variations in $\xi$ for fixed $\theta$, e.g., increasing $\xi_\mathrm{eff}$ increased $c_r$, and vice versa. When taken together, these observations, combined with Equation (\ref{eqn:relate_ocp_to_acp}) and the interpretation of $\xi_\mathrm{eff}$, suggest that the manifold of marginal stability for OCPfs is contained wholly within that for ACPf. Accounting for a non-trivial directionality in the flow affects perhaps only the subset of the latter that is ultimately accessed.

\subsection{Exploring Criticality in oblique Couette-Poiseuille flows}\label{ssec:criticality}

In this section, we present the findings of a comprehensive investigation into the modal stability of oblique Couette-Poiseuille flows. We introduce the critical Reynolds number, denoted $Re_c$, which represents the minimum Reynolds number below which the flow remains linearly stable. Beyond this value, at least one disturbance, characterized by the critical wavenumbers $\left(\alpha_c,\beta_c\right)$, becomes unstable. When analyzing two-dimensional flows, Squire's Theorem \citep{squires} allows us to focus solely on disturbances that are independent of the spanwise direction, that is, $\beta_c = 0$. However, for general three-dimensional profiles, an accurate assessment of stability necessitates the consideration of modes with non-zero $\beta$. Consequently, in the case of an oblique Couette-Poiseuille flow $\left(\xi,\theta\right)$, a thorough exploration of the entire three-dimensional $\left(\alpha, \beta, Re\right)$-space is required.

To reduce the degree of computation, we now consider some important simplifications. First, we note that in the stability literature for ACPf, the analysis for $\xi<0$ is typically neglected, since the modal growth rates are symmetric around $\xi = 0$ (although the corresponding real parts might not be). \cite{potter_1966} rationalized this by adopting the coordinate transformation $y\to -y$. Since $\xi_\mathrm{eff}\left(\theta + \upi\right) = -\xi_\mathrm{eff}\left(\theta\right)$, a similar argument allows us to restrict our attention to $\theta\in\left[0, \upi\right]$. However, a second reduction is also possible and can be achieved by noting that, at a constant $\xi$, if $\left(\alpha_c, \beta_c, Re_c\right)$ is the critical tuple for $\theta=\theta^\prime$, then $\left(\alpha_c, -\beta_c, Re_c\right)$ is necessarily the critical tuple for $\theta = \upi - \theta^\prime$. This result is immediate from the definition of $\xi_\mathrm{eff}$ in Equation (\ref{eqn:effective_xi}), since
\begin{equation}
    \xi_{\mathrm{eff}}\left(\alpha, \beta, \xi,  \theta^\prime\right) = -\xi_{\mathrm{eff}}\left(\alpha, -\beta, \xi,  \upi-\theta^\prime\right),
\end{equation}
where we have assumed $\alpha > 0$. Thus, it suffices to explore the range $\theta\in\left[0, \upi/2\right]$. For the Fourier wavenumbers, we focused on small to intermediate values, in particular, $\left(\alpha, \beta\right) \in \left[-3, 3\right]\times\left[-3, 3\right]$. This is generally the subspace of the wavenumber plane within which linear instability is first encountered in most canonical shear flows, and particularly for ACPfs \citep{potter_1966}. In total, $O\left(10^{10}\right)$ gridpoints were investigated, and our numerical procedure, including our method for traversing such an unwieldy parameter space, is outlined in Appendix \ref{appendix:appA}. In what follows, all results are presented for values of $\theta$ in degrees rather than in radians.

 \begin{figure}
        \centering
        \includegraphics[width=\textwidth]{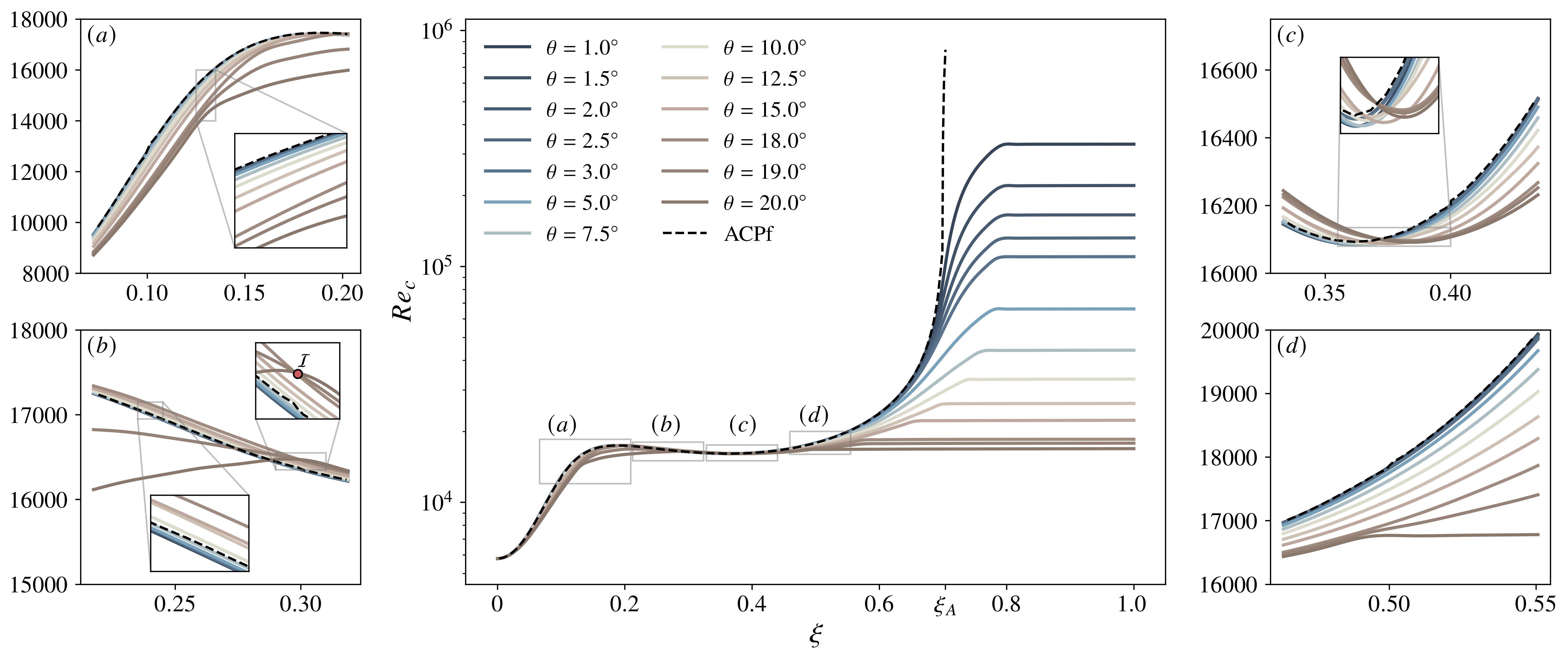}
        \caption{The critical Reynolds number $Re_c$ against $\xi$ for $\Theta_1\equiv\left(0, 20^\circ\right]$. Throughout this figure, the dashed line indicates the equivalent plot for ACPf. The insets magnify regions of particular interest that have been discussed in the text. A circle in inset $(b)$ denotes the crossing point $\mathcal{I}$. In this range of shear angles, a typical $Re_c$-curve mimics that for ACPf when $\xi\lessapprox\xi_A$, but appears to have been ``dragged" down from infinity when $\xi > \xi_A$, yielding a finite $Re_c$ even beyond this threshold wall speed.}
        \label{fig:crit_RE_small_angles}
\end{figure}

In general, \textit{the introduction of skewness in Couette-Poiseuille flows was found to be destabilizing}, at least relative to ACPf. However, two qualitative regimes could still be identified in $\theta$. The first, denoted $\Theta_1$, comprises $0^\circ<\theta\lessapprox 20^\circ$ and is arguably the most interesting of the two, as it exhibits drastic changes in stability throughout its extent. Since oblique Couette-Poiseuille flows reduce to the standard aligned case as $\theta\to 0$, it is natural to expect the stability characteristics of ACPf to continue at least to modest $\theta$. Figure \ref{fig:crit_RE_small_angles} supports this intuition. For all $\theta\in\Theta_1$, a short range of stabilization is followed by an inflection point in the $Re_c$-curves between $0.2\lessapprox \xi\lessapprox 0.4$ and then further growth, a trend that is precisely reminiscent of ACPf \citep{potter_1966}.

Perhaps the most striking feature is the fact that linear instability seems to persist throughout the entire range of wall speeds considered here. This behavior was observed even for ``small" angles, such as $\theta\in\left\{1^\circ, 1.5^\circ\right\}$ (and even down to $\theta\in\left\{0.5^\circ, 0.75^\circ\right\}$, not shown here), where the wall motion is approximately parallel to the pressure gradient. This is in stark contrast to ACPf, which achieves unconditional linear stability, $Re_c\to\infty$, against infinitesimal disturbances beyond the so-called cutoff wall speed $\xi_A\approx 0.7$ \citep{potter_1966, cowley_smith_1985}. A crude explanation for this is that the inclusion of a spanwise velocity makes $\beta$ a relevant stability parameter, providing, in light of the effective wall speed $\xi_{\mathrm{eff}}$ and the analysis outlined at the end of Section \ref{ssec:spectra}, an additional buffer for an OCPf to return to a region of parameter space that is linearly unstable for ACPfs. In fact, in Section \ref{ssec:long_wave}, we argue, using a modified version of the long-wavelength approximation of \cite{cowley_smith_1985}, that all OCPfs for which $\theta\neq 0$ are always linearly unstable: \textit{there is no cutoff wall speed for OCPfs}. The absence of such a threshold for OCPfs seems to manifest itself in terms of the appearance of a limiting regime in the critical parameters. Here, the existence of linear instability seems to become entirely independent of $\xi$, as evidenced, in part, by the flattening of the $Re_c$-curves in Figure \ref{fig:crit_RE_small_angles}.

We note that before achieving the respective asymptotes in their $Re_c$-curves, the destabilization experienced by OCPfs in $\Theta_1$ at any wall speed is not necessarily monotonic with $\theta$. Beginning with inset $(a)$ in Figure \ref{fig:crit_RE_small_angles}, we see that increasing the shear angle is conclusively destabilizing up to $\xi\approx0.2$. However, between approximately $0.2<\xi\lessapprox0.275$, some of the more modest angles, say $\theta\gtrapprox10^\circ$, tend to stabilize, while the values of $Re_c$ for even larger angles, $\theta\gtrapprox18.5^\circ$, remain below those of ACPf; see Figure \ref{fig:crit_RE_small_angles} $(b)$. Around $\xi\approx0.3$ lies the crossing point $\mathcal{I}$, which initiates a region of monotonic stabilization for $\theta\gtrapprox 10^\circ$, a pattern that persists until around $\xi\approx 0.375$. Here, as seen in Figure \ref{fig:crit_RE_small_angles} $(c)$, all $Re_c$-curves experience a turning point, which occurs at increasing wall speeds with $\theta$, and begin to ascend toward their eventual plateaus. Beyond this location, the stabilization becomes monotonic throughout $\theta>0$, as can be verified in Figure \ref{fig:crit_RE_small_angles} $(d)$.

\begin{figure}
    \centering
    \includegraphics[width=\textwidth]{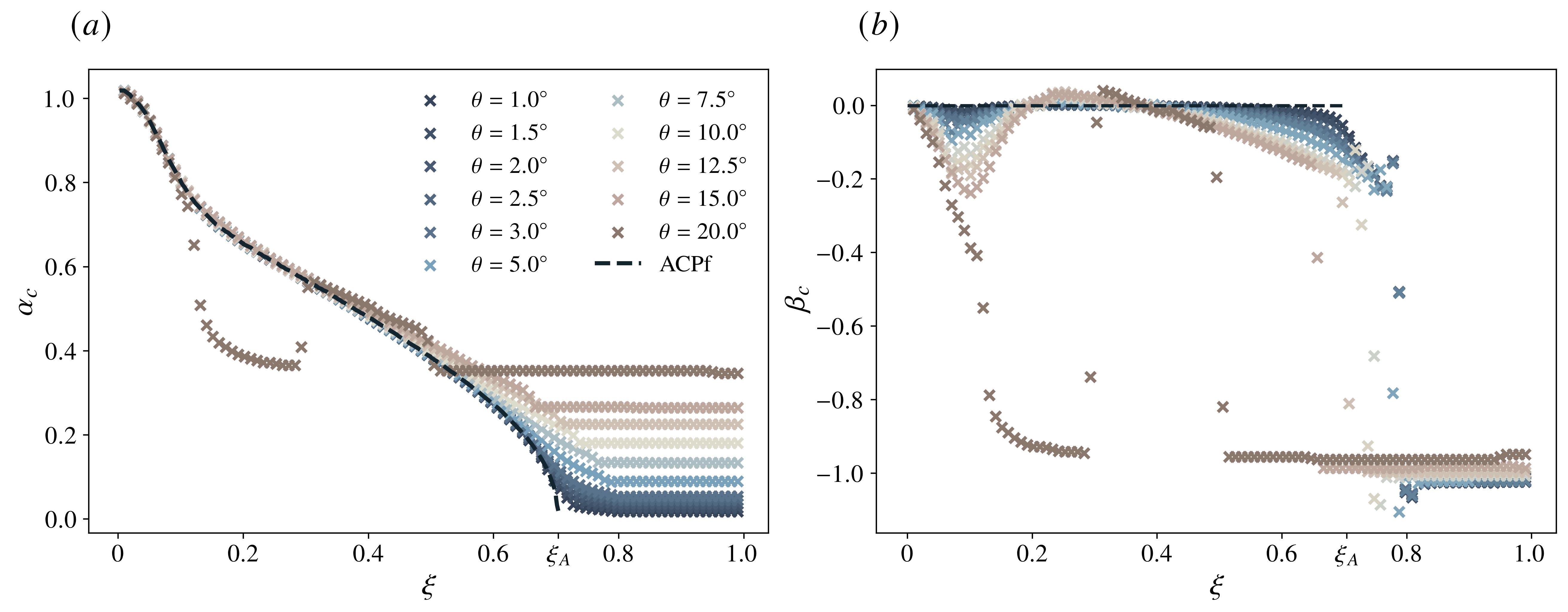}
    \caption{Curves of the critical wavenumbers, $\alpha_c$ and $\beta_c$, versus $\xi$ for $\Theta_1\equiv\left(0, 20^\circ\right]$. As usual, a dashed line represents the equivalent plots for ACPf (note that for the latter, Squire's Theorem implies $\beta_c = 0$ for all linearly unstable wall speeds). Asymptotic behavior similar to the curves for $Re_c$ is observed for all $\theta$. Furthermore, at wall speeds beyond the cutoff value $\xi_A$ for ACPf, the $\alpha_c$-curves appear to once again have been pulled away from $\alpha_c = 0$ as $\theta$ increases.}
    \label{fig:crit_kx_kz_small_angles}
\end{figure}

Figure \ref{fig:crit_kx_kz_small_angles} presents the spatial wavenumbers at criticality for $\theta\in\Theta_1$. Consistent with the trends observed for ACPf, the critical streamwise wavenumber $\alpha_c$ generally displays an initial monotonic decline toward $\alpha_c = 0$, the latter limit corresponding to the complete loss of linear instability in ACPf beyond $\xi = \xi_A$. However, even for the smallest shear angles treated here, we see that the critical streamwise wavenumber for OCPfs is only close to, but never exactly, zero. Furthermore, in conjunction with the critical Reynolds number, the $\alpha_c$-curves also level out at sufficiently high wall speeds, reinforcing the presence of a limiting regime in modal stability. Meanwhile, the critical spanwise wavenumber $\beta_c$ is generally non-zero, a reminder of the three-dimensional nature of the base flow and the consequent inapplicability of Squire's Theorem. An especially interesting behavior is observed in Figure \ref{fig:crit_kx_kz_small_angles} $(b)$ for a short range around $\xi \approx0.3$, where $\beta_c\approx0$. Here, as shown in Figure \ref{fig:crit_RE_small_angles}, the $Re_c$-curves for OCPfs in $\Theta_1$ also experience an inflection point. The latter is a key stability feature in ACPf, and, in this range, \cite{potter_1966} had noted that $\xi\approx c_{r,c}$, the real part of the $x$-phase speed at criticality, hinting at some sort of link between this equality and the accompanying destabilization. We were able to verify this relation for $\Theta_1$ as well, suggesting that its secondary effect here is a preference for spanwise-independent modes (note that this is implicit for ACPf). Finally, just before this inflectional region, for $\theta=20^\circ$, we can resolve a rather dramatic trough in the $\alpha_c$-curve, which seems to temporarily terminate at the corresponding asymptotic ($\xi$-independent value) before recovering to its original trajectory. We interpret this as the first indication of an imminent departure from the modal characteristics of ACPf, which naturally leads to a discussion of $\Theta_2$, the second stability regime defined by $20^\circ < \theta\leq 90^\circ$.

 \begin{figure}
        \centering
        \includegraphics[width=\textwidth]{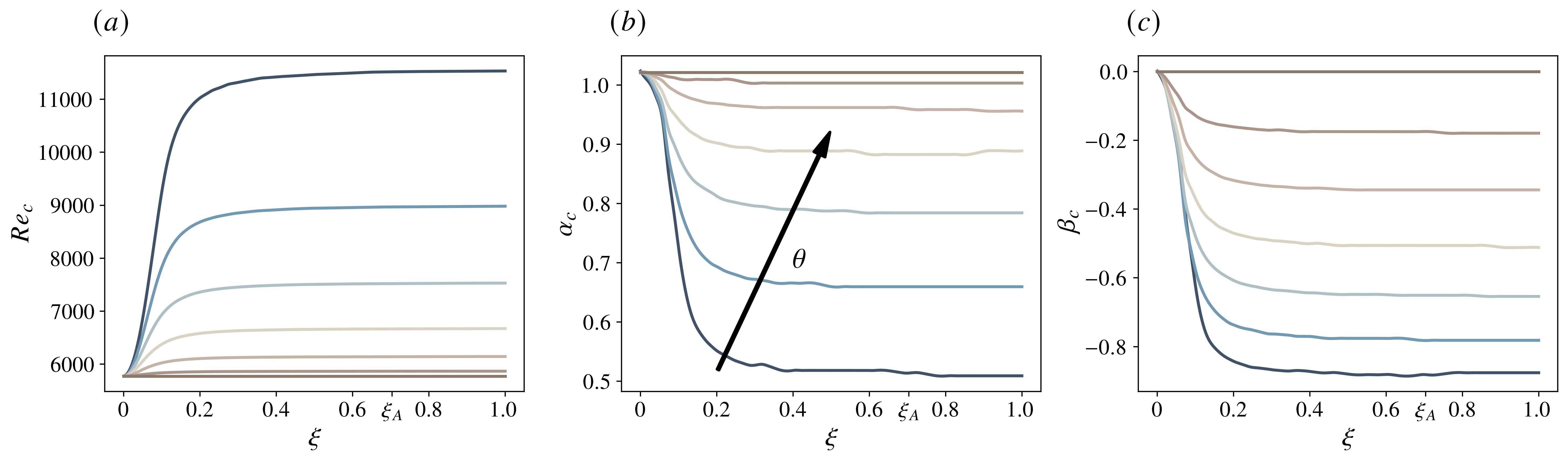}
        \caption{The critical Reynolds numbers and Fourier wavenumbers plotted against $\xi$ for some choices of $\theta\in\Theta_2\equiv\left(20^\circ, 90^\circ\right]$. The black arrow depicts the direction of increasing $\theta$ in increments of $10^\circ$ from $\theta = 30^\circ$ to $\theta = 90^\circ$ (perfect orthogonality). At the latter angle, the critical triplet is constant in $\xi$ and equal to that obtained from an analysis of the two-dimensional Orr-Sommerfeld equation for Poiseuille flow.}
        \label{fig:large_theta_crit_params}
\end{figure}

In particular, Figure \ref{fig:large_theta_crit_params} highlights for $\Theta_2$ a noticeable shift in the stability characteristics of OCPfs. The $Re_c$-curves lose their inflectional nature as in $\Theta_1$, and while increasing the shear angle still induces destabilization, the decrease in $Re_c$ at any $\xi$ is uniform in $\theta$. Furthermore, all critical parameters reach their asymptotic values at smaller wall speeds, doing so by following relatively smoother trajectories (compare, for example, with the $\beta_c$-curves in Figure \ref{fig:crit_kx_kz_small_angles}). From the perspective of the critical Reynolds number, the most unstable OCPf configurations occur as $\theta\to 90^\circ\in\Theta_2$, when the wall motion is perfectly orthogonal to the direction of the pressure gradient. In this limit, the streamwise and spanwise velocities reduce to
\begin{equation}
    U = 1-y^2, \qquad W = \dfrac{\xi}{2}\left(1+y\right),
\end{equation}
which are, respectively, pPf and Couette profiles. At this angle, we found that the critical parameters remained invariant for all the wall speeds studied here, approaching, in fact, the equivalent tuple for pPf. Specifically, we had
\begin{equation}
\label{eqn:ppf_tuple}
    \left(\alpha_c, \beta_c, Re_c\right)_{\theta=90^\circ} = \left(1.02, 0, 5773.22\right),
\end{equation}
effectively indicating, for this $\theta$, a superposition of the stability of the individual velocity components (note that the Couette flow is always linearly stable, see \cite{Romanov1973}). To rationalize this, we recall that the influence of the spanwise crossflow on the OS operator is partially modulated by the wavenumbers, particularly through the effective wall speed $\xi_{\mathrm{eff}}$. Since the corresponding eigenvalue problem can always be mapped to an equivalent one for ACPf, one would wish to somehow negate the Couette contribution, which is known to be stabilizing, in order to ``maximize" instability. This is achieved most optimally in disturbances with $\beta = 0$, immediately reducing the OS operator to that for pPf under Squire's Theorem and yielding the critical tuple in Equation (\ref{eqn:ppf_tuple}). 

For full contour plots of the critical parameters in the $\left(\xi, \theta\right)$-plane as well as a discussion of the critical phase speeds and growth rates, we refer the reader to Appendix \ref{appendix:appB}.

\subsection{The Limiting Regime of Modal Stability}\label{ssec:wavenumber_scaling}
\begin{figure}
    \centering
    \includegraphics[width=\textwidth]{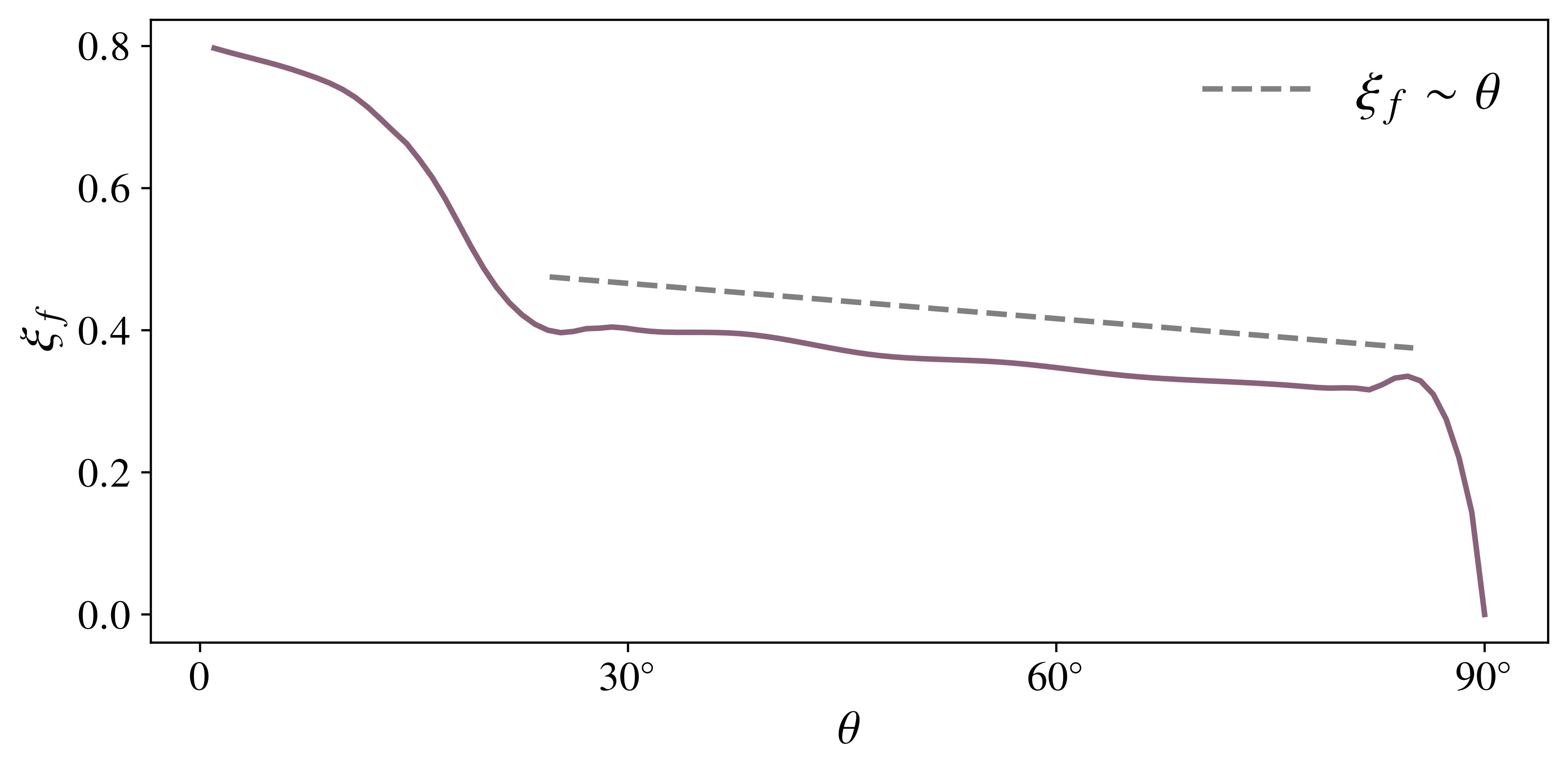}
    \caption{The variation with $\theta$ of $\xi_f$, the wall speed at which the critical parameters asymptote. A dashed line indicates the linear law $\xi_f\sim\theta$, which holds well in $\Theta_2$.}
    \label{fig:xi_asymptote_theta}
\end{figure}

In the previous section, it was observed that when the wall speed is sufficiently high, the stability of oblique Couette-Poiseuille flows becomes independent of $\xi$. The values of $\xi_f$, which represents the approximate wall speed that initiates this limiting regime, are shown in Figure \ref{fig:xi_asymptote_theta}. It can be seen that $\xi_f$ generally decreases with $\theta$, following a roughly linear relationship within the range $\Theta_2$, where OCPfs demonstrate the strongest deviations from the stability characteristics of ACPf. In this section, by adopting a simple juxtaposition with known results on the linear stability of pPf and ACPf, we aim to derive analytical formulae for the asymptotic values of the critical parameters. A small part of the following argument was briefly mentioned earlier when discussing criticality for $\theta=90^\circ$, but will now be elaborated upon.

For purely streamwise velocity profiles, a classic argument due to Squire \citep{squires} is that transversal ($\beta = 0$) modes must become unstable at a lower $Re$ than both longitudinal ($\alpha=0$) and oblique ($\alpha,\beta\neq 0$) modes. The proof proceeds by defining a two-dimensional ($\beta=0$) Orr-Sommerfeld problem and arguing that, at criticality, the corresponding Reynolds number $Re_{c, 2D}$ cannot be larger than that of the three-dimensional ($\beta\neq 0$) case $Re_{c,3D}$. Now, consider ACPf, $\theta=0$, and recall that in the limit $\xi\to 0$, pPf can be recovered. In particular, an application of Squire's Theorem to both these flows yields
\begin{equation}
\label{eqn:squires_acpf}
  Re_{c, 3D}^\mathrm{pPf} > Re_{c, 2D}^\mathrm{pPf},\qquad Re_{c, 3D}^\mathrm{ACPf} > Re_{c, 2D}^\mathrm{ACPf}.
\end{equation}
However, from the work of \cite{potter_1966}, we know that $Re_{c, 2D}^\mathrm{ACPf} \geq Re_{c, 2D}^\mathrm{pPf}$, which immediately allows us to conclude that
\begin{equation}
    Re_{c, 3D}^\mathrm{ACPf} > Re_{c, 2D}^\mathrm{pPf}
\end{equation}
as well. For oblique Couette-Poiseuille flows, due to the mean three-dimensionality, Squire's Theorem is no longer valid. However, Equation (\ref{eqn:relate_ocp_to_acp}) implies that there exists a one-to-one mapping of the OS eigenproblem for an arbitrary OCPf, initialized with any combination of the wavenumbers, to an equivalent (in general) three-dimensional one for ACPf. Therefore, we have that
\begin{equation}
\label{eqn:rec_inequalities}
    Re_c\left(\theta\neq 0, \xi\neq 0; \alpha,\beta\right) = Re_c\left(\theta = 0, \xi = \xi_\mathrm{eff}; \alpha,\beta\right)\geq Re_{c, 2D}^\mathrm{pPf}.
\end{equation}
Here, the merits of the effective wall speed $\xi_\mathrm{eff}$ are once again apparent, as it can be made as large or as small as possible due to its dependence on $\theta$ and, more importantly, on the wavenumbers themselves. Therefore, to optimize the instability for an OCPf, which is equivalent to ``$\geq$'' in Equation (\ref{eqn:rec_inequalities}) approaching equality, the OS operator should degenerate precisely into the two-dimensional analog for pPf. This can happen if and only if
\begin{equation}
\label{eqn:wavenumber_constraints}
    \xi_\mathrm{eff} = \xi\left(\cos\theta+\gamma \sin\theta\right) = 0, \qquad \alpha^2+ \beta^2 = \alpha_{c,\mathrm{pPf}}^2 \qquad \alpha Re = \alpha_{c,\mathrm{pPf}}Re_{c,\mathrm{pPf}}
\end{equation}
\begin{figure}
    \centering
    \includegraphics[width=0.95\textwidth]{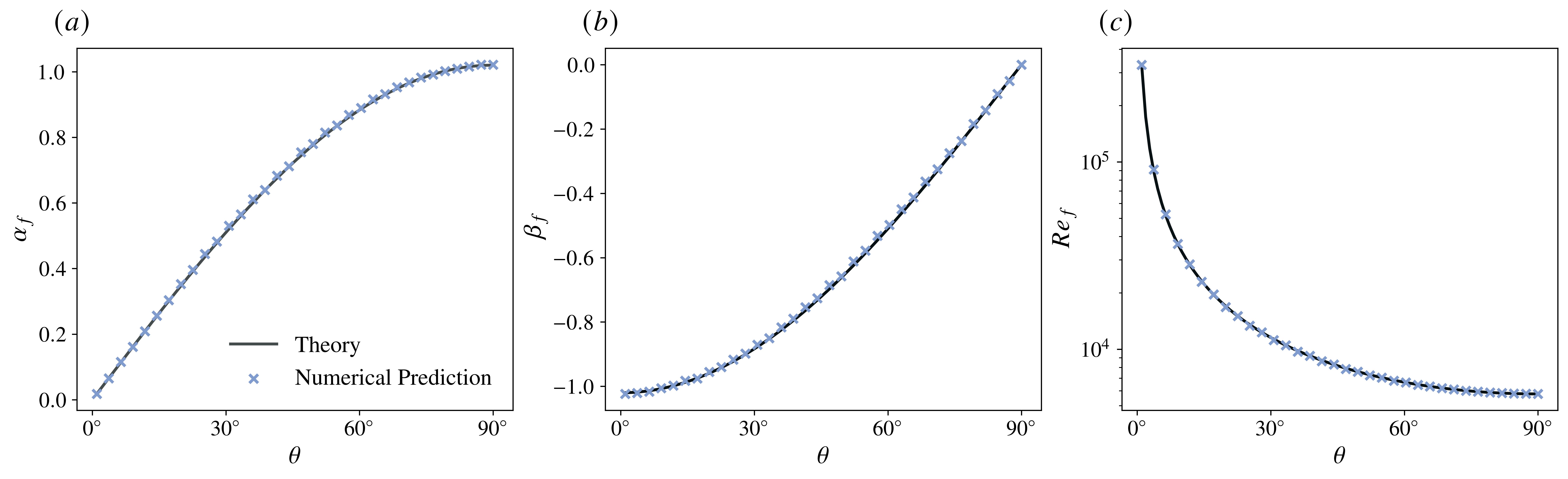}
    \caption{The asymptotic values of $(a$--$b)$, the critical streamwise and spanwise wavenumbers and $(c)$, the critical Reynolds number versus $\theta$. The solid line denotes the theoretical estimate provided in Equation (\ref{eqn:asymptotic_params}).}
    \label{fig:analytical_vs_data}
\end{figure}where $\gamma = \beta/\alpha$ and $\left(\alpha_{c,\mathrm{pPf}}, Re_{c,\mathrm{pPf}}\right)\approx \left(1.02, 5773.22\right)$; see \cite{orszag_1971}. Assuming $\xi\neq 0$, the first of these three equations yields
\begin{equation}
\label{eqn:asymptotic_angle}
    \gamma = -\cot\theta
\end{equation}
a result that can be combined with the remaining constraints in Equation (\ref{eqn:wavenumber_constraints}) to obtain the following closed solutions
\begin{equation}
\label{eqn:asymptotic_params}
    \alpha = \alpha_{c,\mathrm{pPf}}\left|\sin \theta\right|, \qquad \beta = -\alpha_{c,\mathrm{pPf}}\,\mathrm{sgn}\left(\sin \theta\right)\cos \theta \qquad Re = Re_{c,\mathrm{pPf}}\left|\csc\theta\right| 
\end{equation}
where $\mathrm{sgn}$ represents the signum function and we have selected the positive solution for $\alpha$. Figure \ref{fig:analytical_vs_data} presents the asymptotic values of the critical parameters, denoted by the subscript $\left\langle\cdot\right\rangle_f$, obtained from our numerical results overlaid with the analytical solution in Equation (\ref{eqn:asymptotic_params}). An almost exact match is observed up to the resolution error of the domain sweep. Some crucial remarks can now be made. As the shear angle approaches zero, Equation (\ref{eqn:asymptotic_params}) claims that the asymptotic streamwise wavenumber vanishes, while the asymptotic spanwise wavenumber experiences a discontinuity. Although this may seem erroneous at first glance, we note that the critical parameters in ACPf continuously vary with wall speed, showing no limiting behavior, so Equation (\ref{eqn:asymptotic_params}) has no meaning in this limit anyway. On the other hand, Figure \ref{fig:crit_kx_kz_small_angles} $(a)$ seems to suggest that a flattening of the critical parameters for ACPf, if it occurs, should do so for $\alpha\to 0$. Meanwhile, we note that for $\theta\to \upi/2$, an asymptotic spanwise wavenumber $\beta_f=0$ is predicted, which validates our findings in Section \ref{ssec:criticality}.

\begin{figure}
    \centering
    \includegraphics[width=\textwidth]{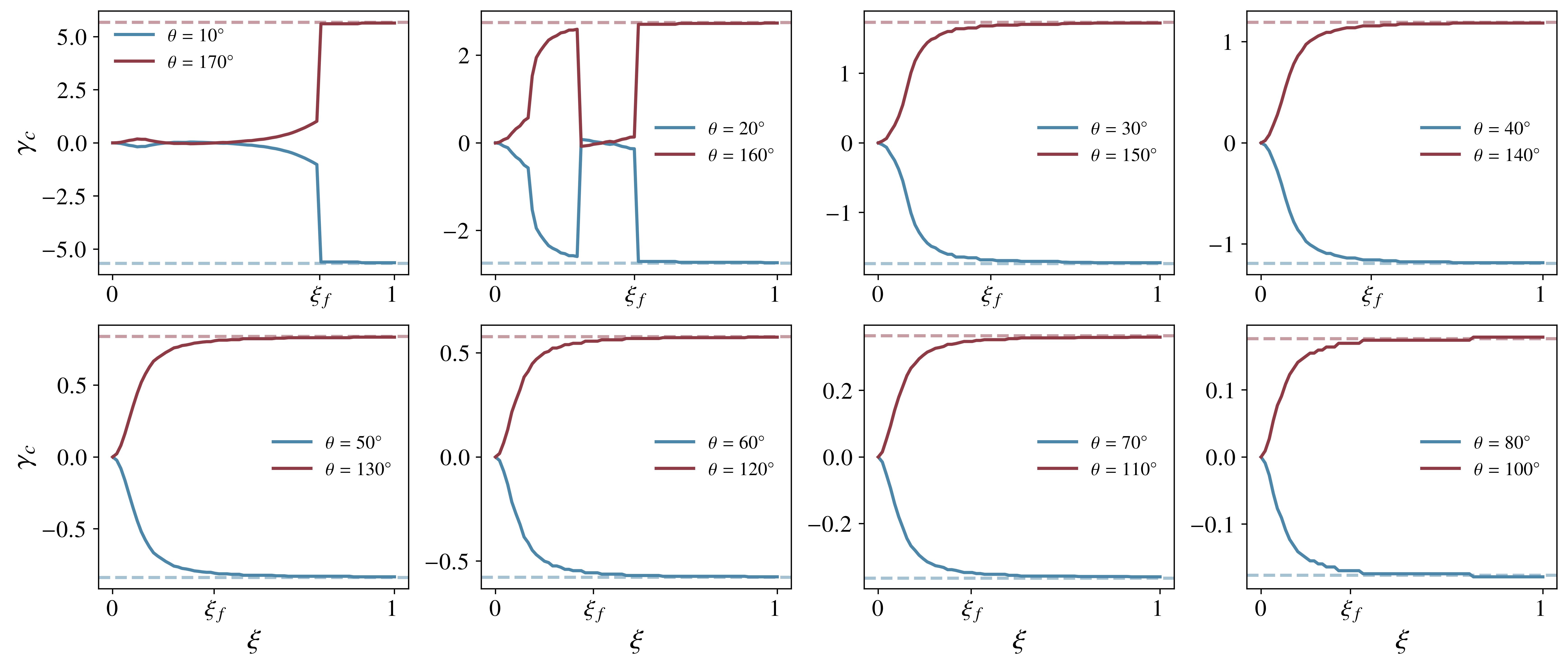}
     \caption{The ratio $\gamma = \beta/\alpha$ at criticality versus the wall speed for various pairs of supplementary angles $\left(\theta, 180^\circ - \theta\right)$. In each case, a dashed line indicates the asymptotic value, $\gamma_c = -\cot\theta$, Equation (\ref{eqn:asymptotic_angle}). Notice that for $\xi\leq \xi_f$, $\gamma_c$ has no discernible order although for the full range of wall speeds, $\gamma_c\left(\upi-\theta\right)$ is a reflection of $\gamma_c\left(\theta\right)$ around the line $\gamma_c = 0$.}
\label{fig:wavenumber_scaling}.
\end{figure}

In Figure \ref{fig:wavenumber_scaling}, we present $\gamma_c$, the ratio $\gamma$ at criticality for various symmetrically chosen shear angles around $\theta = \upi/2$, which should be interpreted in light of Equation (\ref{eqn:asymptotic_angle}). We first see that $\gamma_c\to 0$ as $\xi\to 0$; this is expected, of course, since pPf is recovered in this limit, for which the most unstable disturbance has a vanishing spanwise wavenumber. Furthermore, we observe that $\gamma_c\left(\upi-\theta\right) = -\gamma_c\left(\theta\right)$, which supports the initial restriction of the state space described at the beginning of Section \ref{ssec:criticality}. Finally, for $\xi\geq \xi_f$, the $\gamma_c$-curves in Figure \ref{fig:wavenumber_scaling} plateau precisely at $\gamma_c = -\cot\theta$, as predicted by Equation (\ref{eqn:asymptotic_angle}). An interesting consequence arises when considering the wavenumber vector $\boldsymbol{k} = \begin{pmatrix} \alpha & \beta \end{pmatrix}^\intercal$. In wave theory, the wavenumber vector encodes the direction of wave motion, and a wave with wavenumber vector $\boldsymbol{k}$ propagates at an angle $\psi$ to the positive streamwise direction, where $\tan\psi = \beta/\alpha = \gamma$. From Equation (\ref{eqn:asymptotic_angle}), we can then conclude that the asymptotic eigenmode propagates at an angle $\psi=\theta - \upi/2$ to the pressure gradient, that is, exactly perpendicular to the wall motion.

\subsection{Modified Long-Wavelength Analysis}\label{ssec:long_wave}

A notable feature in the stability analysis of aligned Couette-Poiseuille flow is the unconditional stabilization achieved beyond $\xi = \xi_A$, the so-called \textit{cutoff velocity}. \cite{potter_1966} estimated this wall speed to be $\xi_A\approx 0.7$, which was later validated by \cite{cowley_smith_1985} through a weakly nonlinear analysis using the scaling $\alpha\sim Re^{-1}$ of the lower and upper branches of the neutral curves in the high-$Re$ limit. The authors employed a long-wavelength distinguished limit to the Orr-Sommerfeld equations, allowing $\left(\alpha, Re\right)\to \left(0,\infty\right)$ and fixing the product $\lambda^{-1} = \alpha Re$. Numerically, a parameter sweep can then be performed in the $\left(\xi, \lambda\right)$-plane to determine the maximum $\xi$ for which the asymptotic Orr-Sommerfeld operator displays neutral stability; this is the cutoff value.

In Section \ref{ssec:criticality}, we saw that, unlike ACPf, OCPfs remained linearly unstable beyond $\xi=\xi_A$, even for small but non-zero $\theta$, instead accessing a $\xi$-independent asymptotic regime described by the convergence of the critical parameters. This suggests the absence of an equivalent cutoff threshold when $\theta\neq 0$. To verify this claim, we augment the long-wavelength approach of \cite{cowley_smith_1985} to account for the three-dimensionality of OCPfs. Our modification is simple and involves treating the additional factor of $\gamma\equiv\beta/\alpha$ in $\xi_\mathrm{eff}$ through an appropriate limit for the spanwise wavenumber. In particular, we propose $\beta\to 0$ and allow $\gamma$ to remain finite, supporting this choice as follows.

First, to achieve complete linear stability in aligned Couette-Poiseuille flow, a vanishing spanwise wavenumber is implicit from Squire's Theorem. To establish this, note that for a purely streamwise base flow, Squire's argument proceeds by reducing the three-dimensional Orr-Sommerfeld operator to a two-dimensional one (for which $\beta=0$) by considering, among others, the following relation between the wavenumbers of the two eigenproblems
\begin{equation}
\label{eqn:squire_relation}
    \alpha_{2D}^2 = \alpha_{3D}^2 + \beta_{3D}^2
\end{equation}
For a temporal stability problem as in this study, the wavenumbers are always assumed to be strictly real. Therefore, if $\alpha_{2D}\to 0$ at the cutoff wall speed, the positive semi-definiteness of the right-hand side of Equation (\ref{eqn:squire_relation}) implies that each of $\alpha_{3D}$ and $\beta_{3D}$ must vanish identically. Despite this, we recall that Squire's result has no immediate equivalent for mean three-dimensional base flows. However, our stability results and those of other canonical shear flows that are known to be unstable suggest that marginal stability, if it exists, usually does so only for small to intermediate -- or even vanishing -- values of $\gamma$.

\begin{figure}
    \centering
    \includegraphics[width=\textwidth]{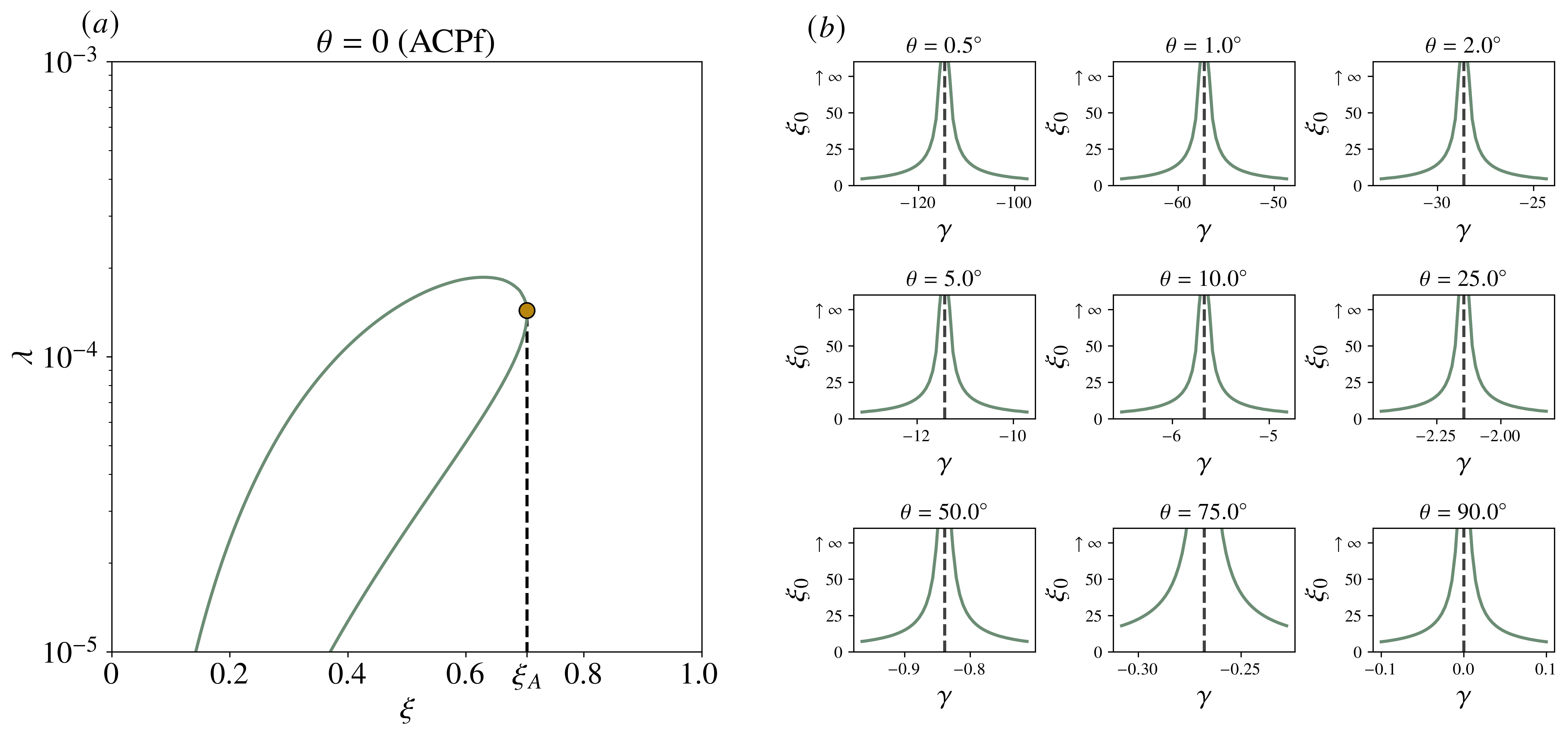}
    \caption{Long-wavelength instability in oblique Couette-Poiseuille flows; $(a)$, the neutral curve in the $\left(\lambda, \xi\right)$-plane for ACPf calculated from the asymptotic Orr-Sommerfeld problem with $\gamma = 0$ -- the cutoff pair is denoted by a yellow circle and the vertical dashed line indicates the associated wall speed $\xi_A$; $(b)$, plots of $\xi_0$ for various $\theta$ centered around the wavenumber ratio of interest $\gamma=-\cot\theta$, again denoted by a dashed line. Although our numerical search was inevitably restricted to an upper limit for $\xi$, we noted that the results remained robust to even higher values.}
    \label{fig:long_wavelength_analysis}
\end{figure}

For the modified version of the long-wavelength analysis, the Orr-Sommerfeld eigenvalue problem becomes
\begin{equation}
\label{eqn:long_wavelength_eq}
    \left[i\lambda\mathcal{D}^4 + \left(U+\gamma W - c\right)\mathcal{D}^2 - \mathcal{D}^2 U - \gamma\mathcal{D}^2W\right]\hat{v} = 0
\end{equation}
Note that as $\theta\to 0$ or $\gamma\to 0$, we recover the standard long-wavelength equations for ACPf as in \cite{cowley_smith_1985}. For OCPfs, the state-space of the asymptotic problem is $\left\{\xi, \theta, \lambda, \gamma\right\}$, and we focus particularly on the quantity $\xi_0$, defined as follows
\begin{equation}
\label{eqn:xi_o}
    \xi_0\left(\theta^\prime, \gamma^\prime\right) = \left\{\max\xi\mid c_i\left(\lambda, \xi, \theta=\theta^\prime, \gamma = \gamma^\prime\right) = 0\right\}
\end{equation}
At a fixed $\theta$, $\xi_0$ represents the largest wall speed capable of sustaining neutral stability for wavenumber ratios $\gamma
$. Numerically, therefore, Equation (\ref{eqn:xi_o}) is equivalent to determining the cutoff wall speeds in two-dimensional $\gamma$-slices of the three-dimensional stability manifold embedded in $\left(\xi, \lambda, \gamma\right)$-space. Figure \ref{fig:long_wavelength_analysis} summarizes the results of this approach. A benchmark run was performed by constructing the long-wavelength neutral curve in the $\left(\lambda, \xi;\gamma=0\right)$-plane for ACPf, as shown in Figure \ref{fig:long_wavelength_analysis} $(a)$ -- the cutoff wall speed was calculated as $\xi_A\approx0.70370$ and compares well with the ground truth. Figure \ref{fig:long_wavelength_analysis} $(b)$, on the other hand, plots $\xi_0$ for some representative values of $\theta$. We see that, irrespective of the direction of wall motion, there exist peaks in $\xi_0$ that appear to extend to infinity. In fact, this divergence occurs precisely when $\gamma = -\cot\theta$, that is, when $\gamma$ obeys the relation provided in Equation (\ref{eqn:asymptotic_angle}). This result should, of course, not be surprising given the nature of the asymptotic regime. As discussed in Section \ref{ssec:criticality}, OCPfs experience a modal instability throughout $\xi\in\left[0, 1\right]$, whose dependence on wall speed ultimately drops completely when $\xi\geq \xi_f$. Since the existence of a cutoff velocity in ACPf is necessarily related to the strength of wall motion and not to any inherent skewness of the flow, we conclude that in this range, likely even as $\xi\to\infty$, instability must always persist given the right wavenumber combination. In other words, for OCPfs, if a cutoff threshold for unconditional stability exists, it cannot do so at finite $\xi$. Note that no formal conclusions can be drawn about the critical Reynolds number under this paradigm; however, while it may very well be extremely large, it must remain finite.

Interestingly, for all $\theta$ considered here, Figure \ref{fig:long_wavelength_analysis} $(b)$ highlights the presence of a long-wavelength instability at the asymptotic wavenumber ratio $\gamma =-\cot\theta$ even for $\xi\leq \xi_f$. At the same time, in this range, Sections \ref{ssec:criticality} and \ref{ssec:wavenumber_scaling} (see, for example, Figure \ref{fig:wavenumber_scaling}) demonstrate that the associated critical wavenumbers do not necessarily follow the asymptotic laws derived in Equations (\ref{eqn:asymptotic_angle}) and (\ref{eqn:asymptotic_params}). Therefore, for these wall speeds, we can infer that although disturbances with wavenumber ratios $\gamma=-\cot\theta$ can potentially suffer exponential modal growth, they are not the fastest to do so (in the sense of a minimal Reynolds number required to achieve a positive growth rate). However, as $\xi$ increases beyond $\xi_f$, while other wavenumber ratios stabilize (again, in the sense of a minimal Reynolds number), the effect, if any, on the asymptotic value is likely negligible -- we discuss this further in Appendix \ref{appendix:appC}.

\subsection{Eigenmodes and Linear Energetics}
\begin{figure}
    \centering
    \includegraphics[width=0.9\textwidth, height=0.93\textheight]{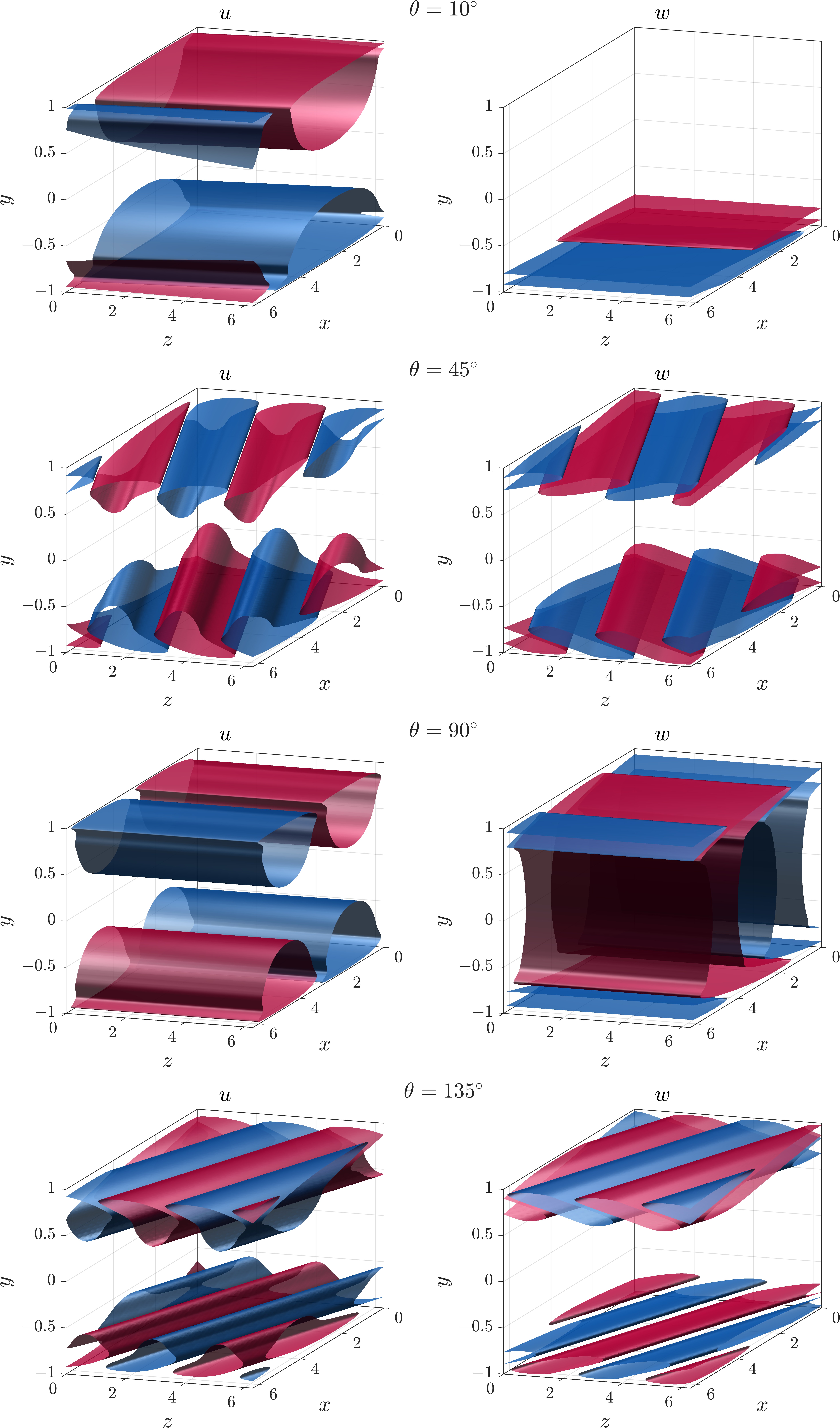}
    \caption{Iso-surfaces of the streamwise $u$ and spanwise $w$ velocity fluctuations for the most unstable eigenmode at $\xi=0.35$ for different $\theta$. For each case, the blue and red contours represent 25\% of the (signed) minimum and maximum values of the perturbations, respectively.}
    \label{fig:eigenmodes}
\end{figure}

Figure \ref{fig:eigenmodes} illustrates for $\xi=0.35$ the spatial distributions of $u$ and $w$, which are, respectively, the streamwise and spanwise velocity perturbations associated with the most unstable eigenmode at criticality. In a two-dimensional shear flow such as ACPf, this instability is instigated by the Tollmien-Schlichting (TS) wave, which comprises spanwise-elongated streamwise-propagating rolls. The TS instability supports the so-called ``classical" route to turbulence, where, under sufficient exponential amplification, secondary instabilities are generated and lead to non-linear breakdown and, thereafter, transition \citep{herbert, schmidstability, cossubrandt}.

For $\theta=10^\circ\in\Theta_1$, it is observed that both disturbance components propagate approximately parallel to the streamwise direction. This can be attributed, in part, to the critical spanwise wavenumber being close to zero, which, as shown in Figure \ref{fig:crit_kx_kz_small_angles} $(b)$, is typically the case for intermediate wall speeds within this range of angles. The streamwise fluctuations bear some resemblance to a TS wave, suggesting a possible similarity between the mechanisms of modal transition in ACPf and weakly skewed OCPfs. However, the spanwise fluctuations are non-zero and consist of weakly parallel flattened structures that are localized near the lower wall. The exact role of these structures in the transition process is not immediately clear and requires further numerical investigation, which is beyond the scope of this paper. On the other hand, for $\theta = 45^\circ\in\Theta_2$ and $\theta = 135^\circ$, $\xi=0.35\approx\xi_f$, indicating that the stability characteristics of both flows are close to the asymptotic regime. Therefore, the most unstable wavenumber pair satisfies Equation (\ref{eqn:asymptotic_angle}) and is oblique. Consistent with the argument presented in Section \ref{ssec:wavenumber_scaling}, we observe that the associated eigenmode propagates exactly perpendicular to the direction of motion of the wall. The streamwise and spanwise fluctuations are qualitatively similar, both consisting of vortices tilted slightly away from each end of the channel. The cross-sections of these vortices for $u$ are somewhat distorted, while for $w$ they have a more regular, elliptical shape. Furthermore, we found that between $\theta$ and $180^\circ-\theta$, the support of these structures moved from the lower to the upper wall, although this effect was not very noticeable.

The most interesting behavior is observed for $\theta = 90^\circ$, when the pressure gradient and the wall velocity vectors are exactly orthogonal. First, we recall from Equation (\ref{eqn:ppf_tuple}) that despite the value of $\xi$ for this configuration, $\beta_c = 0$, and the Orr-Sommerfeld operator can always be identified with that for the plane-Poiseuille flow. Thus, the wall-normal and streamwise components (see Equation (\ref{c_operator})) of the disturbance are identically invariant with the wall speed, precisely reducing to the TS instability found in pPf. However, unlike the latter flow, for which $\beta\to 0$ induces a vanishing normal vorticity in the (resulting) homogeneous Squire equation, the spanwise fluctuations are no longer zero due to the spanwise shear in the off-diagonal forcing term, Equation (\ref{eqn:orr_somm}). They comprise bands of transverse arch-like structures that, to the best of our knowledge, have not been previously recorded in the linearized analysis of any canonical shear flow. Moreover, varying $\xi$ had little effect on the shape of these modes, appearing to change only their energy.

\begin{figure}
    \centering
    \includegraphics[width=\textwidth]{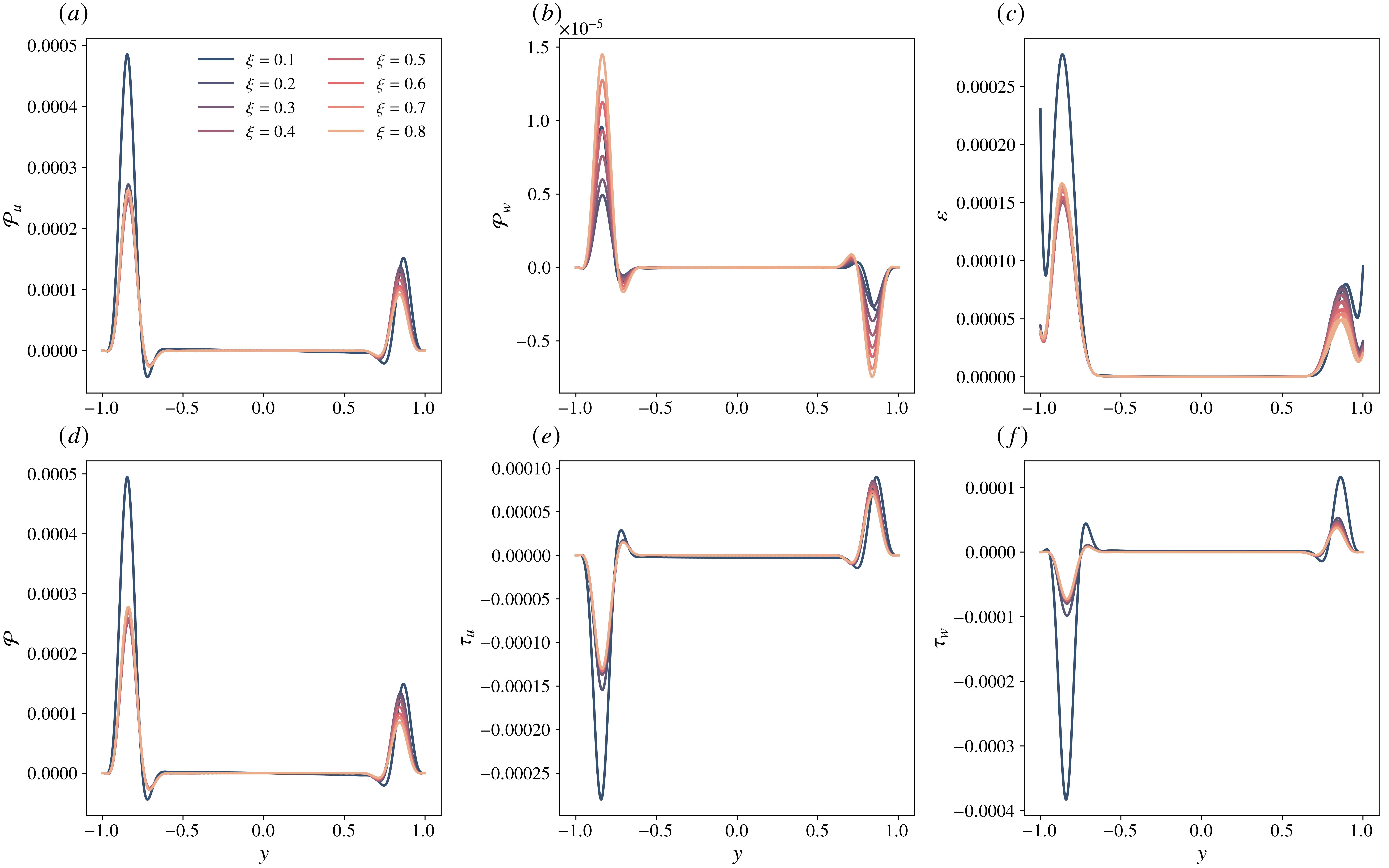}
    \caption{At criticality, the spatial distribution of the perturbation energy budget terms for $\theta=30^\circ$ and wall speeds between $\xi=0.1$ to $\xi=0.8$ in increments of 0.1. From left to right, $(a)$, the streamwise $\mathcal{P}_u$ and $(b)$, the spanwise $\mathcal{P}_w$ production, $(c)$, the viscous dissipation $\varepsilon$, $(d)$, the total production ($\mathcal{P}=\mathcal{P}_u+\mathcal{P}_w$), and the Reynolds stresses, $(e)$, $\tau_u$ and $(f)$, $\tau_w$.}
    \label{fig:energy_budget}
\end{figure}

Figure \ref{fig:energy_budget} shows the linear energy budget at criticality for the angle $\theta = 30^\circ$. As the wall speed increases from $\xi=0.1$ to $\xi=0.2$, the streamwise production $\mathcal{P}_u$ decreases somewhat dramatically in the lower half of the channel, consistent with the initial rise in $Re_c$ seen in Figure \ref{fig:large_theta_crit_params} $(a)$. However, for $\xi\geq \xi_f$, it eventually stagnates near the stationary wall, whereas a continuous, although noticeably slow, decline is observed near the moving wall. On the other hand, while $\mathcal{P}_w$ seems to always suffer from a region of negative production near the upper wall, it operates at least one order of magnitude lower than $\mathcal{P}_u$. As a result, the total production $\mathcal{P}\approx\mathcal{P}_u$ remains positive throughout most of the channel. Viscous dissipation decreases with increasing $\xi$ and, similar to $\mathcal{P}_u$ (and, therefore, $\mathcal{P}$), it converges to some extent for high wall speeds. Therefore, even from the standpoint of the linear energy budget, there is a clear indication of modal stability in OCPfs approaching asymptotic regimes, where the potential for exponential amplification at criticality seems to become entirely agnostic to the wall speed.

\begin{figure}
    \centering
    \includegraphics[width=\textwidth]{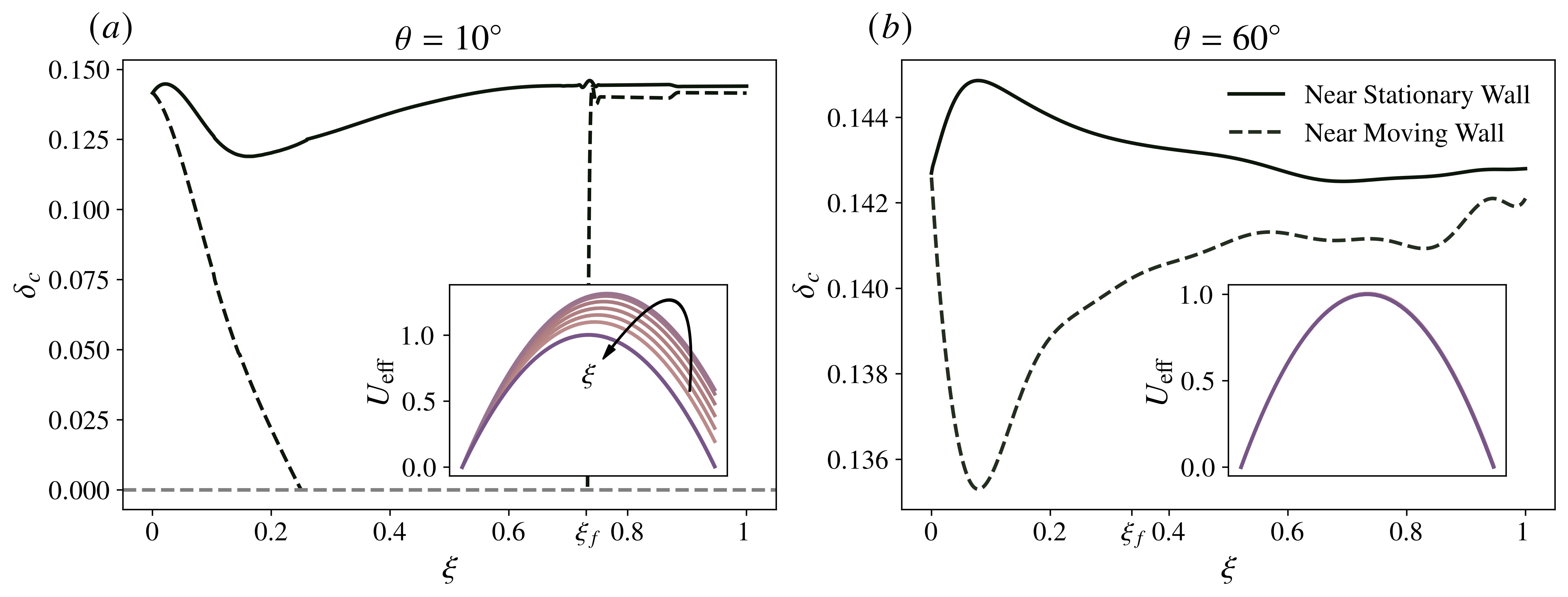}
    \caption{The development of the thicknesses $\delta_c$ of the critical layers for representative angles from each stability regime. A solid versus dashed line is used to distinguish the lower, stationary, wall from the upper, moving, wall. In each case, an inset illustrates the effective mean velocity profiles $U_\mathrm{eff}$ for values of the wall speed between $\xi = 0.2$ to $\xi=1$ in increments of 0.1. For $\theta=10^\circ$, an arrow depicts the direction of increasing $\xi$ (note that $\xi\to\xi_f$ implies $\xi_\mathrm{eff} \to 0$). For $\theta=60^\circ$, while not immediately apparent, the effective velocity profiles for the wall speeds chosen here coincide almost exactly.}
    \label{fig:crit_layers_ocpf}
\end{figure}

The spatial variation of the Reynolds stresses is also shown in Figure \ref{fig:energy_budget} $(e$--$f)$. Regarding $\tau_u$, we see that while it is primarily negative near the stationary wall, it is \textit{always} positive near the moving wall. This is in stark contrast to ACPf and many of its variants, for example, \citep{sadeghihiggins, nouar, GUHA_2010}, for which an increasing wall speed also generates a region of \textit{negative} stress in the upper half of the channel. Since the base streamwise shear in ACPfs is (typically) also negative throughout this region, the latter phenomenon decreases the overall energy production, stabilizing the flow -- see Appendix \ref{appendix:appA}. A related commentary can be made on the changes and eventual disappearance of the \textit{critical layers} at each wall, which, for strictly streamwise flows, are the wall-normal locations where the streamwise velocity matches the real part $c_r$ of the (critical) $x$-phase speed. In ACPf, the critical layer near the moving wall is known to vanish as $\xi$ increases, which is often identified with stabilization. For OCPfs, due to the structure of the Orr-Sommerfeld equation, an analogous argument can be constructed using the effective velocity profile
\begin{equation}
    U_\mathrm{eff} = U + \gamma W = 1-y^2 + \dfrac{\xi_\mathrm{eff}}{2}\left(1+y\right)
\end{equation}
which essentially represents a projection of the base velocity in the direction of the wavenumber vector \citep{schmidstability}. Here, we are interested in the wall-normal location(s) $y_c$ such that $U_\mathrm{eff}\left(y_c\right) = c_r$. In general, since the effective velocity is quadratic in $y$, two such points can exist, associated with each wall, and are explicitly given in closed form as follows
\begin{equation}
\label{eqn:critical_roots}
    y_c = \dfrac{1}{4}\left(\xi_\mathrm{eff} \pm \sqrt{\left(4+\xi_\mathrm{eff}\right)^2 - 16c_r}\right)
\end{equation}
The thickness of each layer can then be expressed as $\delta_c = 1-\left|y_c\right|$, and is shown in Figure \ref{fig:crit_layers_ocpf} for two shear angles, $\theta=10^\circ\in\Theta_1$ and $\theta=60^\circ\in\Theta_2$. For the first case, similar to ACPf, both critical layers initially become thinner. Eventually, near the point of inflection in the associated $Re_c$-curve, the critical layer near the moving wall vanishes, whereas the one near the lower wall continues to thicken and stabilize the flow. Mathematically, this can be attributed to changes in the asymmetry of $U_\mathrm{eff}$, which, in turn, are influenced by variations in the effective wall speed $\xi_\mathrm{eff}$. However, a particularly intriguing behavior is observed when $\xi\geq \xi_f$. In this case, the thickness $\delta_c$ for the moving wall experiences a sudden increase from zero to a roughly constant value approximately equal to that of the lower wall. This can be explained by the constraints on the asymptotic wavenumber pair, Equation (\ref{eqn:wavenumber_constraints}), which enforce $\xi_\mathrm{eff}\approx 0$ in the asymptotic regime and reduce $U_\mathrm{eff}$ to a \textit{symmetric} parabolic profile. As a result, the absolute values of the roots in Equation (\ref{eqn:critical_roots}) coalesce and remain $\left|y_c\right| \leq 1$, which appears to be consistent with the plateauing observed in the critical parameters, Figures \ref{fig:crit_RE_small_angles} and \ref{fig:crit_kx_kz_small_angles}. In a similar vein, for $\theta=60^\circ$, because the asymptotic regime is accessed earlier (that is, at smaller $\xi$), the effective velocity profile is almost always perfectly symmetric, allowing the thicknesses to remain roughly identical and non-zero throughout most of the $\xi$ range explored here. On a separate note, for $\theta=60^\circ$, Figure \ref{fig:crit_layers_ocpf} $(b)$ highlights an initial increase in $\delta_c$ near the stationary wall, which seems to support the monotonicity of the corresponding $Re_c$-curve in this range, Figure \ref{fig:large_theta_crit_params}. Furthermore, considering the behavior of the critical layers when the wall speed approaches the point of inflection in the $Re_c$-curves for ACPf and OCPfs in $\Theta_1$, it is likely that the absence of such a feature for $\Theta_2$ is related to the fact that both critical layers remain intact at these angles.

\section{Non-Modal Analysis}\label{sec:nm_analysis}
Despite the results of an eigenvalue analysis, the transition to turbulence in many canonical flows is often sub-critical, meaning it occurs in the absence of linear instability. A classic example of this is Hagen-Poiseuille flow, which is widely considered to be linearly stable \citep{schmidstability, drazin_reid_2004}, yet is known to transition at $Re\gtrapprox2000$ \citep{reynolds_pipe_exp, wygnanski_champagne_1973, avila_transition}. A potential explanation for this discrepancy lies in non-modal growth, which is a purely linear phenomenon that can be ascribed to the non-normality of the linearized Navier-Stokes operator, $\mathsfbi{S}$. In shear flows, this amplification can be many orders of magnitude stronger than the growth rates predicted by a spectral analysis \citep{tref_pseudospec, brandt_lift_up}. Non-modal theory has accurately predicted the short-term energy growth observed in numerical \citep{bobergbrosa} and experimental \citep{klingmann_1992, reshotko} studies, as well as the role of streamwise vortices in bypass transition \citep{brandt_lift_up}. Such a framework has also been successful in characterizing streamwise streaky structures typical in fully turbulent boundary layers \citep{butler_streaky, cossu}.

\begin{figure}
    \centering
    \includegraphics[width=\textwidth]{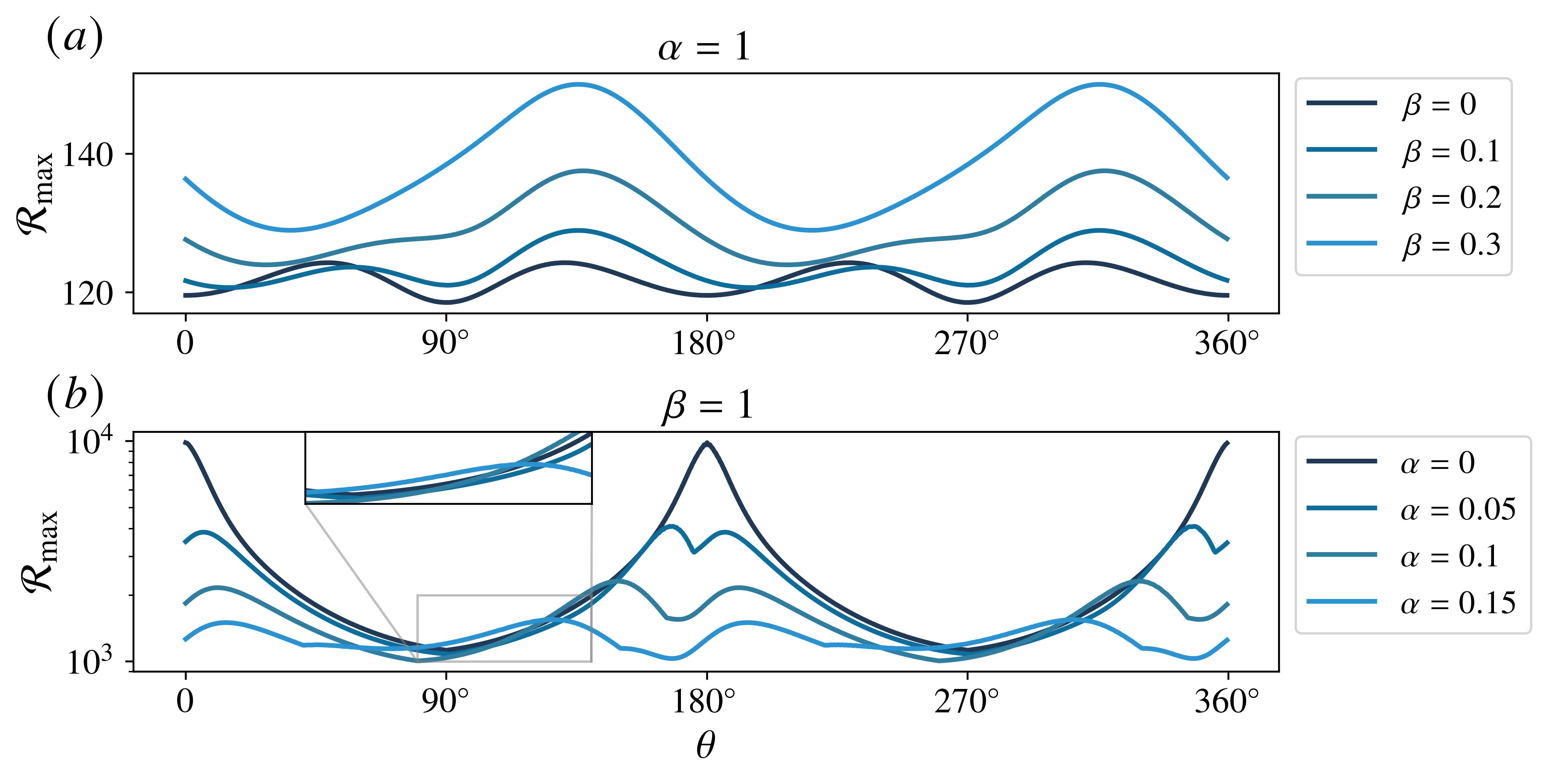}
    \caption{Plots of $\mathcal{R}_{\max}$, the maximum of the resolvent energy norm across all forcing frequencies $\zeta$ for $\xi=0.35$ and $Re = 1500$; $(a)$, $\alpha=1$ and $\beta$ varied in increments of $0.1$; $(b)$, $\beta=1$ and $\alpha$ varied in increments of $0.05$. In $(b)$, the inset zooms in on the region where $\mathcal{R}_{\max}$ appears to increase in conjunction with $\alpha$. Note the periodicity in $\theta$.}
    \label{fig:res_norm}
\end{figure}

We start by investigating the resolvent $\mathsfbi{R}$ and its associated energy norm $\mathcal{R}\equiv\left\lVert\mathsfbi{R}\right\rVert_E$. For some representative pairs of wavenumbers, Figure \ref{fig:res_norm} visualizes $\mathcal{R}_{\max}$, defined as
\begin{equation}
    \mathcal{R}_{\max} = \max_{\zeta} \mathcal{R}\left(\alpha,\beta, Re, \xi, \theta, \zeta\right),
\end{equation}
 for $Re = 1500$, which is sub-critical for all possible OCPf configurations. As is the case with many of the quantities explored in Section \ref{sec:modal_analysis_section}, the periodicity of the base flow in $\theta$ once again embeds itself in $\mathcal{R}_{\max}$, which appears to be $\upi$-periodic. On the other hand, the forcing frequency $\zeta$ that gives rise to $\mathcal{R}_{\max}$ was found to be $2\upi$-periodic, though it is not shown here. Comparing Figures \ref{fig:res_norm} $(a)$ and $(b)$, we immediately see that the response of spanwise-independent disturbances is relatively damped compared to disturbances with $\alpha=0$, a difference that spans approximately an order of magnitude. In the former case, allowing for a weak $z$-dependence by steadily varying $\beta$ produced only a negligible increase. However, for disturbances of the second kind, changes in the streamwise wavenumber appeared to have a more diverse effect. In particular, while $\mathcal{R}_{\max}$ decreased for, say, $\theta\lessapprox90^\circ$, it actually \textit{increased} for $90^\circ\lessapprox\theta\lessapprox165^\circ$. This is significantly different from many classical two-dimensional flows, where it is usually disturbances completely independent of $x$ that elicit the most vigorous response \citep{tref_pseudospec, schmidstability}.

\begin{figure}
    \centering
    \includegraphics[width=\textwidth]{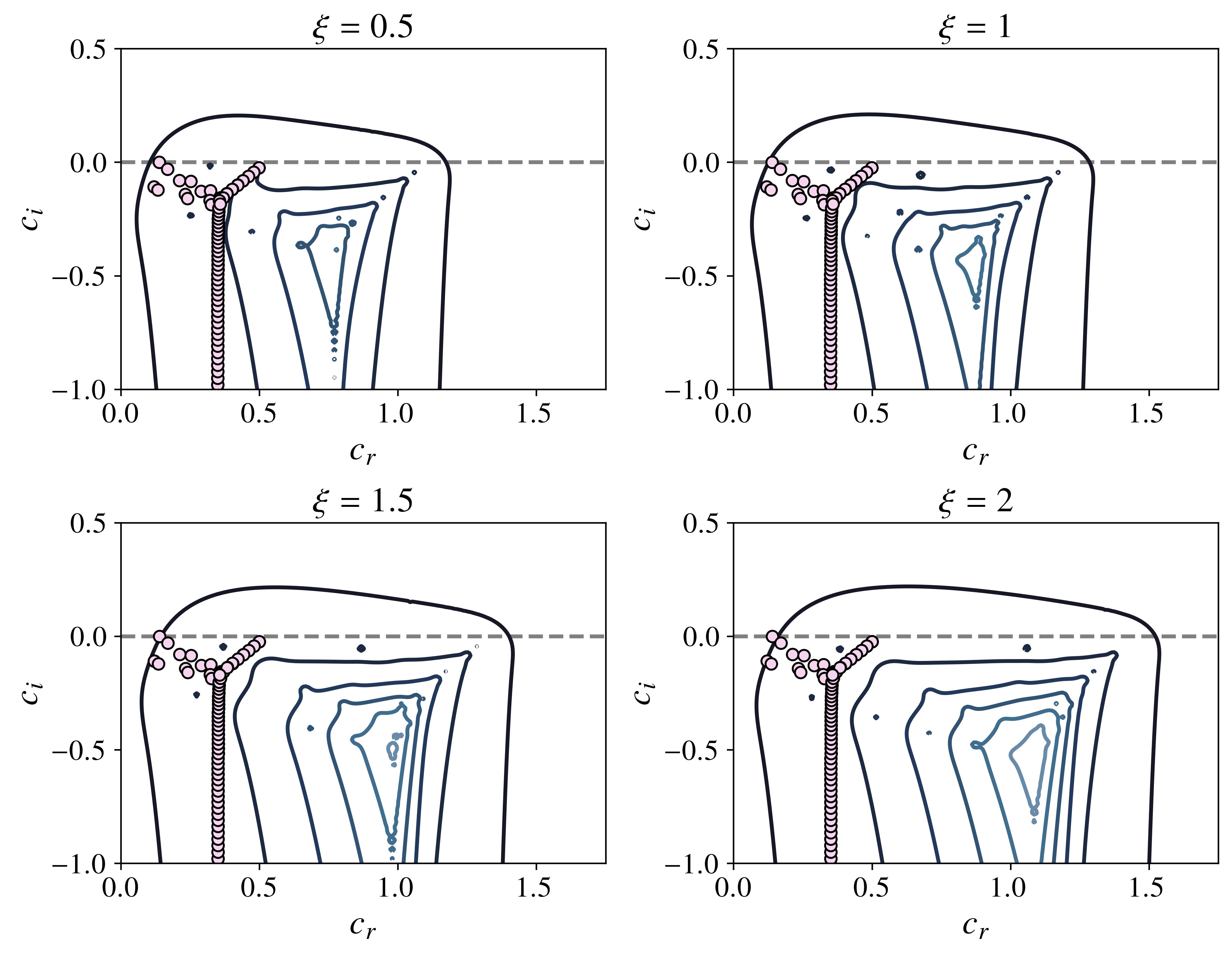}
      \caption{For $\theta=30^\circ$ and various $\xi\geq \xi_f$, the logarithmic level curves at $\left(\alpha_f,\beta_f,Re_f\right)$ for the $\epsilon$-pseudospectra from $\log\epsilon = -1$ to $\log\epsilon = -11$ (outer to inner) in decrements of $-2$. The Orr-Sommerfeld modes in the $x$-phase speed formulation are depicted via circles; notice these appear unchanged with $\xi$, a consequence of the asymptotic triplet enforcing a reduction to the ``same" eigenvalue problem (see Appendix \ref{appendix:appC}). A dashed line indicates the stability boundary $c_i = 0$.}
    \label{fig:pseudo_spec}
\end{figure}

For more general forcing frequencies, $\zeta\in\mathbb{C}$, Figure \ref{fig:pseudo_spec} illustrates for $\theta = 30^\circ$ the contours of $\left\lVert \mathsfbi{R}^{-1}\right\rVert_{-E}$ in the complex plane, where $\left\lVert \cdot\right\rVert_{-E}$ is the ``inverse" energy norm
\begin{equation}
    \left\lVert \mathsfbi{R}^{-1}\right\rVert_{-E} = \sigma_{\min}\left(\mathsfbi{F}\mathsfbi{R}^{-1}\mathsfbi{F}^{-1}\right) 
\end{equation}
and $\sigma_{\min}$ denotes the smallest singular value of the operator $\mathsfbi{F}\mathsfbi{R}^{-1}\mathsfbi{F}^{-1}$. Here, we have chosen to focus on wall speeds $\xi\geq\xi_f$ and the associated asymptotic parameters $\left(\alpha_f, \beta_f, Re_f\right)$, Section \ref{ssec:wavenumber_scaling}. Within the paradigm of the $\epsilon$-pseudospectra, the interpretation of these plots is as follows. Since the properties of the $2$-norm imply
\begin{equation}
    \mathcal{R} = \sigma_{\max}\left(\mathsfbi{F}\mathsfbi{R}\mathsfbi{F}^{-1}\right) = \sigma_{\min}^{-1}\left(\mathsfbi{F}\mathsfbi{R}^{-1}\mathsfbi{F}^{-1}\right) = \left\lVert \mathsfbi{R}^{-1}\right\rVert_{-E}^{-1},
\end{equation}
the set $\Lambda_\epsilon$ in Equation (\ref{eqn:pseudospectra}) admits the alternative definition
\begin{equation}
    \Lambda_\epsilon = \left\{\zeta\in\mathbb{C}\colon\left\lVert \mathsfbi{R}^{-1}\right\rVert_{-E} \leq \epsilon\right\}
\end{equation}
Thus, within the level curve $\left\lVert \mathsfbi{R}^{-1}\right\rVert_{-E}=\epsilon$, $O\left(\epsilon^{-1}\right)$ amplification can be realized. Additionally, the extent to which these contours protrude into the upper half-plane can be connected to the potential for transient energy amplification. In particular, \cite{reddyschmidhenn} showed that such growth cannot occur (that is, $G\leq 1$) if and only if $\beta_\epsilon\leq \epsilon$ for all $\epsilon\geq 0$, where
\begin{equation}
\label{eqn:beta_eps}
    \beta_\epsilon = \sup_{\zeta\in\Lambda_\epsilon\left(\mathsfbi{S}\right)}\Im\left(\zeta\right)
\end{equation}
The significance of the restriction on $\beta_\epsilon$ lies in noting that for a normal operator, the 2-norm $\epsilon$-pseudospectra comprises closed balls of radius $\epsilon$ centered around the eigenvalues \citep{TrefethenEmbree}. Therefore, for a linearly (not necessarily asymptotically) stable normal operator, $\beta_\epsilon$ reaches its maximum at $\beta_\epsilon = \epsilon$, specifically for a marginally stable mode. Returning to Figure \ref{fig:pseudo_spec}, we observe, for all wall speeds, pseudo-resonance down to $\epsilon\approx 10^{-7}$. Increasing $\xi$ appears to expand and widen the pseudo-spectral contours, though their penetration beyond $c_i = 0$ remains essentially unaffected. Nonetheless, level curves admitting even stronger amplification emerge, up to $\epsilon\approx10^{-11}$, indicating that despite the convergence of modal characteristics in this regime, an increase in $\xi$ is still capable of enhancing non-modal mechanisms, given, of course, the appropriate forcing.

\begin{figure}
    \centering
    \includegraphics[width=\textwidth]{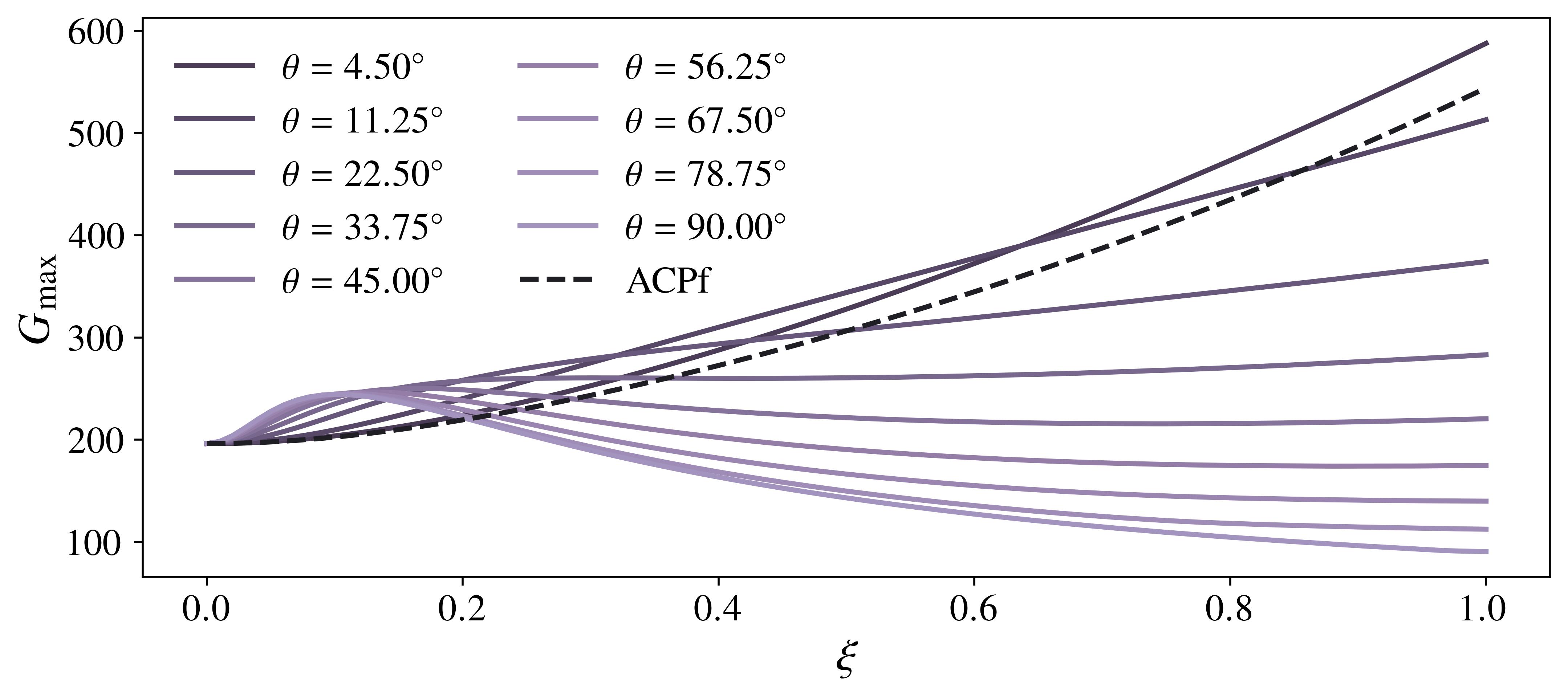}
    \caption{For various $\theta$, curves of $G_{\max}$, the largest possible energy gain exhibited by OCPfs across time and wavenumber space. A black dashed line indicates the equivalent plot for the aligned case, $\theta= 0$. In general, the largest amplification is realized for small but non-zero angles, peaking at $\theta=4.5^\circ$ for most $\xi$. A greater degree of skewness in the flow tends to suppress the amplification, particularly for modest to large wall speeds.}
    \label{fig:gmax_all}
\end{figure}

We now turn our attention to the dynamics of the unforced initial value problem, Equation (\ref{eqn:ivp}). In particular, we are interested in $G_{\max}$, defined for an OCPf configuration $\left(\xi, \theta\right)$ as the maximal amplification in time and wavenumber space
\begin{equation}
    G_{\max}\left(Re, \xi, \theta\right) = \max_{\alpha, \beta, t} G\left(\alpha, \beta, Re, \xi, \theta, t\right)
\end{equation}
Figure \ref{fig:gmax_all} outlines the findings of a large parameter sweep for $G_{\max}$ at $Re=1000$, slightly above the range for transition in aligned Couette-Poiseuille flow as quoted, for example, by \cite{Tsanis_Leutheusser_1988} and \cite{klotz} (note, however, that their Reynolds numbers are based on the wall velocity). As in Section \ref{sec:modal_analysis_section}, all results are presented relative to those of ACPf, $\theta = 0$, which experiences a monotonic increase in $G_{\max}$ with $\xi$. The introduction of a weak misalignment maintains this trend, but allows for greater amplification throughout the full range of wall speeds explored here. This effect was determined to be most pronounced at $\theta\approx4.5^\circ$. At even larger shear angles, two different regimes can be identified in $\xi$. In particular, while $G_{\max}$ continues to grow with $\theta$ (albeit slowly) for $0<\xi\lessapprox0.1$, it tends to decrease quite rapidly for $\xi\gtrapprox0.2$. Furthermore, at least for wall speeds in this range, no asymptotic behavior for $G_{\max}$ was resolved, which is in sharp contrast to our results on modal stability.

\begin{figure}
    \centering
    \includegraphics[width=\textwidth]{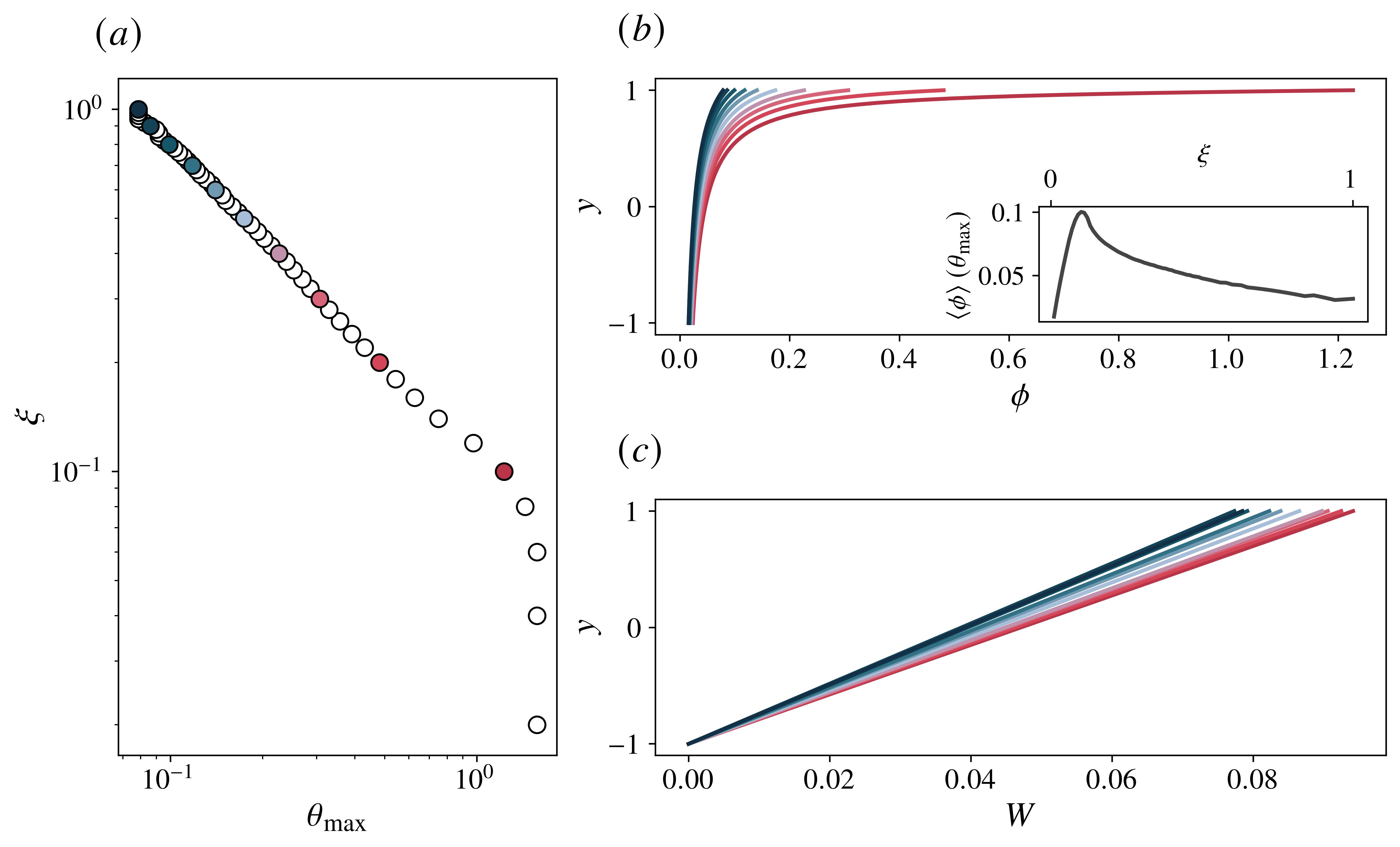}
    \caption{$(a)$, the variation in $\xi$ of $\theta_{\max}$, the shear angle that optimizes $G_{\max}$. Some combinations of $\left(\xi,\theta_{\max}\right)$ have been selected and the associated flow directions and crossflow profiles highlighted with the appropriate color in $(b)$ and $(c)$, respectively. Our conclusions are robust to the choice of these pairs. An inset in $(b)$ shows the $y$-averaged deviation $\left\langle\phi\right\rangle$ of the optimal net base flow from the streamwise direction, Equation (\ref{eqn:avg_skewness}).}
    \label{fig:optimal_angles}
\end{figure}

Interestingly, within the paradigm of transient growth, it is apparent that larger values of $\theta$ are typically the most ``stable", with $\theta = 90^\circ$ providing the strongest reduction in $G_{\max}$ for a wide range of wall speeds. The latter observation stands, of course, in strong opposition to the results presented in the previous section, particularly Equation (\ref{eqn:ppf_tuple}), which claims that a perfectly orthogonal OCPf configuration is, in fact, capable of minimizing $Re_c$ in the $\left(\xi,\theta\right)$-plane. An antagonistic effect, therefore, appears to be at play here, since, individually, both ACPf and the standard Couette flow support strong transient responses, yet for sufficiently skewed OCPfs, $G_{\max}$ can drop to as low as $46\%$ of the equivalent value for pPf at this Reynolds number ($G_{\max}\approx 196$). From a mathematical perspective, one can attribute this to the non-linearity of the operator norm or to the fact that, contrary to modal analysis, we are now investigating the full Orr-Sommerfeld-Squire system, for which $\xi_{\mathrm{eff}}$ no longer constitutes an informative parameter. Physically, however, an intriguing analogy can be drawn to fully turbulent three-dimensional boundary layers, for which increasing skewness, in the mean sense, is known to dampen the generation of Reynolds stresses and, therefore, the production of turbulent kinetic energy relative to the two-dimensional case \citep{eaton,flack,COLEMAN1996333, lozano-durán_giometto_park_moin_2020}. Although the physical mechanisms involved are not yet well understood, it is often believed that the addition of a mean spanwise strain detracts large momentum-carrying eddies from their optimal alignment \citep{van_den_berg_elsenaar_lindhout_wesseling_1975, bradshaw_pontikos_1985}. In the context of laminar OCPfs as treated here, one can partially quantify the existence of such an ideal configuration by considering $\theta_{\max}$
\begin{equation}
    \theta_{\max}\left(Re, \xi\right) = \{\theta^\prime\mid G_{\max}\left(Re, \xi, \theta^\prime\right) = \max_{\theta}G_{\max}\left(Re, \xi, \theta\right)\}
\end{equation}
which, at a given $\xi$, represents the angle of wall motion that achieves the most vigorous non-modal amplification. Figure \ref{fig:optimal_angles} $(a)$ highlights that $\theta_{\max}$ decays primarily as a power law. More importantly, as shown in Figures \ref{fig:optimal_angles} $(b)$ and $(c)$, the crossflow component $W$ associated with $\theta_{\max}$ is quite weak, allowing the flow direction $\phi$ to collapse throughout most of the channel and experience rapid variation only near the upper wall. To further visualize this, an average skewness $\left\langle \phi\right\rangle$, defined as
\begin{equation}
\label{eqn:avg_skewness}
    \left\langle \phi\right\rangle = \int_{-1}^1 \phi\left(y\right)\,\mathrm{d}y
\end{equation}
is also plotted in the inset of Figure \ref{fig:optimal_angles} $(b)$ and remains small ($\lessapprox 6^\circ$ at best) for all $\xi$. Thus, the optimal configuration for energy growth in OCPfs appears to be an approximately collateral boundary layer, with a flow direction roughly constant in $y$ and almost aligned with the streamwise axis.

\begin{figure}
    \centering
    \includegraphics[width=\textwidth]{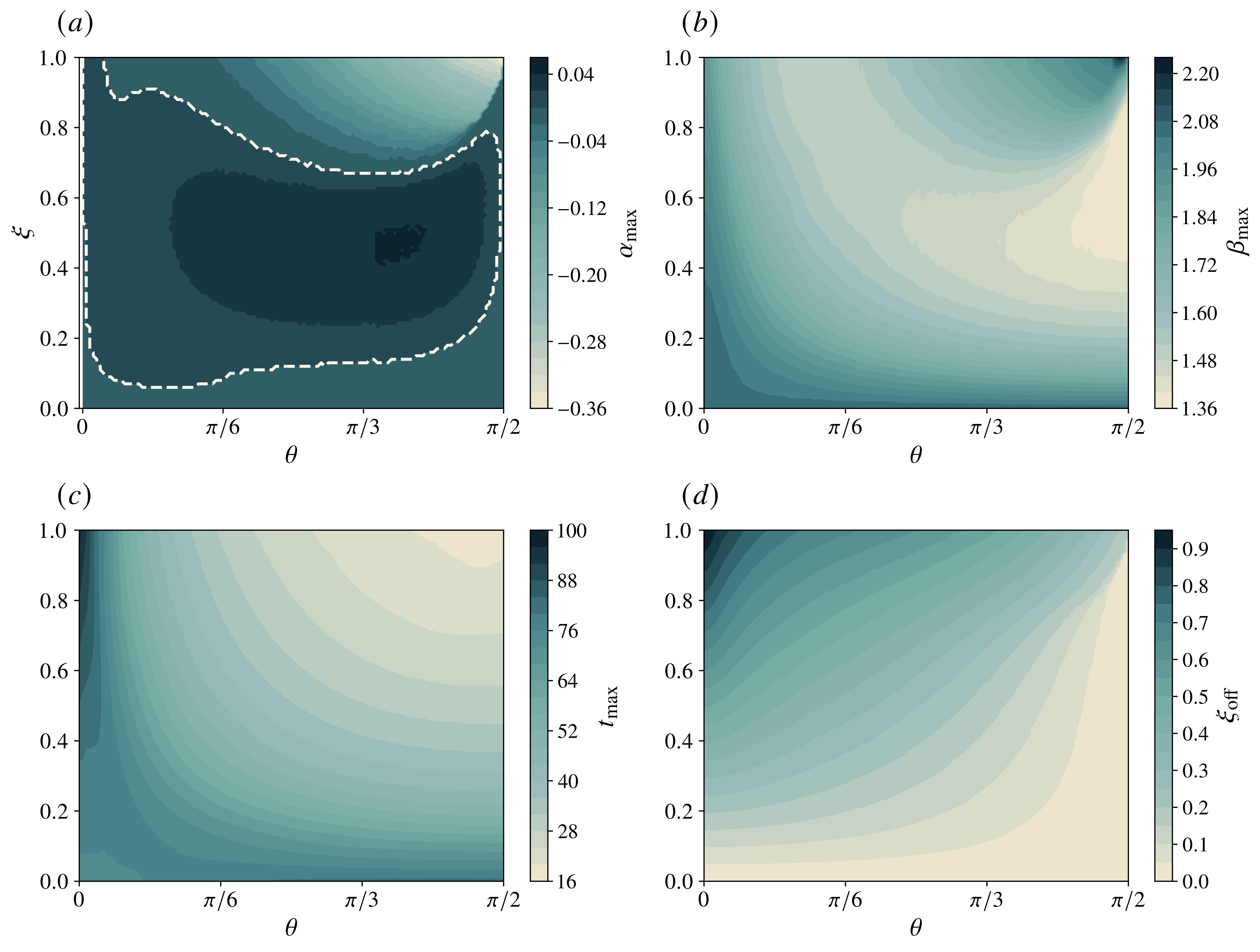}
    \caption{The variation in $\left(\xi, \theta\right)$-space of $(a)$, the streamwise wavenumber $\alpha_{\max}$, $(b)$, the spanwise wavenumber $\beta_{\max}$, and $(c)$, the time $t_{\max}$ at which the maximum energy amplification $G_{\max}$ is attained. In the case of $\alpha_{\max}$, a dashed line indicates the level curve $\alpha_{\max} = 0$. $(d)$ depicts contours of $\xi_{\mathrm{off}}$, which seem to be positively correlated through the non-normality in the OSS operator to $G_{\max}$.} 
    \label{fig:gmax_kx_kz_t}
\end{figure}

In Figure \ref{fig:gmax_kx_kz_t}, we present the contours of the wavenumbers and the time $t$ that achieve the maximum amplification $G_{\max}$. For purely streamwise flows, this optimal gain is generally observed for longitudinal modes, $\alpha_{\max} = 0$ \citep{tref_pseudospec, schmidstability}. However, this may or may not be the case for three-dimensional flows. Indeed, for OCPfs, we found the maximum amplification to occur for small but often non-zero streamwise wavenumbers, with little overall variation in the $\left(\xi, \theta\right)$-space. In some cases, streamwise-invariant disturbances remained optimal, but these were the exception rather than the rule. On the contrary, the optimal spanwise wavenumbers fluctuated more strongly, varying from $\beta_{\max} \approx 2.05$ to $\beta_{\max}\approx 2.25$ and even down to $\beta_{\max}\approx 1.3$. A rudimentary explanation for these trends can be obtained by recalling that the non-normality of the OSS operator $\mathsfbi{S}$ is tied to the mean shear(s) coupling the velocity perturbations in the Squire equation through the operator $i\beta\mathcal{D} U - i\alpha \mathcal{D}W$; see Equations (\ref{eqn:sq}) and (\ref{eqn:orr_somm}). Thus, in the same spirit as the effective wall speed $\xi_{\mathrm{eff}}$, one can attempt to split this operator into its ``Poiseuille" and ``Couette" constituents
\begin{equation}
\label{eqn:xi_off}
    i\beta\mathcal{D} U - i\alpha \mathcal{D}W = i\left(-2\beta y + \xi_\mathrm{off}\right)
\end{equation}
where we have defined 
\begin{equation}
    \xi_{\mathrm{off}} = \dfrac{\xi}{2}\left(\beta\cos\theta - \alpha\sin\theta\right)
\end{equation}
Since the Poiseuille contribution in Equation (\ref{eqn:xi_off}) is agnostic to the wall motion and does not, at any rate, favor either half of the channel, maximizing the effective mean shear is equivalent to maximizing $\left|\xi_{\mathrm{off}}\right|$. Then, assuming $\beta>0$, smaller shear angles would be biased toward larger spanwise wavenumbers because $\sin\theta\approx0$. On the contrary, for larger $\theta$, $\cos\theta$ is small and $\alpha$ should ideally become increasingly negative, though Figure \ref{fig:gmax_kx_kz_t} $(a)$ shows that this preference seems to emerge only at higher wall speeds. In Figure \ref{fig:gmax_kx_kz_t} $(d)$, we see that $\xi_{\mathrm{off}}$ somewhat emulates the changes in $G_{\max}$ at this Reynolds number. For example, while its maximum in the $\left(\xi, \theta\right)$-plane occurs in the purely parallel, high-$\xi$, limit, its minimum is realized as $\theta\to 90^\circ$. Of course, however, this correlation is bound to be imperfect, particularly because $\xi_{\mathrm{off}}$, in this case, cannot be a definitive statistic. The nature of the remaining blocks in $\mathsfbi{S}$ is equally important and does not admit a simple interpretation, calling for a more complex analysis that is outside the scope of this article. Finally, the time $t_{\max}$ taken to achieve the maximum growth appears to be the longest when an OCPf is weakly oblique and generally decreases in the direction of increasing $\xi$ and $\theta$. Note that, for the latter combination of flow parameters, $G_{\max}$ also tends to a minimum, indicating that transient phenomena for these configurations operate on shorter time scales and are likely suppressed by the viscosity before reaching sufficient amplitudes to trigger further instability. We remark in passing that while a search was conducted for $\theta\in\left(90^\circ, 180^\circ\right]$, $G_{\max}$ and all associated optimal parameters were found to be symmetric around $\theta=90^\circ$. An exception to this was $\alpha_{\max}$, which was determined to be anti-symmetric, a result that is not suggested by any operator-level symmetries.

\begin{figure}
    \centering
    \includegraphics[width=\textwidth]{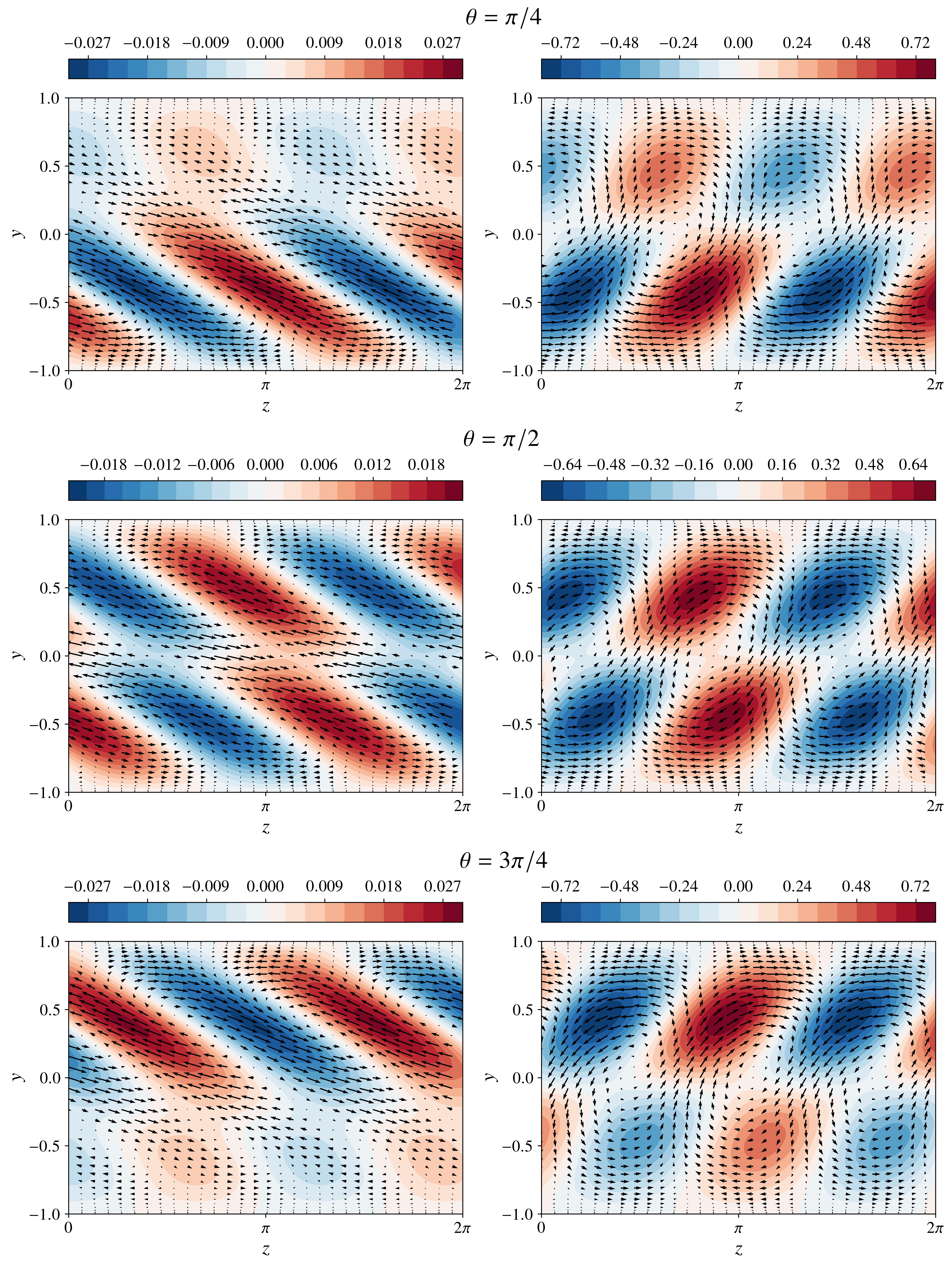}
    \caption{The optimal initial condition (left) and response (right) pair for representative values of $\theta$ at $\xi = 0.25$. Color denotes $u$ and the quiver arrows denote the cross-stream velocity perturbations, $v$ and $w$.}
    \label{fig:optimal_in_and_out}
\end{figure}

Figure \ref{fig:optimal_in_and_out} concludes this section by presenting for various $\theta$ the initial condition and response pair associated with $G_{\max}$. Respectively, these are the first right and left singular functions of the state transition operator $\Phi\left(t,0\right)$. In two-dimensional flows, the optimal initial field, typically observed for $\alpha = 0$, is characterized by weak counter-rotating streamwise vortices that evolve through a redistribution of horizontal momentum by the normal velocity fluctuations to form high-energy streaks at $t = t_{\max}$. Originally proposed in the works of \cite{ellingsenpalm} and \cite{landahl1, landahl2}, this process is commonly referred to as the lift-up effect, in which a linear amplification in time proportional to the streamwise shear can be achieved for a streamwise-independent disturbance, at least in the inviscid limit. In the viscous alternative, this growth would continue to persist, but only to the leading order before decaying due to viscosity \citep{brandt_lift_up}. Although the three-dimensionality of our flow introduces additional nuance, \cite{ellingsenpalm} had suggested that the lift-up process could remain viable even in skewed boundary layers, arguing, however, that streak growth would substantially decrease, particularly so in the case of oblique Couette-Poiseuille flows because the streamwise shear itself decreases as $\theta\to \pi/2$. Indeed, \cite{corbett_bottaro_2001}, for example, found that streamwise streaks developing via lift-up remained the optimal disturbance in Falkner-Skan-Cooke boundary layers. 

Therefore, it is not surprising that the optimal initial conditions in Figure \ref{fig:optimal_in_and_out} comprise weak streamwise vortices whose amplification at the optimal time decreases in response to an increase in flow obliqueness (note that $\alpha_{\max}\approx0$ for the $\xi$ considered here, Figure \ref{fig:gmax_kx_kz_t}, although our results remained unchanged even when this was not the case). However, as captured by both \cite{Blesbois_Chernyshenko_Touber_Leschziner_2013} and \cite{Hack_Zaki_2015} for their respective base flows, these vortices also initially oppose and eventually tilt in the direction of the spanwise shear, reminiscent of the inviscid down-gradient Reynolds stress mechanism proposed by Orr \citep{orr1907, butler_3d_opt_pert}. Therefore, we conclude that at least up to the stage of primary instability, the route to transition for oblique Couette-Poiseuille flows is likely dominated by a lift-up process induced by an Orr-type mechanism. Furthermore, since the Orr mechanism is enhanced by the presence of increasing (spanwise) shear, it is likely that the trends in $G_{\max}$ observed in Figure \ref{fig:gmax_all} can be attributed to a decrease in the effectiveness of the lift-up process.

\section{Conclusion}\label{sec:conclusion}

We performed a comprehensive modal and non-modal stability analysis in oblique Couette-Poiseuille flows, which are described by a wall motion at an angle $\theta$ to the pressure gradient. These are generalizations of the traditional aligned case and, to the best of our knowledge, have not received prior attention in the stability literature. 

We derive the corresponding Orr-Sommerfeld-Squire system, identifying by a simple analogy an effective wall speed that completely characterizes modal solutions. A large-scale numerical sweep reveals that, in general, a misalignment between the pressure gradient and the wall velocity is destabilizing, at least relative to the aligned case. Considerations of symmetry and periodicity allow for a restriction of the parameter space to $\theta\in\left[0^\circ, 90^\circ\right]$ and, in this range, two regimes are identified. For shear angles $0^\circ<\theta\lessapprox20^\circ$, almost all stability features in OCPfs continue from the aligned case. On the other hand, the range $20^\circ\lessapprox\theta\leq90^\circ$ demonstrates sharper differences, including, in particular, the lack of a trademark inflection point in the $Re_c$-curves as observed for ACPf. The linear energy budget and the movement of the critical layers generated by the effective velocity profile seem to confirm these trends. Modal instability is optimized by the perfectly orthogonal configuration, $\theta = 90^\circ$, which exhibits a constant critical tuple for all $\xi$. For all $\theta\neq 0$, we find that unstable modes persist throughout $\xi\in\left[0, 1\right]$, notably distinct from ACPf, where $\xi \approx 0.7$ marks the transition to a regime of unconditional linear stability. This behavior is accompanied by a convergence of the critical parameters starting at the threshold $\xi = \xi_f$, which appears to decrease with $\theta$. A simple theoretical analysis explains the latter phenomenon and derives the exact asymptotic values of the critical parameters in the limiting regime. Separately, a modified long-wavelength analysis is used to confirm that non-trivial OCPfs have no cutoff wall speeds and are always linearly unstable.

On the topic of non-modal disturbances, OCPfs, through the non-normality endowed to the OSS operator by the streamwise and crossflow shears, exhibit trends that conflict with those commonly quoted for two-dimensional flows. For example, the resolvent norm is not necessarily maximized for disturbances with $\alpha = 0$. Meanwhile, the $\epsilon$-pseudospectra reveal that even if modal stability converges beyond $\xi = \xi_f$, non-modal mechanisms might continue to be amplified by changes in the wall speed. Finally, considering the unforced initial value problem, the maximum energy amplification $G_{\max}$ appears to decrease strongly with the skewness of the base profiles, implying that the imposition of three-dimensionality is generally detrimental to energy growth. This is reminiscent of fully turbulent three-dimensional boundary layers, where increased skewness is known to suppress turbulent energy production. Note that while \cite{Hack_Zaki_2015} observed similarly declining energy gains for their Blasius-Stokes flow, \cite{corbett_bottaro_2001} calculated a stronger transient growth relative to the two-dimensional case in their study on swept boundary layers, suggesting a strong dependence on the particular mechanism that enforces the skewness in the base flow. At all wall speeds, however, the configuration that optimizes energy amplification appears to be a weakly three-dimensional collateral boundary layer. Finally, the most energetic initial perturbations seem to develop via a lift-up process enhanced by an Orr-like mechanism driven by the spanwise shear, the latter being absent in streak amplification for two-dimensional flows.


A natural extension of this article seems to be through an investigation of oblique Couette-Poiseuille flows in the turbulent regime, which is perhaps where a majority of practical applications reside. Although there have been some previous studies, they have primarily focused on the perfectly aligned case \citep{KIM2018288, KIM2020108687, Cheng_Pullin_Samtaney_Luo_2023} or the perfectly orthogonal case \citep{COLEMAN1996333, sandham, LE2000480, kannepalli_piomelli_2000}, with little or no attention devoted to intermediate $\theta$. Some initial work on the latter configurations has been conducted, for example, \cite{aps_ocpf}, but additional effort is needed and will likely contribute well to our overall understanding of the physics in three-dimensional boundary layers.


\backsection[Acknowledgements]{We would like to acknowledge high-performance computing time on Leavitt, Bates College, and Anvil, Purdue University, the latter obtained via the Advanced Cyberinfrastructure Coordination Ecosystem: Services \& Support (ACCESS) allocation MCH230042.}

\backsection[Funding]{This research was sponsored by a University of Pennsylvania faculty startup grant.}

\backsection[Declaration of interests]{The authors report no conflict of interest.}


\backsection[Author ORCIDs]{George Ilhwan Park, https://orcid.org/0000-0002-8242-8515; Muhammad Abdullah, https://orcid.org/0000-0001-9338-2631.}


\appendix\section{Numerical Methods and Validation}\label{appendix:appA}

To discretize the Orr-Sommerfeld-Squire system, a Chebyshev pseudo-spectral method was written in \texttt{Python}. The clamped boundary conditions were incorporated as discussed in \cite{tref_spec_methods}. An initial convergence check allowed us to choose $N = 128$ Gauss-Lobatto points for collocation, resulting in a $\left(2N+2\right)\times\left(2N+2\right)$ matrix problem. We found this to be sufficient to achieve precision up to eight decimal places. To efficiently traverse the large parameter space, we scaled to an embarrassingly parallel workload on many CPUs using the open-source \texttt{Python} module \texttt{Ray} \citep{ray_paper}.

Modal solutions were computed by solving the generalized eigenvalue problem using the LAPACK wrapper in \texttt{SciPy}. For the singular value decomposition, we used a built-in sparse solver based on the implicitly restarted Arnoldi method \citep{arnoldi}. We remark that this choice was motivated not by the size of the matrices being created, which is rather small and enables reasonably fast dense solutions, but by the number of gridpoints being investigated. The $\epsilon$-pseudospectra, specifically, were created using Eigentools \citep{eigentools}, a high-level eigenvalue module that implements the economy method of \cite{embreekeller}. Eigentools is wrapped over Dedalus, a general-purpose sparse spectral solver capable of handling nearly arbitrary partial differential systems and boundary conditions \citep{dedalus}.

\subsection{Aligned Couette-Poiseuille flow}
\begin{figure}
    \centering
    \includegraphics[width=\textwidth]{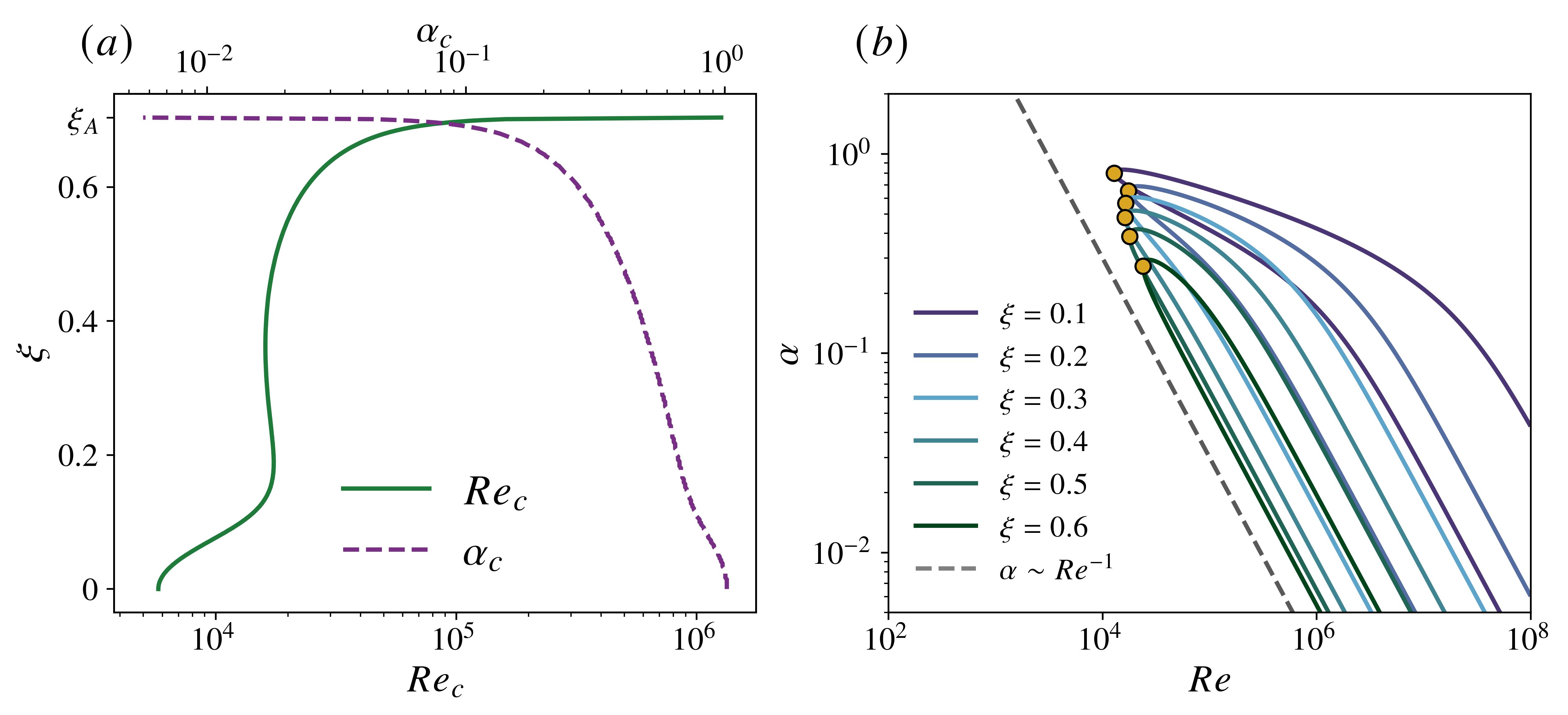}
    \caption{$(a)$, the variation of $\alpha_c$ and $Re_c$ with the wall speed $\xi$. Note where either curve suddenly ``disappears"; this marks $\xi_A$, the cutoff wall speed beyond which the flow is always asymptotically stable. $(b)$, the neutral curves in the $\left(\alpha,Re\right)$-plane for ACPf at $\xi = 0.1$ to $\xi=0.6$ in increments of 0.1. In each case, the critical pair is denoted by a circle. The dashed line here represents the distinguished limit $\alpha\sim Re^{-1}$.}
    \label{fig:acp_crit_params}
\end{figure}

In the limit $\theta\to 0$, the wall movement becomes perfectly parallel to the pressure gradient and the \textit{aligned} Couette-Poiseuille flow is recovered
\begin{equation}
    U\left(y\right) = 1-y^2 + \dfrac{\xi}{2}\left(1+y\right)\qquad W\left(y\right) = 0
\end{equation}
This is a classic base flow that has received numerous treatments in the literature. \cite{potter_1966} was the first to determine that the imposition of a wall shear is generally stabilizing. He found that while the critical Reynolds number $Re_c$ increased, albeit non-monotonically, the critical streamwise wavenumber $\alpha_c$ approached zero (note that Squire's Theorem allows $\beta_c = 0$). Above a threshold value, $\xi_A$, of the non-dimensional wall speed, this stabilization was found to be unconditional. Figure \ref{fig:acp_crit_params} $(a)$ summarizes the stability of ACPf and is consistent with the findings of, for example, \cite{potter_1966, nouar, pof_pc_budget}. Initially, $Re_c$ increases before experiencing an inflection point between $0.2\lessapprox\xi\lessapprox0.4$, and then continues to increase until $\xi_A\approx 0.70370$ -- see Section \ref{ssec:long_wave}. The movement of the neutral curves in the $\left(\alpha, Re\right)$-plane with wall speed is illustrated in Figure \ref{fig:acp_crit_params} $(b)$. As described in \cite{cowley_smith_1985}, their upper and lower branches scale as $\alpha\sim Re^{-1}$ as $Re\to\infty$, which is instrumental in formulating the long-wavelength Orr-Sommerfeld problem. The most unstable eigenmode is the usual Tollmien-Schlichting wave, streamwise-propagating and uniform in $z$.

\begin{figure}
    \centering
    \includegraphics[width=\textwidth]{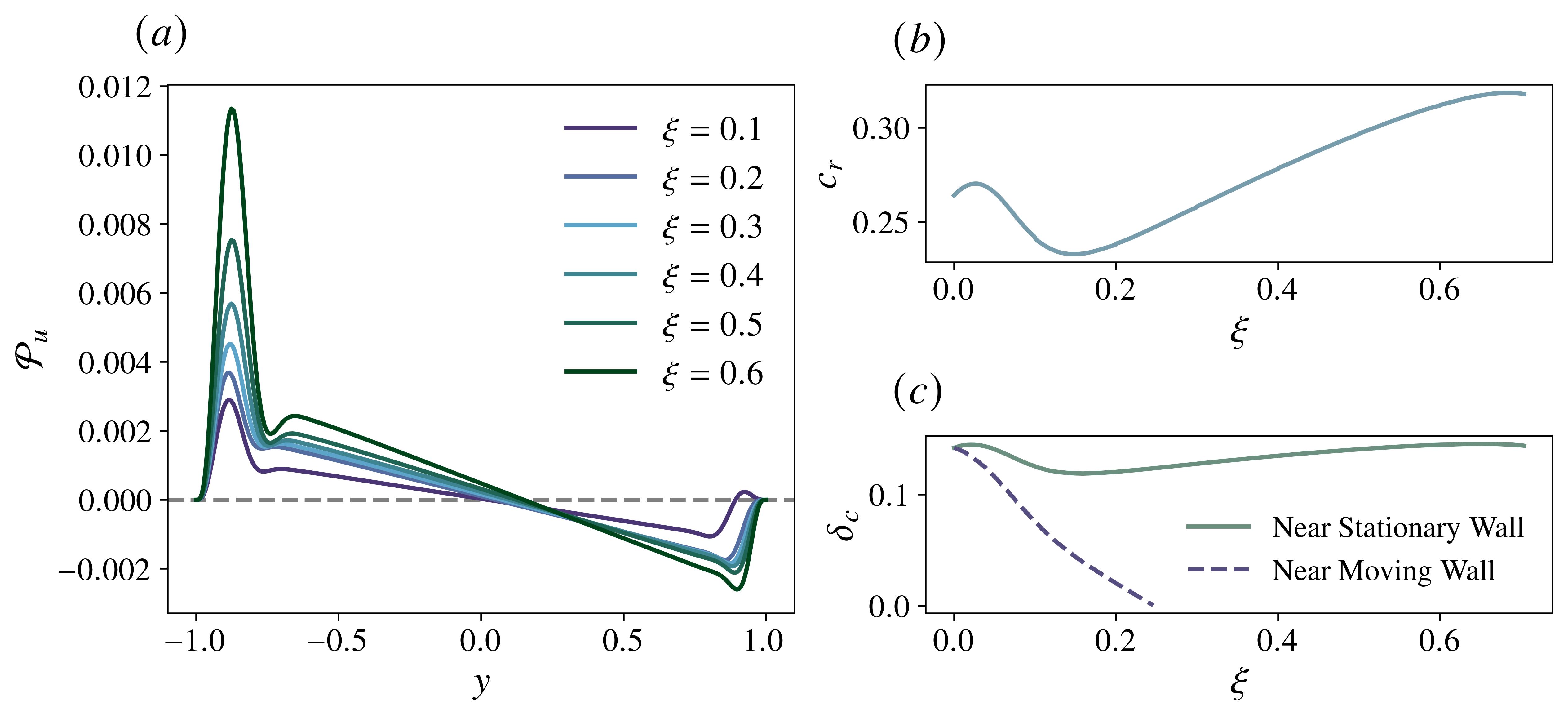}
    \caption{$(a)$, the energy production $\mathcal{P}_u$ against the mean streamwise shear; as $\xi$ increases, a large region of negative production appears near the upper wall. $(b$--$c)$, respectively, the real part of the $x$-phase speed $c$ at criticality and the thicknesses of the critical layers in each half of the channel. The stationary wall is, of course, the lower one.}
    \label{fig:acp_crit_cr_and_cl_thickness}
\end{figure}

Figure \ref{fig:acp_crit_cr_and_cl_thickness} $(a)$ presents an expedited analysis of the linear energetics at criticality for ACPf. The inflectional region in the $Re_c$-curve is typically discussed in the context of the thicknesses $\delta_c$ of the critical layers, the wall-normal location(s) $y_c$ such that $U\left(y_c\right) = c_r$, where energy production is often localized. In principle, the quadratic nature of $U$ allows the existence of two such critical points, associated, respectively, with each wall. Referencing Figures \ref{fig:acp_crit_cr_and_cl_thickness} $(b)$ and $(c)$, we see that, initially, as $\xi$ increases, $\delta_c$ decreases near both walls, supporting destabilization. Eventually, the critical layer near the moving wall vanishes completely due to the asymmetry of $U$ in the upper half of the channel at higher values of $\xi$. Simultaneously, in the same region, $\tau_u$ (note that $\tau_w = 0$) becomes increasingly negative, and since $\mathcal{D}U < 0$ here as well, energy is extracted from the perturbation field back to the base flow. Furthermore, the critical layer near the lower fixed wall gradually thickens, which has a stabilizing effect \citep{potter_1966, GUHA_2010}. Viscous dissipation (not shown here), while confined primarily to thin layers near each wall, appears to increase as well.

\section{Summarizing Modal Stability in Oblique Couette-Poiseuille flows}\label{appendix:appB}

\begin{figure}
    \centering
    \includegraphics[width=\textwidth]{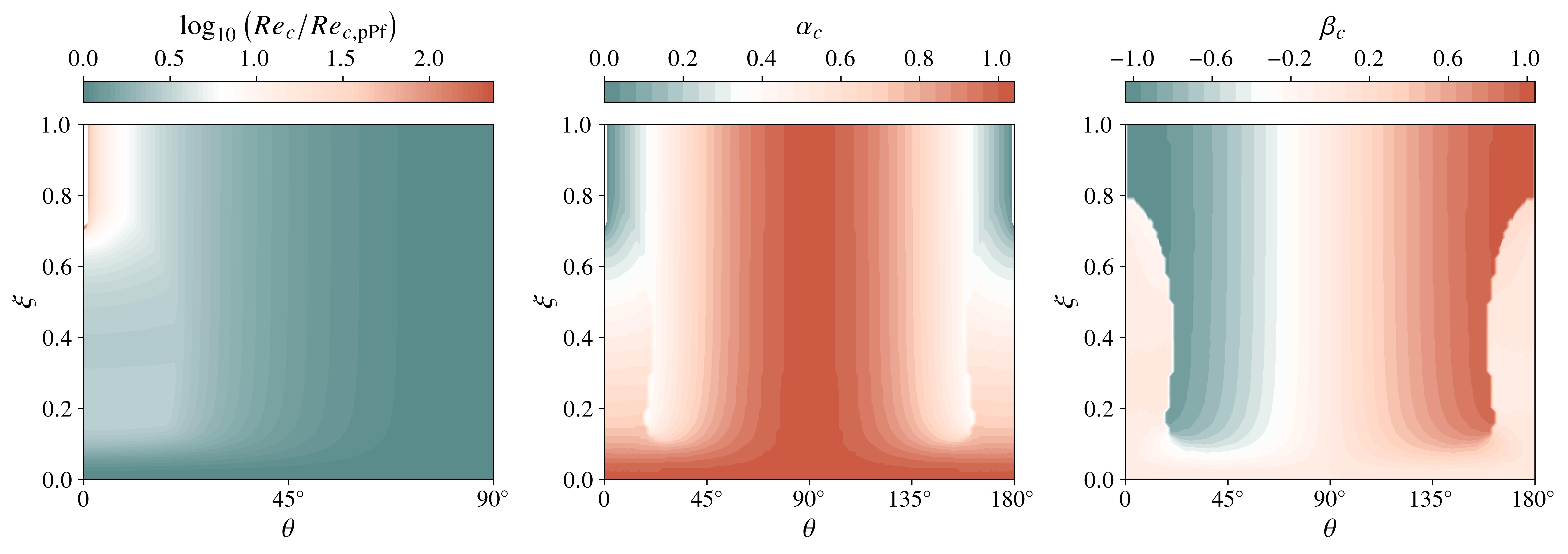}
    \caption{Plots of $\left(\alpha_c, \beta_c, Re_c\right)$ in the $\left(\xi, \theta\right)$-plane. For visual clarity in the contours of the critical Reynolds number, we have restricted the $\theta$-axis to $\theta\in\left[0, 90^\circ\right]$ and opted for a logarithmic scale normalized by $Re_{c,\mathrm{pPf}} \approx 5773.72$, the equivalent threshold for linear instability in plane-Poiseuille flow.}
    \label{fig:full_contour_plots_crit_params}
\end{figure}

In Figure \ref{fig:full_contour_plots_crit_params}, we provide complete data on the critical flow parameters $\left(\alpha_c, \beta_c, Re_c\right)$ in the $\left(\xi, \theta\right)$-plane for OCPfs. As predicted in Section \ref{ssec:criticality}, $\beta_c$ is anti-symmetric about $\theta=\upi/2$, whereas both $\alpha_c$ and $Re_c$ are symmetric. Stability is maximized, at least in the sense of the critical Reynolds number, when $\theta\to 0$. On the other hand, oblique Couette-Poiseuille flows are the most unstable either when $\xi \to 0$ (pPf) or when $\theta\to\upi/2$.

\begin{figure}
    \centering
    \includegraphics[width=\textwidth]{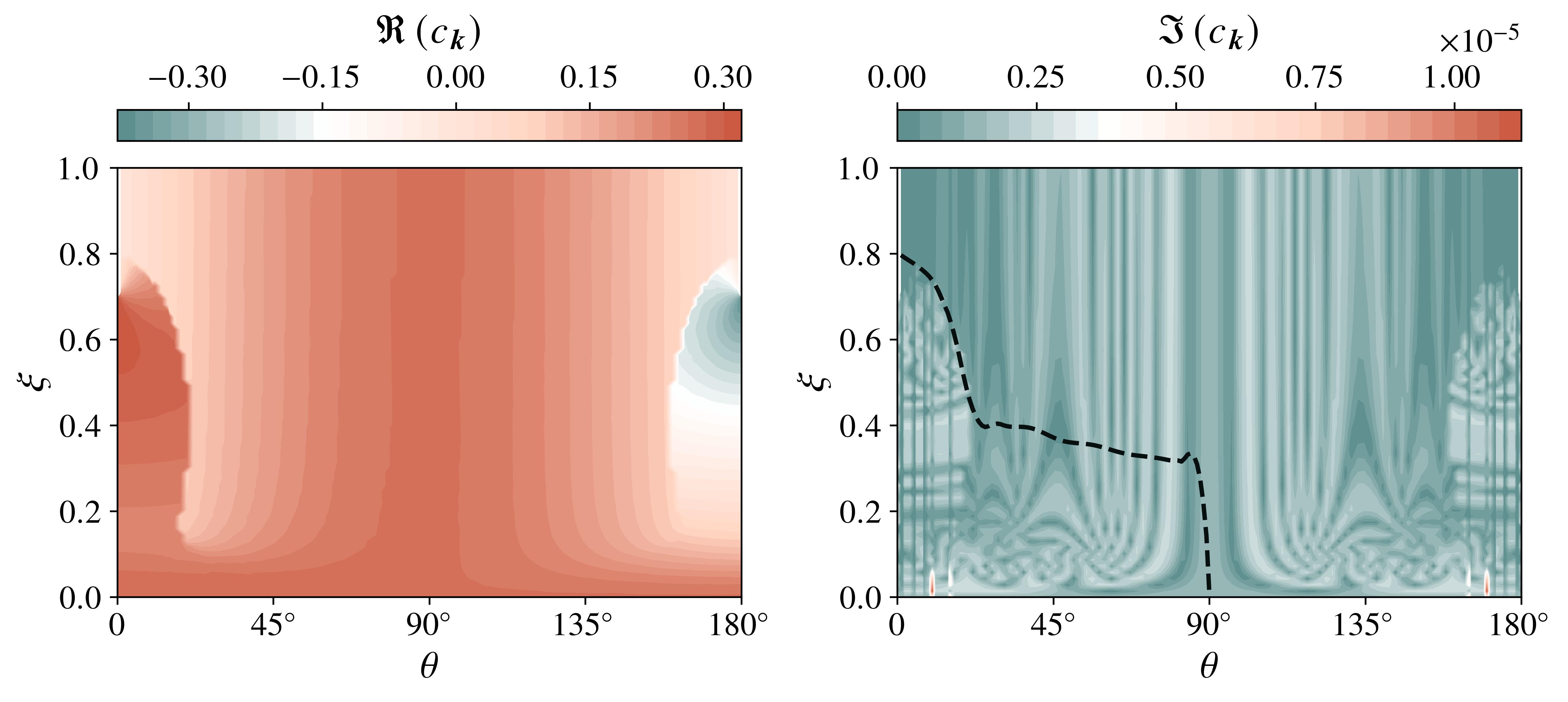}
    \caption{Contours of the real and imaginary components of the $\boldsymbol{k}$-phase speed $c_{\boldsymbol{k}}$; the dashed line indicates $\xi_f$, the wall speed initiating the asymptotic regime. In general, the growth rates are feeble, and while it might not be immediately evident from the modified definition of the phase speed, the most unstable eigenmode is indeed a wall mode ($A$-branch) as for ACPf.}
    \label{fig:full_contour_plots_grs_and_phs}
\end{figure}

Figure \ref{fig:full_contour_plots_grs_and_phs} highlights the variation of the most unstable eigenvalue at criticality. Here, $c_{\boldsymbol{k}}$ denotes the complex $\boldsymbol{k}$-phase speed, defined by means of the following dispersion relation
\begin{equation}
    c_{\boldsymbol{k}}= \dfrac{\omega}{\left\lVert\boldsymbol{k}\right\rVert_2} = \dfrac{\omega}{\sqrt{\alpha^2+\beta^2}}
\end{equation}
where $\omega$ is the complex frequency and $\boldsymbol{k}$ is the wavenumber vector \citep{RUMPF2015213}. While its imaginary component $\Im\left(c_{\boldsymbol{k}}\right)$ represents a (scaled) exponential growth rate, the real part $\Re\left(c_{\boldsymbol{k}}\right)$ characterizes the wave propagation speed in the direction of $\boldsymbol{k}$. Note that this quantity is a generalization of the $x$-phase speed $c = \omega/\alpha$ that occurs somewhat organically in the Orr-Sommerfeld problem and that is typically adopted for the study of two-dimensional flows. Indeed, under Squire's Theorem, $\left\lVert\boldsymbol{k}\right\rVert_2 = \alpha$ and $c_{\boldsymbol{k}} = c$. For mean three-dimensional flows with non-trivial spanwise wavenumbers, the $\boldsymbol{k}$-phase speed is more physically informative and capable of providing a better collapse, since the most unstable waves are generally oblique. Once again, we see that $\Im\left(c_{\boldsymbol{k}}\right)$ is precisely symmetric around $\theta=\upi/2$, although this symmetry is broken for $\Re\left(c_{\boldsymbol{k}}\right)$. The latter is especially prominent for smaller angles, that is, $\Theta_1$ and $180^\circ - \Theta_1$, where the stability characteristics of OCPfs essentially continue from the aligned case, Section \ref{ssec:criticality}. Such asymmetries have also been captured for $c$ when comparing positive versus negative wall speeds in ACPf; see, for example, Figure 3 in \cite{pof_pc_budget}.

\section{The Behavior of the Asymptotic Critical Parameters for $\xi\leq \xi_f$}\label{appendix:appC}

\begin{figure}
    \centering
    \includegraphics[scale=0.5]{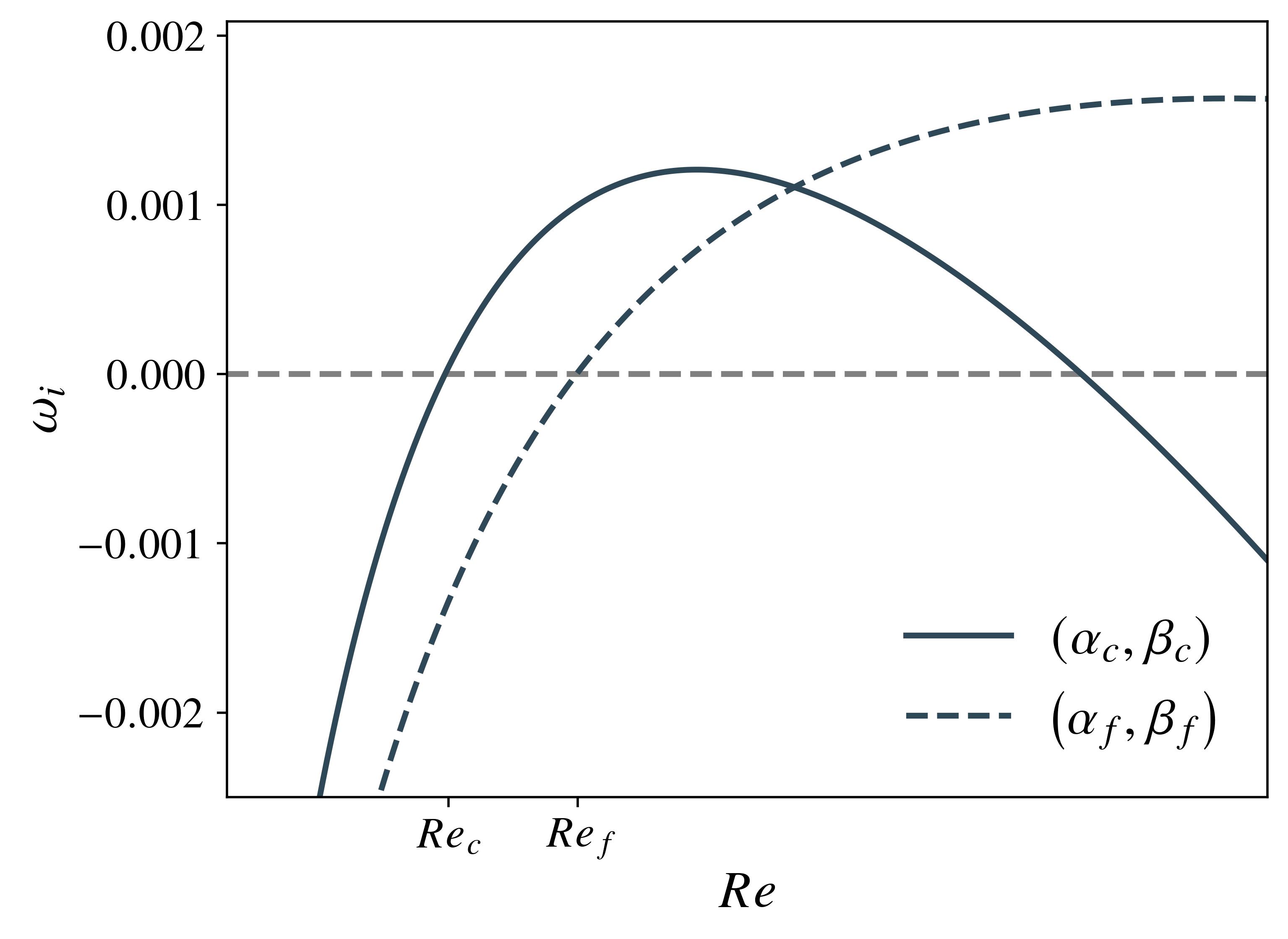}
    \caption{The growth rates of the most unstable mode for the critical versus asymptotic wavenumbers at $\theta= 25^\circ$ and $\xi = 0.1 <\xi_f$. The dashed line indicates the boundary $\omega_i =0$.}
    \label{fig:comparing_asymp_params_for_non_xi_f}
\end{figure}

In Section \ref{ssec:criticality}, it was observed that the critical triplet $\left(\alpha_c,\beta_c, Re_c\right)$ for $\xi\leq \xi_f$ may or may not coincide with the asymptotic values $\left(\alpha_f,\beta_f, Re_f\right)$ at a given $\theta$. Despite this, the conclusion of our long-wavelength analysis in Section \ref{ssec:long_wave} seems to suggest that the latter parameters, particularly the wavenumbers, could still be of relevance within this range of wall speeds. To investigate this, we set $\theta = 25^\circ$ and $\xi = 0.1 <\xi_f$, exploring in Figure \ref{fig:comparing_asymp_params_for_non_xi_f} the growth rates $\omega_i$ of the most unstable mode for the critical (that is, as predicted by a standard parameter sweep) versus the asymptotic pair of wavenumbers. An interesting behavior is captured; whereas the former pair becomes unstable, as expected, at $Re_c$, the latter also eventually destabilizes, doing so precisely when $Re = Re_f$. One can rationalize this by noting that $\left(\alpha_f,\beta_f\right)$ induces a vanishing $\xi_{\mathrm{eff}}$ in the Orr-Sommerfeld problem, ensuring that any dependence on the wall speed drops completely. Thus, the asymptotic wavenumber pair remains, in a sense, unaffected by changes in $\xi$, and eventually coincides with the critical pair beyond $\xi\geq\xi_f$, when all other disturbances stabilize.

\bibliographystyle{jfm}

\bibliography{jfm}


\end{document}